\documentclass[titlepage,12pt,twoside]{article}
\usepackage{amssymb,epsfig,pslatex}
\usepackage{cite}  
\usepackage{float}       
\usepackage{wrapfig}  
\usepackage{hhline} 
\usepackage{dcolumn} 
\restylefloat{figure} 
\usepackage{graphicx}

\makeatletter
\def\@sect#1#2#3#4#5#6[#7]#8{\ifnum #2>\c@secnumdepth
  \def\@svsec{}\else
  \refstepcounter{#1}\edef\@svsec{\csname the#1\endcsname.\hskip0.5em}\fi
  \@tempskipa #5\relax
  \ifdim \@tempskipa>\z@
    \begingroup
      #6\relax
      \@hangfrom{\hskip #3\relax\@svsec}{\interlinepenalty \@M #8\par}%
    \endgroup
    \csname #1mark\endcsname{#7}\addcontentsline
      {toc}{#1}{\ifnum #2>\c@secnumdepth \else
        \protect\numberline{\csname the#1\endcsname}\fi #7}%
  \else
    \def\@svsechd{#6\hskip #3\@svsec #8\csname #1mark\endcsname
      {#7}\addcontentsline{toc}{#1}{\ifnum #2>\c@secnumdepth \else
        \protect\numberline{\csname the#1\endcsname}\fi #7}}%
  \fi \@xsect{#5}}
\@addtoreset{equation}{section}
\makeatother
\renewcommand\thesection{\arabic{section}}
\renewcommand\theequation{\ifnum \value{section}>0
 \thesection.\arabic{equation}%
\else
\arabic{equation}%
\fi}

\def\shat{\hat{s}}

\def\ttbar{{t \bar t}}
\newcommand{\mtt}{M_{t \bar t}}
\newcommand{\be}{\begin{equation}}
\newcommand{\ee}{\end{equation}}
\newcommand{\bea}{\begin{eqnarray}}
\newcommand{\eea}{\end{eqnarray}}
\newcommand{\as}{\alpha_s}

\newcommand{\st}{{\mathbf S_t}}
\newcommand{\stb}{{\mathbf S_{\bar t}}}
\newcommand{\bfig}{\begin{figure}}
\newcommand{\efig}{\end{figure}}
\newcommand{\bc}{\begin{center}}
\newcommand{\ec}{\end{center}}

\def\bhline{\noalign{\ifnum0=`}\fi\hrule \@height  
\boldarrayrulewidth \futurelet \@tempa\@xhline}

\newcommand{\Or}{\mathord{\mathrm{O}}} %changed from \mathop 20/1/95

\newcommand{\rmi}{\mathrm{i}}
\newcommand{\rmd}{\mathrm{d}}
\newcommand{\rpv}{R\!\!\!\!/}
\renewcommand{\thefootnote}{\small\fnsymbol{footnote}}

\begin{document}
\begin{titlepage}
  \begin{flushright}
    PITHA 08/09  
  \end{flushright}        
\vspace{0.01cm}
% \vspace{2cm}
\begin{center}
{\LARGE {\bf Top quark physics at the LHC }}  \\
%:
\vspace{2cm}
{\large{\bf Werner Bernreuther\footnote{Email:
{\tt breuther@physik.rwth-aachen.de}}
}}
\par\vspace{1cm}
Institut f\"ur Theoretische Physik, RWTH Aachen University, 52056 Aachen, Germany\\
\par\vspace{1cm}
{\bf Abstract}\\
\parbox[t]{\textwidth}
{The physics perspectives of the production
and decay of single top quarks  and top quark pairs 
at the CERN Large Hadron Collider (LHC) are reviewed
from a phenomenological point of view.
}
\end{center}
\vspace*{2cm}

PACS number(s): 13.85.t, 13.90.i, 14.65.Ha 

\end{titlepage}
%
%\newpage
%
\setcounter{footnote}{0}
\renewcommand{\thefootnote}{\arabic{footnote}}
\setcounter{page}{1}
%%%%%%%%%%%%%%%%%%%%%%%%%%%%%%%%%%%%%%%%%%%%%%%%%

\section{Introduction}
\label{secintro}

The top quark, the heaviest known fundamental particle, was discovered
thirteen years ago \cite{Abe:1995hr,Abachi:1995iq} at the Tevatron 
proton antiproton collider. At a hadron collider like the Tevatron, top quarks are
predominantly produced together with their antiquarks. Quite recently, the D0  \cite{Abazov:2006gd}
and the CDF  \cite{CDFsinglet} experiment reported also evidence for the observation of
singly produced $t$ and $\bar t$ quarks. To date the Tevatron is the
only source of these quarks. However, this should change soon when the
CERN Large Hadron Collider (LHC) will start operation. Millions of top
 quarks will be produced already in the low luminosity phase $L\sim
 10\, {\rm fb}^{-1}$ of this collider -- see table~\ref{tab-topxsec}.
This will open up the possibility to explore this quark with hitherto
unprecedented precision.

What is interesting about the physics of top quarks? Although it is
almost as heavy as a gold atom it seems to behave as a pointlike particle,
at least at length scales $\gtrsim 10^{-18}$ m, according to
experimental findings. So far the Tevatron results are in accord with
expectations and predictions  within the standard
model (SM) of particle physics. While the mass of this particle has
already been precisely measured, other properties and its production
and decay dynamics could, however,  not be investigated in great detail  so far.
Hopefully this will
change in the years to come. There are exciting physics topics to be
explored -- we mention here only a few of them.
In view of its large mass the top quark is an excellent probe of the
mechanism that breaks the electroweak gauge symmetry and should
therefore play a key role in clarifying the nature of the 
force(s)/particle(s) responsible for this phenomenon. 
The top quark is also 
good probe for possible new parity-violating and/or non-SM $CP$ violating 
interactions which could be induced, for instance, by non-standard Higgs bosons.
Are there new top-quark decay modes, for instance to supersymmetric
particles? So far, experimental data are consistent with the SM
  prediction that $t \to W^+ b$ is the dominant mode --
but its branching ratio and the structure of the $tbW$ vertex is not yet
measured directly with high accuracy. 
Does the pointlike behaviour of the top quark continue once it can be probed
at distance scales significantly below $10^{-18}$ m? 
These and other topics, while being addressed also at the
Tevatron, will be in the focus of the future experiments at the LHC.

The top quark is unique among the known quarks in that it decays
before it can form hadronic bound states. This has important
consequences, as will be discussed in the following sections.
Above all, it offers the possibility to 
explore the  interactions  of a {\it bare} quark at energies of a few
hundred  GeV to several TeV. Furthermore,  it is an important asset of top
quark physics  that not only the effects of the electroweak interactions,
 but also of the strong  interactions of these particles 
can, in most situations,  be reliably predicted.
Needless to say,  this is  necessary
for the analysis and interpretation of
present and future experimental data.

There are already a number of excellent reviews of hadronic top-quark 
production and decay 
\cite{Beneke:2000hk,Chakraborty:2003iw,Wagner:2005jh,Quadt:2006jk}.
In this exposition the  top-quark physics perspectives at
the LHC are  discussed from a phenomenological point of view, of
course taking the results and insights gained at the Tevatron into
account. We shall first sketch the profile of this quark
 in section~\ref{secprofile}. In
 sections~\ref{sectdec},~\ref{secttbar},
 and~\ref{secsingletop} we discuss
top-quark decay,  $\ttbar$ pair production, and finally
single-top-quark production. In each section, we first review
the presently available standard model predictions and discuss then
possible new physics effects.
Moreover, experimental results from the Tevatron and 
measurement perspectives at the LHC will be briefly outlined. 
As usual in particle phenomenology, values of particles masses and decay
widths are given in natural units putting $\hbar = c =1$.

\begin{table}[h!]
\caption{\label{tab-topxsec} Upper part: number of $\ttbar$ events
produced at the Tevatron and expected $\ttbar$ production rates
at the LHC and at a future $e^+e^-$ linear collider (ILC), where $L$ is
the integrated luminosity of the respective collider in units of 
${\rm fb}^{-1}$.   Lower part: Number of $t$ and $\bar t$ events
at the Tevatron and expected number at the LHC produced in 
single top reactions.}
\begin{center}
\renewcommand{\arraystretch}{1.2}
\begin{tabular}{ccc}  
\hline
$\ttbar$ pairs  & dominant reaction & $N_{\ttbar}$  \\ 
\hline
Tevatron: $p{\bar p}$ (1.96 TeV)  & $q{\bar q} \to t{\bar t}$  & $\sim 7
\cdot 10^4 \times L$ \\ 
LHC: $pp$  (14 TeV) &  $ g g  \to t{\bar t}$  &  $\sim 9\cdot 10^5
\times L$ \\ 
ILC: $e^+ e^-$ (400 GeV) &  $e^+ e^-  \to t{\bar t} $  &
$\sim 800 \times L$
\\ \hline
single top & dominant reaction & $(N_{t} +N_{\bar t})$ \\ 
\hline
Tevatron: & $ u + b {\stackrel{ W}{\longrightarrow}} d +t $ & $\sim
 3 \cdot 10^3 \times L$ \\
LHC: & $ u + b {\stackrel{ W}{\longrightarrow}} d +t $ & $\sim 3.3
\cdot 10^5 \times L$ \\ \hline
\end{tabular}
\end{center}
\end{table}

\section{The profile of the top quark}
\label{secprofile}

The top quark couples to all known
fundamental interactions. Because of its large mass, it is expected to couple
strongly to the forces that break the electroweak gauge
symmetry. While the interactions of the top quark have not been
explored in great detail
so far, its mass has been experimentally determined very precisely. 
In this section we briefly describe what is known about the properties
of the top quark, i.e., its mass,
lifetime, spin, and its charges.  Because its mass plays a central
role in the physics of this quark, we shall first discuss the meaning
of this parameter.

\subsection{Mass}
\label{sub-secmass}

The top mass is a
convention-dependent parameter, like the  other parameters of the SM.
As the top quark does not hadronize (see section~\ref{sub-secli}),
it seems natural to exploit the picture of the top quark being a 
highly unstable bare fermion. This suggests to use the concept of
on-shell or  pole mass,
which is defined to be  the real part of the complex-valued
 pole of the quark propagator $S_t(p)$. This is a purely
 perturbative concept. A quark is  unobservable due to colour
 confinement, so its full propagator has no pole. 
In  finite-order perturbation theory 
the propagator of the top quark has a pole
at the  complex value $\sqrt{p^2} = m_t -\rmi \Gamma_t/2$, where
$m_t$ is the pole or on-shell mass and $\Gamma_t$ is the decay width of
the top quark. However,  the
all-order resummation of a class of diagrams, associated with
so-called infrared renormalons, implies that the pole mass has an
intrinsic, non-perturbative
 ambiguity of order $\Lambda_{QCD} \sim $ a few hundred MeV
\cite{Beneke:1994sw,Bigi:1994em,Ball:1995ni,Smith:1996xz}.
So-called short distance masses, for instance the
quark mass $\overline{m}_q(\mu)$ 
defined in the  $\overline{\rm MS}$ renormalization scheme, 
are free from such ambiguities. Here $\mu$ denotes the renormalization
scale. The relation between the pole mass and the 
 $\overline{\rm MS}$ mass is known in QCD to $\Or(\as^3)$
 \cite{Gray:1990yh,Chetyrkin:1999ys,Melnikov:2000qh}. Evaluating
this relation for the  top quark at $\mu=\overline{m}_t$  it reads
\be
\overline{m}_t(\overline{m}_t) = m_t \left(1+\frac{4}{3} \frac{\as}{\pi}
+8.2364 \, (\frac{\as}{\pi})^2 + 73.638 \, (\frac{\as}{\pi})^3 + \Or(\as^4) 
\right)^{-1} \, ,
\label{eq-mpomsb}
\ee
where $\as(\mu= \overline{m}_t)$ is the  $\overline{\rm MS}$ coupling
of 6-flavour QCD. One should remember that the relation
(\ref{eq-mpomsb}) has an additional uncertainty of $\Or(\Lambda_{QCD})$. 
Using $\as=0.109$ we get $m_t/{\overline{m}_t} = 1.06$. Thus, the 
 $\overline{\rm MS}$ mass is 10 GeV lower than the pole mass;
 $m_t = 171$ GeV (see below) corresponds to ${\overline{m}_t}=161$
 GeV.

The present experimental determinations of the top mass at the
Tevatron use global fits to data which involve  a number of
 top-mass dependent
(kinematical) variables. The modeling
involved uses lowest-order matrix elements and parton showering 
 (see section~\ref{sub-topmass}).  Recently, the value
\be
m_t^{exp} = 172.6 \, \pm \, 1.4 \, {\rm GeV}
\label{eq-mtdocdf}
\ee
was obtained from the combined measurements of the
CDF and D0 collaborations \cite{:2007bxa}. The relative error
 of $0.8\%$ is smaller than that of any other quark mass. In view of
this precision the question arises how (\ref{eq-mtdocdf}) is to be 
interpreted. 
How is it related to a well-defined Lagrangian mass parameter?
As most of the measurements use kinematic variables, it
seems natural to identify (\ref{eq-mtdocdf}) with the pole mass. 
However, one should be aware that the present experimental determinations
of the top mass cannot be related to observables which have been calculated in higher-order
perturbative QCD in terms of a Lagrangian mass parameter. 

As is well known, the value of the top-quark mass plays a key role in SM
fits to electroweak precision data which yield
constraints on the SM Higgs mass\footnote{For a discussion of the
role of the top-quark mass in electroweak precision physics, see, e.g. 
\cite{Beneke:2000hk}.}. The common practice is to interpret 
(\ref{eq-mtdocdf}) as the
on-shell mass $m_t$. With the world average of last year,
$m_t^{exp} = 170.9 \pm 1.4$ GeV \cite{:2007tmass}, the upper limit
$m_H < 182$ GeV ($95\%$ C.L.) was  obtained \cite{Alcaraz:2007ri}.
With (\ref{eq-mtdocdf}) the upper limit on $m_H$ increases by about
$18$ GeV.
This limit should be taken with a grain of salt in view of the
uncertainty in interpreting (\ref{eq-mtdocdf}).

In the following sections, $m_t$ always refers to the pole mass. With
no better alternative at present, we shall  stick to interpreting
(\ref{eq-mtdocdf}) in terms of this mass parameter\footnote{Below we shall
  mostly use $m_t=171$ GeV -- the world-average value at the time
when this article was written --  when theoretical results are evaluated
 for a definite top mass.}.

\subsection{Lifetime}
\label{sub-secli}

The top quark is an extremely elusive object. Because its
mass is so large, 
it can decay into on-shell $W$ bosons, i.e., 
the two-particle decay mode $t\to b \, W^+$ is kinematically
possible.  The SM predicts the top quark to  decay  almost 
exclusively into this mode (see section~\ref{sectdec}).
This process is CKM-allowed\footnote{CKM is the acronym for the 
  Cabibbo-Kobayashi-Maskawa matrix $(V_{qq'})$, the $3 \times 3$ unitary matrix
  that parameterizes the strength of the interactions of quarks with $W^\pm$
  bosons. The unitarity of this matrix implies that the modulus of
  the matrix element $V_{tb}$ is close to one, $|V_{tb}| \simeq 0.999$
  \cite{Yao:2006px}.}, leading to the prediction that the average
proper lifetime of the $t$ quark is extremely short, $\tau_t
=1/\Gamma_t \simeq 5 \times 10^{-25}$ s. For comparison, the mean
lifetime of $b$ hadrons, which involve the next heaviest quark, is
almost 13 orders of magnitude larger,  $\tau_{b\, hadron} \simeq 1.5
\times 10^{-12}$ s. The lifetime $\tau_t$ is an order of magnitude
smaller the  hadronization time $\tau_{had} \simeq
1/\Lambda_{QCD} \approx 3 \times 10^{-24}$ s, which characterizes the
time it takes for an (anti)quark produced
in some reaction to combine with other produced (anti)quarks and form a colour-neutral
hadron due to confinement. Thus, top quarks are unable to form top
mesons $t{\bar q}$ or baryons $t q q'$. In particular there will be no
spectroscopy of toponium $\ttbar$ bound states \cite{Bigi:1986jk}. 
To illustrate this central feature further, we consider the production of
a $\ttbar$ pair at the Tevatron or at the LHC, $p {\bar p}, \,  p p \to \ttbar X$. The
distance the $t$ and $\bar t$ quarks are able 
to move away from their
production vertex before they decay is on average $0.1$ fm, a length
scale much smaller than the typical hadron size of $1$ fm. At distances
not more than about $0.1$ fm even the strong interactions of $t$ and 
$\bar t$ quarks are still weak owing to the asymptotic
freedom property of QCD. Therefore the top quark behaves
like a highly unstable particle which couples only weakly
to the other known quanta.

The decay width of the top quark manifests itself in the Breit-Wigner
line-shape, i.e., the width of the invariant mass distribution of the
top-decay products, $\rmd\sigma/\rmd M_t$, $M_t^2=(\sum
p_f)^2$. Unfortunately, the top width $\Gamma_t\simeq 1.3$ GeV is much
smaller than the experimental resolution at the Tevatron or at the LHC.
At present, no sensible method is known how to directly determine
$\Gamma_t$ at a hadron collider. In conjuction with some 
assumptions, an indirect determination is possible from the
measurement of the $\ttbar$ and the single-top production 
cross sections \cite{Chakraborty:2003iw}.

An important consequence of the distinctive property of the top quark
not forming hadrons pertains top-spin effects.
The spin polarization and/or spin-spin correlations
which are imprinted upon an ensemble of single top quarks or 
$\ttbar$ pairs by the production dynamics are not diluted by hadronization,
but result in characteristic angular distributions and 
correlations of the final state particles/jets into which the top quarks
decay  \cite{Kuhn:1983ix} (see sections~\ref{subang},~
\ref{sub-tspcorr}, and~\ref{sub-singtpol}). 
Gluon radiation off a top quark may flip its spin, but
such a chromomagnetic $M1$ transition is suppressed by the large mass $m_t$. The 
spin-flip transition rate of an off-shell $t$ quark decaying into an on-shell gluon with
energy $E_g$ and a $t$ quark may be estimated
as  $\Gamma(M1) \sim \alpha_s E_g^3 /m_t^2$, where 
$\alpha_s(m_t)= 0.1$.  Thus the spin-flip time 
$\tau_{flip} = 1/ \Gamma(M1)$ is on average
much larger than the top-lifetime $\tau_t$.
In any case, top-spin effects can be computed reliably in perturbation theory
because the QCD coupling $\alpha_s$ is small at energies of the order
of $m_t$. 
This is in contrast to the case of lighter quarks, for instance $b$
quarks, where the transition of $b$ quarks to $b$ hadrons is governed by 
non-perturbative strong interaction
dynamics, and such transitions cannot be computed so far with ab initio methods. 
A large fraction of quark hadronization results in
spin-zero mesons. Thus there will be 
 no information left in the meson decay products on
the quark spin at production. 
In this respect top quarks offer a richer phenomenology than lighter quarks:
``good observables'' -- i.e., observables that are both measurable
and reliably predictable -- are not only quantities like production
cross sections, decay rates, transverse
momentum and rapidity distributions, but also final-state angular distributions
and correlations that are caused by top-spin polarization and $\ttbar$ spin 
correlations. 

\subsection{Spin}
\label{sub-spin}

There is no doubt that the top quark, as observed at the Tevatron, is a
spin $1/2$ fermion -- although a dedicated experimental verification has so far not
been made. The observed decay $t\to b \, W$, the known 
spins of $W$ and $b$, and the conservation
of total angular momentum  imply that the top quark is a fermion. If the spin of the
top quark were $3/2$, the $\ttbar$ cross section at the Tevatron would be much larger
than the measured one. A direct experimental evidence for the top quark having
spin  $1/2$ would be the observation of the resulting polarization and 
spin-correlation effects (see sections~\ref{sub-tspcorr}
and~\ref{sub-singtpol}).  Definite measurements are expected to be feasible at the LHC
\cite{Hubaut:2005er,Ball:2007zza}. Another possibility is the
measurement of the differential cross section $\rmd \sigma/\rmd\mtt$
near the $\ttbar$ production threshold \cite{Berger:2000zu} ($\mtt$
denotes the invariant mass of the $\ttbar$ pair). As is well-known the
behaviour  of the near-threshold cross section as a function of the
particle velocity is characteristic of the spin of the produced
particle and antiparticle.

\subsection{Colour and electric charge} 
\label{sub-charge}

Top quarks, like the other  quarks,
carry colour charge -- they transform as a colour triplet under the
$SU(3)_c$ gauge group of the strong interactions.  Colour-confinement precludes
the direct measurement of this quantum number; but indeed,
measurements of the $\ttbar$ production cross section are consistent
with the SM predictions for a colour-triplet and antitriplet 
quark-antiquark pair.

The top quark is the $I_3=1/2$ weak-isospin partner of the $b$ quark,
assuring the consistency of the SM at the quantum level.
The electric charge of the top quark, which is therefore $Q_t =2/3$ in units
of the positron charge $e>0$  according to the SM, has so far not been measured.
The observed channel $\ttbar \to b {\bar b}W^+W^-$ does a priori not preclude the
possibility that the observed  top resonance is  an exotic heavy quark
with charge $Q=-4/3$ decaying into $b W^-$ \cite{Chang:1998pt}. However, this 
has been excluded in the meantime 
by the D0 and CDF experiments at the Tevatron \cite{Abazov:2006vd,Beretvas:2006sw}.
The top-quark charge can be directly determined by measuring the production
rate of $\ttbar$ 
plus a hard photon and taking the ratio
$\sigma(\ttbar \gamma)/\sigma(\ttbar)$. At the LHC this ratio is 
approximately proportional to $Q_t^2$  because $\ttbar$ and $\ttbar\gamma$
production is dominated by gluon fusion.
This will be discussed in more detail in section~\ref{sub-assprod}.

\section{Top-quark decays}
\label{sectdec}

Because the top quark is an extremely short-lived resonance, only
its decay products can be detected by experiments. Thus for comparison 
with data, theoretical predictions must entail, in general, top-production and
decay. However, this resonance is narrow, as $\Gamma_t/m_t \simeq 0.008$. 
Thus one can factorize, to good approximation, the theoretical description
of these reactions into the production of on-shell single top quarks
or $\ttbar$ pairs (being produced in a certain spin configuration) and the decay of $t$ and/or
$\bar t$. We treat  top-quark
decays first while the survey of hadronic production of these quarks
is postponed to the following sections.  
We shall review (polarized) 
top-quark decays in the SM, then discuss effects of possible
anomalous couplings in the $tbW$ vertex, and finally consider several 
new decay modes which are possible in various SM extensions.

\subsection{SM decays}
\label{subsmdec}

In the SM, which involves three generations of quarks and leptons, the only
two-particle decays of the top quark\footnote{Unless stated 
otherwise, the discussion of
this section applies analogously also to $\bar t$ decays.} 
which are possible to lowest order in the 
(gauge) couplings are $t\to b W^+$, $t\to s W^+$, and $t\to d W^+$. Their rates
are proportional to the squares of the CKM matrix elements $|V_{tq}|^2$, $q=b,s,d$,
respectively. The rate of $t\to X$, i.e. the total decay width
$\Gamma_t$ of the top quark, is given by the sum of the widths of these 
three decay modes, as 
the branching ratios of the loop-induced  
flavour-changing neutral current decays are negligibly small in the SM (see
section \ref{subfcnc}).
The analysis of data from weak decays of hadrons yields
$0.9990 <|V_{tb}|<0.9992$ at $95\%$ C.L. \cite{Yao:2006px}, using
the unitarity of the CKM matrix. From the recent observation of the oscillation
of $B_s \leftrightarrow {\bar B}_s$ mesons by the D0 and CDF experiments at the
Tevatron and from 
analogous data on  $B_d \leftrightarrow {\bar B}_d$
oscillations one can extract the ratio $0.20 <|V_{td}/V_{ts}|<0.22$ 
\cite{Yao:2006px}.
The unitarity relation  $|V_{tb}|^2+|V_{ts}|^2+|V_{td}|^2=1$ implies that
the total decay rate is completely dominated by $t\to b W^+$, and one gets
for the branching ratios
\be
B(t \to b W^+) = 0.998,\qquad B(t \to s W^+) \simeq 1.9 \times
10^{-3}, \qquad B(t \to d W^+) \simeq 10^{-4} .
\label{eqtbranch}
\ee
There is direct information from the Tevatron which implies that
$|V_{tb}|\gg |V_{td}|,\,|V_{td}|$, without using the unitarity constraint. 
The CDF and D0 collaborations measured
\be
R \equiv \frac{B(t \to b W)}{\sum_{q=b,s,d}B(t \to q W) } =
\frac{|V_{tb}|^2}{|V_{tb}|^2+|V_{ts}|^2+|V_{td}|^2} 
\label{eqcdfr}
\ee
by comparing the number of $\ttbar$
candidates with $0$, $1$, and $2$ tagged $b$ jets.
The right-hand side of (\ref{eqcdfr}) is the standard-model
interpretation of this ratio. 
A collection of CDF and D0 results on $R$ is given in 
\cite{Leone:2007mk,Abazov:2008yn}; the
recent D0 result is $R=0.97^{+0.09}_{-0.08} $ \cite{Abazov:2008yn}.
The D0 \cite{Abazov:2006gd} and the CDF \cite{CDFsinglet} experiments
reported evidence for single top quark
production. The agreement of the measured production cross section with the
SM expectation was used by these experiments for a direct determination of
the CKM matrix element $V_{tb}$ with the result
$0.68 < |V_{tb}|\leq 1$   \cite{Abazov:2006gd} and 
$|V_{tb}|=0.88\pm 0.14\pm 0.07$ \cite{CDFsinglet}.
 (See also section \ref{secsingletop}).

\subsubsection{The total decay width:}  
\label{subtotdecw} \quad
As just discussed, the total decay width of the top quark 
is given in the SM, to the precision 
required for 
interpreting the Tevatron or forthcoming LHC experiments, by the partial
width of the decay $t\to b W^+$ -- more precisely, of the sum of the
widths of the decays
\be
t \to b \, W^+ \,  \to \, b\,\ell^+ \nu_{\ell}, \quad  b \,  u \, {\bar d}, \quad
b\, c \, {\bar s},\quad \cdots
\label{smtdecay}
\ee
where $\ell=e,\mu,\tau$ and the ellipses indicate 
CKM-suppressed decays of the $W$ boson. Taking
the intermediate
$W$ boson to be on-shell, one
gets to Born approximation:
\be
\Gamma_t^B \equiv \Gamma_t^B(t\to b W) = \frac{G_F}{8\pi \sqrt{2}}
m_t^3 |V_{tb}|^2 \left (1 -\frac{m_W^2}{m_t^2}\right)^2 \left(1 + 
2 \frac{m_W^2}{m_t^2}\right )
= 1.44 \, {\rm GeV},
\label{twidth}
\ee
where $G_F$ is the Fermi constant, $m_t=171$ GeV and $m_W=80.40$ GeV has
been used, and the mass of the $b$ quark ($m_b \simeq 4.9$ GeV) has
been neglected. The order $\alpha_s$ QCD corrections \cite{JezKu89}, 
the order $\alpha$ weak and electromagnetic\footnote{As usual, $\alpha = e^2/4\pi$.
Recall that the weak and electromagnetic couplings are related by
$g_w = e/\sin\theta_w$.} corrections  \cite{Denner:1990ns,Eilam:1991iz}, and the corrections 
due to the width of off-shell $W$ bosons
\cite{Jezabek:1993wk} and to $m_b\neq 0$ were determined quite some time ago.
Moreover, the order $\alpha_s^2$ QCD corrections
were computed as an expansion
in $(m_W/m_t)^2$  \cite{Czarnecki:1998qc,Chet99}. While the
$\Or(\alpha)$ corrections are positive and rather small ($+2\%$)
and are almost compensated by the finite $W$-width corrections,
the  $\Or(\as)$ and $\Or(\as^2)$ corrections are negative. Thus,
essentially only the QCD corrections matter, and the SM prediction
for the radiatively corrected width may be represented by the 
following expression  \cite{Czarnecki:1998qc,Chet99}:
\be
\Gamma_t = \Gamma_t^B \, (1 - 0.81 \as - 1.81 \as^2) \, ,
\label{eq-twidthqcd}
\ee
where $\as = \as(m_t)$ is the QCD coupling in the ${\overline{\rm MS}}$ 
renormalization scheme. With $\as(m_t)=0.108$ the correction factor
(\ref{eq-twidthqcd}) to the lowest-order width is $0.89$.
Thus for $m_t=171$ GeV we have\footnote{The higher-order QCD corrections
to the top-decay width should be evaluated using
${\overline m}_t$ rather than the pole mass $m_t$, in order to
avoid renormalon ambiguities.} $\Gamma_t=1.28$ GeV.

It should be noted
that gluon radiation, $t\to bW g$, can make up a fair
fraction of the total rate if the cut on the minimal gluon energy $E_g$
is relatively low. For gluon
energies $E_g\gtrsim 10$ GeV the branching ratio 
$B(t\to bW g)\simeq 0.3$.

For the dominant semileptonic and nonleptonic decay modes (\ref{smtdecay})
the branching ratios are, for $\Gamma_W, m_b \neq 0$ and including the
$\Or(\as)$ QCD corrections:
\be
B(t \to b \ell^+ \nu_{\ell}) = 0.108 \quad (\ell=e,\mu,\tau), \qquad
B(t \to b q {\bar q}') = 0.337 \times |V_{q {\bar q}'}|^2  \,.
\label{eqbrtallow}
\ee
Using the central values of the respective CKM matrix elements \cite{Yao:2006px}
we have then
$B(t \to b u {\bar d}) = B(t \to b c {\bar s}) = 0.328$,
$B(t \to b u {\bar s})= 0.017,$ and $B(t \to b c {\bar b})= 6 \times 10^{-4}.$

\subsubsection{W-boson helicity:} 
\label{subWbhel} 
\begin{figure}[ht]
\begin{center}
\includegraphics[width=5cm]{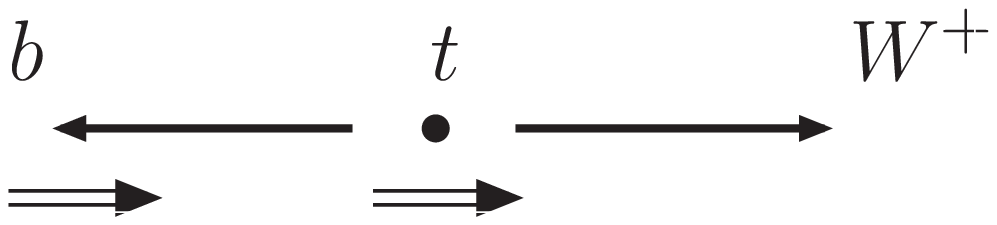} 
%\vspace*{0.4cm}
\includegraphics[width=5cm]{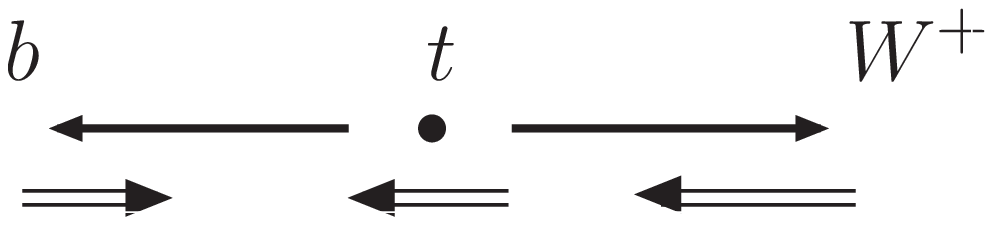}
%\vspace*{0.4cm}
\includegraphics[width=5cm]{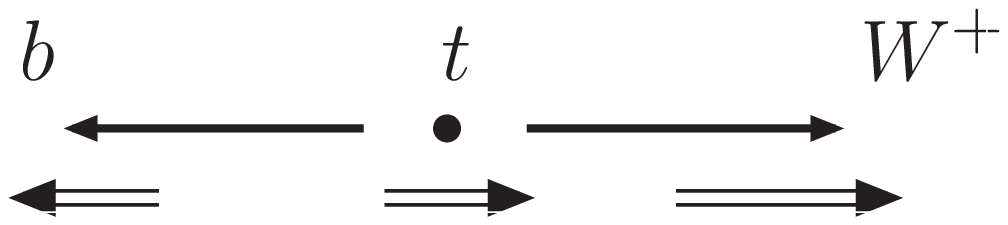}
\end{center}
\caption{\label{fig-Whel} Illustration of top-quark decay into a $b$
  quark and a $W^+$ boson with $\lambda_W =0,\mp 1$. For 
 $W^+(\lambda_W =+ 1)$ the $b$ quark must have positive helicity (to
lowest order), which has vanishing probability for $m_b \to 0$.}
\end{figure}

\quad
In the SM the strength and structure of
the $tbW$ vertex is determined (up to $V_{tb}$) 
by the universal $V-A$ charged-current interaction. A basic test of
the structure of this vertex is the measurement of the decay fractions 
$F_0=B(t\to b W(\lambda_W =0))$, $F_\mp =B(t\to b W(\lambda_W =\mp 1))$ 
into $W^+$ bosons of helicity $\lambda_W =0,\mp 1$. 
By definition $F_0 + F_- + F_+ = 1$. The  $V-A$ structure and
angular momentum conservation allow the decay  into a zero-helicity and negative
helicity $W$ boson, but the decay amplitude into $W(\lambda_W = +1)$ is suppressed
by a factor $m_b^2/m_W^2$. This is due to the fact that the $V-A$ law forces
the $b$ quark, if it were massless, to have negative helicity -- but this is 
in conflict with angular momentum conservation. The three cases are illustrated in
figure \ref{fig-Whel}.  
For the decay fractions one obtains at tree level, putting $m_b=0$,
and using  $m_W=80.40$ GeV:
\bea
F_0^B = \frac{m_t^2}{m_t^2 + 2 m_W^2} =  0.6934 - 0.0025 \times (171 - m_t\, [{\rm GeV}]) 
\, , \nonumber \\
F_-^B = \frac{2 m_W^2}{m_t^2 + 2 m_W^2} = 0.3066 + 0.0025 \times (171 - m_t\,
[{\rm GeV}])
\, ,\qquad F_+^B = 0 \, .
\label{eqfborn}
\eea
Once gluon (and photon) radiation is taken into account, $F_+ \neq 0$ even in the
limit $m_b=0$.  The $W$-helicity
fractions $F_{0,\mp}$ were computed in \cite{Fischer:1998gsa,Do:2002ky},
taking the $\Or(\as)$ QCD and 
$\Or(\alpha)$ electroweak corrections, and the corrections
due to the finite $W$ width and  $m_b\neq 0$ into account. 
These corrections are very small; in particular they generate a small fraction $F_+$.  
The result of \cite{Do:2002ky} is
\be
F_0 = 0.99 \times F_0^B \, ,
\qquad F_- = 1.02 \times F_-^B  \, ,\qquad F_+ = 0.001\, .
\label{eqf1loop}
\ee
% Recently, also the $\Or(\as^2)$ contributions to  $F_{0,\mp}$ were
% determined \cite{}; they decrease $F_0$ and increase $F_-$ and   $F_+$
% by a few per mille.
For ${\bar t} \to {\bar b} W^-$ we have 
${\bar F}_0=F_0$, ${\bar F}_-=F_+$, and  ${\bar F}_+=F_-$ in the SM.
Violations of these relations  due to the $CP$-violating
KM phase $\delta_{KM}$ are negligibly small.  

The large fraction $F_0 \simeq 0.7$ signifies that top-quark decay is a source of
longitudinally polarized $W$ bosons -- in fact, the only significant one
at the LHC. (Almost all  $W$ bosons produced in QCD reactions are
transversely polarized.) 
Recall that, in the SM, 
 the longitudinally polarized state of the $W$ boson is generated by the charged
component of the $SU(2)$ Higgs doublet field. If the dynamics of
electroweak symmetry breaking is different from the SM Higgs
mechanism,  one may expect deviations of the $tbW$ vertex from its SM
structure, and $F_0$ should be sensitive to it. The fraction $F_+$ is obviously
sensitive to a possible $V+A$ admixture in the charged weak current
involving the top quark. These issues will be addressed
in sections \ref{subang} and \ref{subnaom}. 

Information about the polarization of the $W$ boson is obtained from 
the angular distributions of one of its decay products, 
$W^+ \to \ell^+ \nu_\ell, q {\bar q}'$. As a $u$-type jet cannot
be distinguished experimentally from a  $d$-type jet, the best choice 
is to consider a charged lepton
$\ell^+ = e^+, \mu^+$. Consider
 the decay $t\to b W^+ \to b \ell^+ \nu_\ell$ and
 define $\psi^*$ to be the angle between the direction of
$\ell^+$ in the $W^+$ rest frame and the  $W^+$ direction in the $t$
rest frame. Then  one obtains for the
distribution of this angle:
\be
\frac{1}{\Gamma}\frac{\rmd \Gamma}{\rmd \cos\psi^*} = \frac{3}{4} F_0 \sin^2 \psi^*
+ \frac{3}{8} F_- (1-\cos\psi^*)^2 + \frac{3}{8} F_+ (1+\cos\psi^*)^2 \, ,
\label{eqhelpsi}
\ee
Thus, in the one-dimensional distribution (\ref{eqhelpsi}),
interference terms due to  different polarization states of the
intermediate
$W$ boson do not contribute\footnote{Higher dimensional
distributions, which are sensitive to such interferences,
that is, to non-diagonal terms in the $W$-boson spin density
matrix $(\rho^W)_{ij}$, can also be considered \cite{Fischer:2001gp}.}.
The helicity fractions can be obtained from a fit of 
(\ref{eqhelpsi}) to the measured $\cos\psi^*$ distribution
 and from the constraint $F_0 + F_- + F_+ = 1$.
In addition one may employ the forward-backward asymmetry $A_{FB}$
with respect
to $\cos\psi^*$,  $A_{FB}=3(F_+ - F_-)/4$. A generalization of this
asymmetry has been suggested in \cite{AguilarSaavedra:2006fy}:
\be
A_z = \frac{N(\cos\psi^*>z)  - N(\cos\psi^*<z)}{N(\cos\psi^*>z)
  +N(\cos\psi^*<z)} \, ,
\label{eq-afbz}
\ee
with $-1<z<1$. The fractions $F_{0,\mp}$ can be obtained from appropriate
combinations of $A_z$. The use of these asymmetries helps to
 reduce measurement uncertainties.

At the Tevatron the CDF and D0 experiments measured the $W$-boson
 helicity  fractions in semileptonic
 \cite{Abulencia:2006ei,Abulencia:2006iy,Abazov:2006hb} 
and recently also in nonleptonic \cite{Abazov:2007ve} top-quark
decays\footnote{Because
in $W \to q {\bar q}'$ the flavour of the jets cannot
be tagged, a $W$ daughter jet was chosen at random in
\cite{Abazov:2007ve}. The sign ambiguity in the calculated
$\cos\psi^*$ was avoided by considering the distribution of
$|\cos\psi^*|$.}.
The results  given in Table~\ref{tab-Whelex} for $F_0$ and $F_+$ were
obtained by putting $F_+$ and  $F_0$ to their SM values, respectively.
Recently, these assumptions were dropped in 
a simultaneous measurement of $F_0$ and $F_+$ by the D0 collaboration
\cite{Abazov:2007ve}, with the result: $F_0=0.425 \pm 0.166 \pm 0.102$
and $F_+ = 0.119 \pm 0.090 \pm 0.053$. 

For comparison, table~\ref{tab-Whelex} contains also
the expected systematic measurement uncertainties
for $F_{0,\pm}$ at the LHC, as estimated in the study of
\cite{Hubaut:2005er}. Statistical errors should not be a problem at
the LHC once a sufficiently large sample of top quark events will have been
recorded. After selection of $\sim 10^6$  $\ttbar$  events in the
dileptonic and lepton + jets  decay channels the statistical error
on the helicity fractions will be an order of magnitude below the
systematic one \cite{Hubaut:2005er}.
The analysis \cite{AguilarSaavedra:2007rs} arrived at similar results.
Thus $F_{0,\pm}$ should be measurable at the LHC with a precision
of about $2\%$. This will allow a precise determination of the structure of the
the $tbW$ vertex, as will be discussed in section \ref{subnaom}. 

\begin{table}[h!]
\caption{\label{tab-Whelex} First and second row: Tevatron results for the $W$ helicity
  fractions in $t$ quark decay. Third row: estimated systematic error
  in future LHC measurements.}
\begin{center}
\renewcommand{\arraystretch}{1.2}
\begin{tabular}{llllll}
\hline
 & $F_0$ & & $F_+$ & & $F_-$ \\
\hline
Tevatron: \, CDF & $0.85^{+0.16}_{-0.23}$ & \cite{Abulencia:2006ei} &
$-0.02 \pm 0.08$ & \cite{Abulencia:2006iy} & \\
\hspace*{1.9cm} D0 & $0.62\pm 0.10 $ & \cite{Abazov:2007ve} & $-0.002 \pm
0.07$ & \cite{Abazov:2007ve} & \\
LHC ($\Delta_{\rm syst}$) \cite{Hubaut:2005er} & $\pm 0.015$ & & $\pm
0.012$ & & $\pm 0.024$ \\
\hline
\end{tabular}
\end{center}
\end{table}

\subsubsection{Distributions for semileptonic and nonleptonic decays:} 
\label{subang} \quad
Apart from (\ref{eqhelpsi})
there are other energy and angular distributions that are useful for
studying the structure of the top-decay vertex in
 semi- and nonleptonic  decays of top quarks,  $t \to b \ell \nu_\ell, \; b 
q \bar{q}'$  $(q{\bar q}'$ = $u{\bar d}$, $c{\bar s}, ...)$. A number of
distributions were calculated for these decays, including radiative
corrections. 
  Here we discuss in detail only angular
distributions for polarized top-quark decays. They are important for
determining top-spin effects -- see below and 
sections~\ref{sub-tspcorr} and~\ref{sub-singtpol}.
 
An ensemble of top quarks 
self-analyzes its spin polarization via its
weak decays.  Consider the decay $t\to f+ \cdots$ of a polarized top   in the 
top quark rest frame.  
Information about the top spin vector  is encoded 
in the distribution of $\cos \theta_f$, where $\theta_f$
is the angle between the direction of the particle/jet $f$ (used as
$t$ spin analyzer) in the $t$ rest frame
 and the polarization vector of the top quark.
It has the a priori form
\be
\frac{1}{\Gamma_f}\frac{\rmd\Gamma_f}{\rmd\,{\cos} \: \theta_f} =
  \frac{1}{2}(1 + p\, c_f \;{\cos} \: \theta_f) \, ,
\label{dcost}
\ee
where $\Gamma_f$ denotes the partial decay width, $p$ is the
polarization degree of the ensemble, and $c_f$ is the $t$ spin-analyzing
power of $f$. Obviously, $|c_f|\leq 1$. 
In the SM the charged lepton or the $d$-type quark from $W$-boson
 decay are the best $t$ spin analyzers. We have 
$c_{\ell^+} = c_{\bar d}= 1$ at tree level, while $c_{\nu_\ell}=c_u =
-0.30$ and
$c_b = [2(m_b/m_t)^2-1]/[2(m_b/m_t)^2+1] =  - c_{W^+}= - 0.39$ (assuming
the reconstruction  of the $W^+$ boson direction of flight). That is, for an ensemble of
100 $\%$ polarized top quarks, the probability for the $\ell^+$ being
emitted in the direction of the $t$ spin is maximal, while it is zero
for the emission opposite to the $t$ spin. The circumstance that the
charged lepton has a larger $t$ spin-analyzing power than its mother,
the $W$ boson, seems curious at first sight. But this is due to the
fact that the $\ell^+$ distribution is generated by the amplitudes
with intermediate $W^+(\lambda_W =0)$ and $W^+(\lambda_W =-1)$ bosons
which interfere, and this leads to constructive and completely
destructive interference in the direction parallel and opposite to the $t$-spin,
respectively. This information about the $t$ spin contained in the
interference terms is missing in the distributions (\ref{dcost}) 
for $f=W^+$ and $f=b$. 

To order $\as$ the
CKM allowed final states are  (i)
$\ell \nu_\ell + b$ jet and $\ell \nu_\ell \, +\, b$ jet +
gluon jet in semileptonic decays
and (ii) a $b$ jet plus two or three non-$b$ jets  in nonleptonic
decays. The  spin-analyzing
power of a final state particle/jet $f$ decreases slightly due to gluon radiation.

\begin{table}[h!]
\caption{\label{tab-spinpo}Spin-analyzing power
$c_f$ to lowest order and to $\Or(\as)$ for semileptonic \cite{Czarnecki:1990pe}
and nonleptonic  \cite{Brandenburg:2002xr} top quark  decays for
$m_t=171$ GeV.}
\begin{center}
\renewcommand{\arraystretch}{1.2}
\begin{tabular}{llllllll}
\hline
 & $\ell^+$ & ${\bar d}$ & $u$ & $b$ & $j_<$ & ${\bf T}$ & $j_>$ \\ 
\hline
{\rm LO:} & 1& 1& -0.32 & -0.39 & 0.51 & -0.32 & 0.2 \\
{\rm NLO:}& 0.999 & 0.97 & -0.31 & -0.37 & 0.47 & -0.31 & \\
\hline
\end{tabular}
\end{center}
\end{table}
The  correlation coefficients $c_f$ were
computed to order $\as$  for the semi- and nonleptonic channels in
\cite{Czarnecki:1990pe} and in \cite{Brandenburg:2002xr},
respectively,
and are collected, together with the lowest-order values 
 in table \ref{tab-spinpo}. 
For the nonleptonic channels,  $j_<$ and $j_>$ denote
the least energetic and
most energetic non-$b$ jet, respectively, defined by the Durham clustering 
algorithm. As the identification of the flavours  of
the quark  jets from CKM allowed $W$ decay is not possible -- or inefficient in the
case of the $c {\bar{s}}$ final state --, table \ref{tab-spinpo}  shows that,
in the case of nonleptonic decays, the 
least energetic  non-$b$ jet is the most efficient top-spin analyzer.
Again, this is a consequence of $V-A$ and
angular momentum conservation. 
The vector ${\bf T}$ denotes the oriented thrust
axis for nonleptonic final states, defined by the requirement ${\bf T\cdot p}_b>0$.
The coefficients 
$c_f \neq c_\ell$  in table \ref{tab-spinpo} depend slightly 
 on the value of $m_t$.

The analogous angular distributions for the decays of
antitop quarks, ${\bar t}\to {\bar f}+ \cdots$, are
\be
\frac{1}{\Gamma_{\bar f}}\frac{\rmd\Gamma_{\bar f}}{\rmd\,{\cos} \:
      \theta_{\bar f}} =
  \frac{1}{2}(1 + p\, c_{\bar f} \;{\cos} \: \theta_{\bar f}) \, ,
  \quad  c_{\bar f} = - c_{f} \, .
\label{dcostbar}
\ee
The last relation is valid if  $CP$
invariance holds. Violation of this relation requires
that the respective decay amplitude
has a $CP$-violating absorptive part \cite{Bernreuther:1992be}.
Within the SM such an effect is negligibly small. 

If the $t\to b$ transition, i.e., the  decays $t \to b f_1 f_2$ are
affected by new interactions, the  values of the $c_f$ given in table
\ref{tab-spinpo} will change in general -- see
sections~\ref{subnaom} and~\ref{subchH}. 

SM predictions including QCD corrections
for lepton-energy and energy-angular distributions in (polarized)
semileptonic top-quark
decay are also 
available \cite{Czarnecki:1990pe,Fischer:1998gsa,Fischer:2001gp}.
The shapes of the $\ell$ and $\nu_\ell$ energy distributions are
good probes of the chirality of the current which induces the $t \to
b$ transition. For a small $V+A$ admixture to this current
 these distributions were
determined in \cite{Jezabek:1994zv,Bernreuther:2003xj}.

\subsection{Top-quark decays in SM extensions} 
\label{subtext}
In the SM all decay modes other than $t\to W b$ are rare, as their 
branching
ratios are $\Or(10^{-3})$ or less. Nevertheless, exotic decay modes
with branching ratios of the order of a few percent are still
possible. Several examples will be reviewed below. If new
particles/interactions affect top quarks, they
may not lead to new decay modes, but they should, in any case, have an effect 
on the $tbW$ vertex. LHC experiments should be able to determine
this vertex quite precisely, as will be discussed now.

\subsubsection{Anomalous couplings in the $tbW$ vertex:}
\label{subnaom} \quad
A model-independent analysis of the structure of the
 $tbW$ vertex can be made as follows.
The amplitude ${\cal M}_{tbW}$ of the decay $t(p) \to b(k) \, W^+(q)$,
where all particles are on-shell ($p, k$ and $q=p-k$ denote four-momenta)
has the general form-factor decomposition \cite{Bernreuther:1992be,Ma:1991ry}
\be
{\cal M}_{tbW} = - \frac{g_W}{\sqrt 2} 
\epsilon^{*\mu} {\bar u}_b \left[(V^*_{tb} +f_L) \gamma_\mu P_L +
f_R\gamma_\mu P_R + \rmi \sigma_{\mu\nu}  q^\nu 
(\frac{g_L }{m_W} P_L  + \frac{g_R }{m_W} P_R)
\right] u_t \, . 
\label{eq-ffdecomp} 
\ee
Here $P_{L,R}=(1\mp\gamma_5)/2$ and the two chirality conserving 
and  flipping form factors\footnote{If the $W$ bosons are off-shell,
  two additional form factors appear in the matrix element. However, they
do not contribute in the limit of vanishing masses of the fermions
into which the $W$ boson decays.}
$f_{L,R}$ and  $g_{L,R}$, respectively,
 are  dimensionless (complex) functions of the squared $W$ boson
 four-momentum $q^2$. The parameterization in (\ref{eq-ffdecomp})
is chosen such that non-zero values of  $f_{L,R}$ and  $g_{L,R}$
signify deviations from the structure of the tree-level Born vertex. 
An equivalent description of the $t\to bW$ vertex is obtained using
an effective Lagrangian approach \cite{Kane:1991bg}. In this context,
the above form factors evaluated at $q^2=m_W^2$ correspond to
anomalous couplings. 

In SM extensions corresponding to renormalizable theories,
 $f_{L,R}\neq 0$ can appear at tree-level while $g_{L,R}\neq 0$ must
be loop-induced.
The form factors are gauge-invariant, but in general
not infrared-finite. They should be used to parameterize only possible new
``infrared safe'' short-distance contributions to the $tbW$ vertex,
caused for instance by the exchange of new heavy virtual particles. 
A search for anomalous couplings  in $t \to b f {\bar f}'$ decay-data 
should proceed as follows. One computes decay distributions within the
SM including radiative corrections, and adds the contributions linear
in the anomalous  form factors  $f_{L,R}$ and $g_{L,R}$ which are
generated by the  interference of (\ref{eq-ffdecomp}) with the SM Born
amplitude. This assumes that these anomalous effects are small, which
can be checked a posteriori.

For a small $V+A$ admixture to the SM current, energy
and higher-dimensional distributions were computed
in this fashion in \cite{Jezabek:1994zv,Bernreuther:2003xj}.
In this case neutrino energy-angular distributions
turn out to be most sensitive to $f_R \neq 0$.
It is important to take the
QCD corrections into account in (future) data analyses, as gluon radiation  can mimic
a small $V+A$ admixture. 

There are tight indirect constraints on some of the anomalous couplings
from the measured branching ratio  $B({\bar B}\to X_s \gamma)$, 
in particular on $f_R$ and $g_L$,
as the contributions of these couplings to $B$ are enhanced by
a factor $m_t/m_b$  \cite{Fujikawa:1993zu,Cho:1993zb}.
A recent analysis \cite{Grzadkowski:2008mf}
arrives at the bounds given in table \ref{tab-anobound}. (For earlier work,
 see \cite{Larios:1999au,Burdman:1999fw}.) These bounds were obtained
 by allowing only one coupling to be non-zero at a time.

\begin{table}[h!]
\caption{\label{tab-anobound} Current 95 $\%$ C.L. upper and lower
bounds on anomalous couplings in the $tbW$ vertex from  
$B({\bar B}\to X_s \gamma)$ \cite{Grzadkowski:2008mf}. The couplings
are assumed to be real.}
\begin{center}
\renewcommand{\arraystretch}{1.2}
\begin{tabular}{lllll}
\hline
  & $f_L$ & $f_R$ & $g_L$ & $g_R$ \\
\hline
${\rm upper \, bound}$ & 0.03& 0.0025& 0.0004 & 0.57 \\
${\rm lower \, bound}$ & -0.13 & -0.0007& -0.0015 & -0.15  \\
\hline
\end{tabular}
\end{center}
\end{table}
One should keep in mind that these bounds are not rock-solid as the effects of different
couplings might cancel among each other or might be off-set by
other new physics contributions.

The possible size of these anomalous couplings was investigated in
several SM extensions.  While in the 
minimal supersymmetric extension of the SM the radiative corrections
to the lowest order $V-A$ $tbW$ vertex due to supersymmetric particle exchanges 
 are only at the level of $1\%$
or smaller \cite{Brandenburg:2002xa,Cao:2003yk}, effects can be
somewhat larger in some alternative models of electroweak 
symmetry breaking  \cite{He:1999vp}. 

The level of precision with which the helicity fractions are presently
known from Tevatron experiments (see table \ref{tab-Whelex}) do not
imply constraints on the anomalous couplings which can compete
with those given in table~\ref{tab-anobound}.
However, future high statistics data on top quark
decays at the LHC can provide
information on the couplings $f_R, g_L$, and $g_R$ at the level of
a few percent -- i.e., there is the prospect of directly 
determining these couplings with good precision.
 Simulation studies for the LHC 
analyzed $\ttbar$ production and decay into 
lepton plus jets channels  \cite{Hubaut:2005er,AguilarSaavedra:2007rs}
and also  
dileptonic channels
\cite{Hubaut:2005er}. Basic observables for determining
the anomalous couplings are the $W$-boson helicity fractions
(whose estimated measurement uncertainties are given in table \ref{tab-Whelex})
and the forward-backward asymmetries (\ref{eq-afbz}), which
were used in \cite{AguilarSaavedra:2007rs}.
These observables are not sensitive to the absolute strength of
the $tbW$ vertex and to $f_L$. 
Both studies assume real form factors.
The parametric dependence of the observables
on the anomalous couplings yields estimates for the expected
confidence intervals. Assuming that only one non-standard coupling is
nonzero at a time, 
\cite{AguilarSaavedra:2007rs} concludes that $f_R$, $g_L$, or $g_R$
should be either detected or excluded at the 2 s.d. level (statistics
plus systematics) if their values lie outside
the following intervals:
\be
f_R\,(2\sigma): \:\: [-0.055, 0.13] \, , \quad 
g_L \,(2\sigma): \: \:  [-0.058,0.026]\, , \quad
g_R \,(2\sigma): \: \: [-0.026, 0.031] \,.
\label{eq-expanomc}
\ee
The analysis of  \cite{Hubaut:2005er} arrived, as far as $g_R$ is
concerned, at a sensitivity level of the same order. Thus the
sensitivity to $g_R$ expected at the LHC is an order of magnitude
better
than the current indirect bound given in table \ref{tab-anobound}.

Single top-quark production and decay at the
LHC will also provide a sensitive probe of these anomalous
couplings \cite{Hikasa:1998wx,Boos:1999dd}.
If the single-top-production cross sections for the $t$-channel
and $s$-channel processes (see section \ref{secsingletop})
will be measured  at the
LHC with reasonable precision, then this additional
information will allow to determine/constrain also $f_L$ and
the  absolute strength of the $tbW$ vertex \cite{Chen:2005vr}.

In general the form factors $f_{L,R}$ and  $g_{L,R}$ can be complex, which
need not necessarily be due to $CP$ violation.
Because the form factors are in the timelike region, $q^2>0$, they can
have absorptive parts. CP invariance implies, apart
from the requirement of a real CKM matrix,  that the following relations
hold between the form factors $f_{L,R}, g_{L,R}$ 
 and the corresponding form factors $f_{L,R}',
g_{L,R}'$ in the  ${\bar t}\to {\bar b} W^-$ 
decay amplitude \cite{Bernreuther:1992be}:
\be
 f_i = f_i', \quad  g_i =  g_i', 
\label{eq-cpffaki}
\ee
where $i=L,R$. Thus, absorptive parts due to $CP$-invariant
interactions satisfy (\ref{eq-cpffaki}) while dispersive (and
absorptive) parts
generated by $CP$-violating interactions 
violate these relations.
The $T$-odd triple correlation 
${\cal O}= \st \cdot ({\bf \hat p}_{\ell^+}\times {\bf \hat p}_b)$ 
in polarized semileptonic $t$ decay, where $\st$
denotes the top spin, is sensitive to $CP$ violation and
$CP$-invariant absorptive parts in the $tbW$ vertex. Measuring ${\cal
  O}$ and the the corresponding correlation $\bar{\cal O}$ in $t$ and
$\bar t$ decay and taking the difference would be a clean
$CP$-symmetry test in top decay. One finds that 
$\langle{\cal O}\rangle -  \langle{\bar{\cal O}}\rangle \propto {\rm Im}(g_R- g_R')$
\cite{Bernreuther:1992be,Bernreuther:1993xp}. Other $CP$ asymmetries in
top-quark decay were discussed in \cite{Ma:1991ry}.

A general analysis of the semi- and nonleptonic (anti)top-quark
 decays via $W$ exchange was made in terms of helicity parameters
in \cite{Nelson:1997xd,Nelson:1998pu,Nelson:2000dn}.

While the $b$-jet, $\ell$, $\nu_ell$ energy distributions
and the $b$-jet  and $\nu_ell$ angular distributions   
from top-quark decay can be used to probe anomalous couplings
 (\ref{eq-ffdecomp}) in the $tbW$ vertex, 
the angular distribution
(\ref{dcost}) of the
charged lepton is insensitive 
to small anomalous couplings $f_{L,R}$, $g_{L,R}$ \cite{Grzadkowski:1999iq,Rindani:2000jg}.
This holds for the secondary lepton distribution $\sigma^{-1}\rmd \sigma/\rmd
\cos\theta_\ell \rmd\phi_ell$ irrespective of the  top-quark
 production process \cite{Godbole:2006tq}.

\subsubsection{Decays to charged Higgs bosons:}
\label{subchH} \quad
Many SM extensions which involve a larger Higgs sector  predict
the existence of charged Higgs
bosons $H^\pm$, apart
from the existence of more 
 than one neutral Higgs state. One of the
simplest extensions of the SM results from the addition of a second 
Higgs doublet field. This  leads
to three physical neutral $(h, H, A)$ and a pair of charged $(H^\pm)$
spin-zero bosons. The non-supersymmetric two-Higgs doublet models (2HDM)
are conventionally classified into three types: In the type I model only one
of the two Higgs doublet fields  $\Phi_1, \Phi_2$ 
is coupled to the quarks and leptons at tree level,
while in the type II 2HDM the Higgs doublets  $\Phi_1$ and $\Phi_2$
couple only  to right-handed
down-type fermions $(d_{iR},\, \ell_{iR})$ and 
up-type fermions $(u_{iR},\, \nu_{iR})$, respectively. Type III 2HDM allow for
tree-level couplings to both up-type and down-type fermions for each of the Higgs
doublets \cite{Gunion:1989we}.  
In type III models the exchange of neutral Higgs bosons can mediate
transitions between quarks (and leptons) of the same charge at tree level.
Such couplings are strongly constrained for transitions between $b,s,d$ quarks (and
between charged leptons) \cite{Yao:2006px} -- contrary to transitions $t\to c$. This will be
discussed in section \ref{subfcnc}.

In the phenomenology
of 2HDM extensions of the SM discussed here, 
two new parameters enter, which are taken to
be the mass $m_{H^+}$ of the charged Higgs boson 
 and $\tan\beta =v_2/v_1$, where $v_1,\, v_2$ is the
vacuum expectation value of $\Phi_1$ and $\Phi_2$, respectively.

In 2HDM the mass of $H^\pm$ is strongly constrained by the data 
on the radiative decays ${\bar B}^0 \to X_s \, \gamma$ from
the $B$ meson factories.
The measured inclusive branching ratio agrees very well  with the SM
prediction, 
leaving only a 
small margin for possible new physics contributions. For the 2HDM
type-II models this implies
the lower bound $m_{H^{\pm}} > 350$ GeV \cite{Gambino:2001ew}. Thus,
within these models, 
a charged Higgs boson can affect top quark
decays only by mediating, in addition to $W$ boson exchange, 
three-body decays like  $t\to b c {\bar s},\; b \tau^+ \nu_\tau$.
However, because of the strong lower bound on $m_{H^+}$ the possible changes
of the respective branching ratios are very small compared to the values 
(\ref{eqbrtallow}) predicted by the SM. 
 
A model-independent lower bound, $ m_{H^{\pm}} > 79.3$ GeV at $95\%$ C.L.,
was obtained at the LEP2 collider \cite{Yao:2006px} from the non-observation of
$e^+e^- \to H^+ H^-$.
 
The minimal supersymmetric extension of the SM (MSSM) includes 
a two-Higgs doublet sector of type II. Within the MSSM the constraint
from $B(b\to s \gamma)$  on the mass
of $H^\pm$ does not apply, as the one-loop contributions from $H^-$ exchange
to the $b\to s \gamma$ amplitude can be compensated to a large
extent by the contributions from
the exchange of supersymmetric particles (in particular of charginos and top
squarks). Thus, within the MSSM, $m_{H^+} < m_t$ and 
 the decay of a top
quark into a $b$ quark and an on-shell $H^+$ is still possible.
To Born approximation the decay rate is (putting $m_b=0$ in the phase space function, but
not in the coupling to $H^+$):
\be
\Gamma^B(t\to b \, H^+) = \frac{G_F}{8\pi \sqrt{2}}
m_t^3 |V_{tb}|^2 \, \left(1-\frac{m_{H^+}^2}{m_t^2}\right)^2 \; \left(\frac{m_b^2}{m_t^2}\tan^2\beta
+ \cot^2\beta \right) \,. 
\label{eqgacharH}
\ee
The $\Or(\as)$ QCD corrections  to (\ref{eqgacharH})
are also known \cite{Li:1990cp,Czarnecki:1992zm} and are such that the
ratio of the rates for $t\to b \, H^+$ and $t\to b \,
W^+$ remains almost unaffected by these corrections. 
Comparing (\ref{eqgacharH}) with (\ref{twidth}) one sees that the decay rate 
for $t\to b \, H^+$ becomes comparable in size to $t\to b \, W^+$ for small
and large values
 of $\tan\beta$ if $m_{H^+}$ is not too close to the phase space limit.
For fixed $m_{H^+}$ the rate (\ref{eqgacharH}) is smallest
at $\tan\beta =\sqrt{m_t/m_b}\sim 6$. For instance, putting
$m_{H^+}=140$ GeV, we have $B(t\to b \, H^+) \simeq 0.01$ and $\simeq
0.1$ for $\tan\beta =6$ and $30$, respectively. 
The quantum corrections to
(\ref{eqgacharH}) within the MSSM 
are also known \cite{Coarasa:1996qa}. The corrections depend strongly
on the parameters of the model; they can be large, especially 
for $\tan\beta\gg 1$ \cite{Carena:1999py}.
Values of $\tan\beta \lesssim 1 - 3 $ are not tolerable within the MSSM, because
this would push the mass of the lightest neutral Higgs boson of this model
below its present experimental lower bound \cite{Yao:2006px}. Moreover,
 values of $\tan\beta\gtrsim
\Or(m_t/m_b) \sim 50$ are both theoretically and experimentally disfavored.

The main two-body
decays of a  charged Higgs boson with mass $79 \, {\rm GeV}< m_{H^+} < m_t$ are 
$H^\pm \to \tau \nu_\tau, \, cs,\, c b$. The tree-level partial widths are
given by
\be
\Gamma^B(H^+\to f_u \, {\bar f}_d) = N_c \, \frac{G_F}{ 4\pi \sqrt{2}}
m_{H^+} |V_{f_u f_d}|^2 \, \left(m_{f_d}^2 \tan^2\beta + m_{f_u}^2 \cot^2\beta \right) \,. 
\label{eqgacHdec}
\ee
Here $f_u {\bar f}_d = c {\bar s},\, c {\bar b},\,  \nu_\tau \tau^+$,
$N_c= 3 \, (1)$ for quarks (leptons),  and the CKM matrix elements 
involved are $|V_{cs}| \simeq 0.96 $ and $|V_{cb}| \simeq 4\times 10^{-2}$. In (\ref{eqgacHdec})
the fermion masses were again neglected  in the phase space function, but
not in the couplings to $H^+$. 
Thus for $\tan\beta > 1$, $H^\pm \to \tau \nu_\tau$ is the dominant
channel; we have $B(H^+ \to \tau^+ \nu_\tau)/B(H^+ \to c{\bar s})> 10$
for $\tan\beta > 2$, and  $B(H^+ \to c {\bar b})/B(H^+ \to c{\bar s})> 1$
for $\tan\beta \gtrsim 2.6$.
For $ m_{H^+}> 130$ GeV the decay $H^+ \to W^+ b{\bar b}$ (via a virtual $t$ quark)
becomes also relevant.
Therefore, one expects  the appearance of the decay modes 
 \be
t \to b \, H^+ \to b \tau \nu_\tau, \; b c{\bar s},\; c b {\bar b},
\; W^+ b b{\bar  b} \, ,
\label{eqchHfs}
\ee
with  branching ratios
being considerably larger than the respective SM predictions.

The existence of a charged Higgs boson would not significantly change
the  total  $t\bar t$ 
or single-top cross sections at the Tevatron and 
LHC. Instead one searches for
the appearance of any of the signatures from $t\to H^\pm b$ decay just mentioned --
i.e., for  violations of the (CKM modified) universality of the charged weak current
interactions which is reflected in the SM predictions for the branching ratios
into the different dilepton, single-lepton, and all-jets final states from $t \bar t$
decay. At the LHC such investigations should also be possible 
 for  single top production and  decay.

Searches  by the D0 and CDF
experiments \cite{Abazov:2001md,Abulencia:2005jd,Grenier:2007xj}
for $t \to H^+ b$ in $\ttbar$ events at the Tevatron were negative
so far. Resulting exclusion limits on the mass of  $H^\pm$ and its
couplings to fermions are model-dependent. Analyses are mostly done
within the framework of the MSSM, assuming the appearance of the
above-mentioned final states from top decay. Exclusion limits in the
$\tan\beta$, $m_{H^+}$ plane $(m_{H^+}< m_t)$ are given in  \cite{Abulencia:2005jd}.
Assuming that $H^\pm$ decays exclusively to $\tau \nu_\tau$ then
$B(t \to H^+ b)< 0.4$ at $95\%$ C.L.  \cite{Abulencia:2005jd}.

The discovery potential of the LHC for this top-quark decay mode
has been investigated in simulation studies both by the ATLAS and CMS
collaborations, considering $t {\bar t} \to
H^\pm W^\mp b {\bar b}$ events with decays $W\to \ell \nu_\ell$
 $(\ell=e,\mu)$, $H^\pm \to \tau \nu_\tau$
\cite{Ball:2007zza}, and also  $H^\pm \to c s$ \cite{:1999fr}. These studies
indicate that almost all of the region in the $\tan\beta$, $m_{H^+}$ plane 
 $(m_{H^+}< m_t)$ not yet excluded so far can be covered by the LHC.

For the decay of polarized top quarks $t\to H^+ b$ the 
angular distribution $\Gamma_{H}^{-1}\rmd \Gamma_{H}/\rmd\cos\theta_H$
is of the form  (\ref{dcost}), where $\theta_H$ is the angle between the
top-spin vector and the $H^+$ direction of flight in the $t$ rest frame.
For $ m_{H^+}$ not too close to $m_t$, the correlation coefficient
$c_{H^+}$ is given in type II 2HDM to good approximation by 
\be
 c_{H^+} = \frac{1-(m_b/m_t)^2 \tan^4\beta}{1+(m_b/m_t)^2 \tan^4\beta} \,.
\label{eqspinH}
\ee
The $\Or(\as)$ corrections to (\ref{eqspinH}) are also known
\cite{Korner:2002fx}.
For the corresponding $b$-jet angular distribution we have $ c_b =
-c_{H^+}$.
The observation of this decay mode
and the measurement of these distributions  would allow 
the determination of the parameter $\tan\beta$.

\subsubsection{Decays into supersymmetric particles:}
\label{subsusyd}
\quad 
In the MSSM the lightest neutralino $({\tilde \chi}^0_1)$ is likely
to be the lightest supersymmetric particle, which is stable
and is a candidate for the non-baryonic dark matter of the universe. The reason
for the  lightest supersymmetric particle being stable in the MSSM
is $R$-parity conservation. 
The model is constructed such that a discrete 
symmetry, $R$ parity, is conserved
in any reaction. Each particle is assigned the quantum number
$R\equiv (-1)^{3B+L+2S}$, where $B,L,S$ denotes baryon number, lepton number,
and spin, respectively -- i.e., $R=1$ for the known particles and the
Higgs bosons, and $R=-1$ for their superpartners.

Moreover, one of the two spin-zero superpartners
${\tilde t}_{1,2}$ of the top quark -- ${\tilde t}_1$ by convention --
may also be relatively light. This is 
because quantum corrections involving the large top mass lead to a large
splitting of the masses of ${\tilde t}_1$ and ${\tilde t}_2$
\cite{Ellis:1983ed}. The non-observation of top-squark pair production
at the LEP2 collider yields the (almost model-independent) lower
bound of about $90$ GeV on the masses of these squarks \cite{Yao:2006px}. 

If $m_t > m_{{\tilde \chi}^0_1} + m_{{\tilde t}_1}$ then 
the top quark can 
have the new decay mode
\be
 t \to {\tilde t}_1 \, {\tilde \chi}^0_1 \, .
\label{eq-susytdec}
\ee
A priori, this tree-level
process can have a rather large
branching ratio. 
The subsequent decay channels
of ${\tilde t}_1$ depend on the masses and couplings
of the supersymmetric particles.

1a)  If  the lightest chargino ${\tilde \chi}^+_1$ is the second lightest
supersymmetric particle and its mass satisfies 
$m_{{\tilde \chi}^+_1} <
  m_{{\tilde t}_1}- m_b$ then 
the tree-level decay ${\tilde t}_1 \to b {\tilde \chi}^+_1$ dominates,
 and  ${\tilde \chi}^+_1$ 
decays via  ${\tilde \chi}^+_1 \to {\tilde \chi}^0_1 \, \ell^+
\, \nu_\ell$ and ${\tilde \chi}^+_1 \to {\tilde \chi}^0_1 \, q \,
{\bar q}'$. Thus we have  a new set of final states in top-quark decay,
$t\to b f {\bar f}' {\tilde \chi}^0_1 {\tilde \chi}^0_1$.
In this case hadronic $\ttbar$ production, with one top quark
decaying into $W b$, leads to the new final states
\be
 \ttbar \to W^+ \, b  \, {\bar b}\,  q \, {\bar q}' \,{\tilde \chi}^0_1 \, {\tilde \chi}^0_1, 
\quad W^+ \, b \, {\bar b} \, 
\ell^- \, {\bar \nu}_\ell \,{\tilde \chi}^0_1 \, {\tilde \chi}^0_1 
\label{eqttchiplus}
\ee
plus the charge-conjugated channels, with the two neutralinos escaping undetected.

1b) If $m_{{\tilde \chi}^+_1} >
  m_{{\tilde t}_1}- m_b$, the chargino is virtual and  ${\tilde t}_1$
will decay primarily via the three-body decays 
 ${\tilde t}_1 \to W^+ b {\tilde \chi}^0_1, \,  H^+ b {\tilde
   \chi}^0_1$, and 
${\tilde t}_1 \to b \ell^+ {\tilde\nu}_\ell, \, b {\tilde \ell}^+
{\nu}_\ell$ 
 \cite{Hikasa:1987db,Baer:1991cb,Porod:1998yp}, 
provided the sleptons ${\tilde\nu}_\ell$,  ${\tilde
  \ell}$ and the charged Higgs boson $H^+$ are light enough. 
In this case we have, for instance, 
\be
 \ttbar \to W^+ \, b  \, W^- {\bar b}  \,{\tilde \chi}^0_1 \, {\tilde \chi}^0_1, 
\quad W^+ \, b \, {\bar b} \, 
\ell^- \, {\tilde \nu}_\ell \,{\tilde \chi}^0_1 \, . 
\label{eqttchi1plus}
\ee

2) If the both the lightest chargino ${\tilde \chi}^\pm_1$ and the
scalar leptons are heavier than 
${\tilde t}_1$, the following  situations can arise, depending
on the parameters of the model.
a) The dominant decay of ${\tilde t}_1$ is
 ${\tilde t}_1 \to c {\tilde \chi}^0_1$, a
process induced at one-loop \cite{Hikasa:1987db}.
In this case, hadronic $\ttbar$ production yields the 
final state 
\be
 \ttbar \to W^+ \,  b\, {\bar c} \,{\tilde \chi}^0_1 \, {\tilde \chi}^0_1 \, 
\label{eqtt2chi0}
\ee
plus the charge-conjugated channel. 
The signature of the 
decay $t \to  c \, {\tilde \chi}^0_1 \, {\tilde \chi}^0_1$ is similar to
the flavour-changing neutral current decay mode 
$t \to c\, Z \to c \, \nu\, {\bar \nu}$  which
will be discussed in section \ref{subfcnc}. 
b) Actually in this kinematic situation, the four-body decay modes
${\tilde t}_1 \to b {\tilde \chi}^0_1 f {\bar f}'$ can have
larger rates than  ${\tilde t}_1 \to c {\tilde \chi}^0_1$ in a wide
range of the MSSM parameter space \cite{Boehm:1999tr}. In this case 
the final states from top-decay are identical to those in 1a) and to a
subset of 1b), and the final states from $\ttbar$ decay are identical
to those in (\ref{eqttchiplus}) and to the first set in
(\ref{eqttchi1plus}).

The signals associated with  the final states (\ref{eqttchiplus})
and (\ref{eqttchi1plus}) -- charged leptons (if $W\to \ell \nu_\ell)$,
two or four jets including two $b$ jets, and substantial missing 
transverse energy/momentum -- are identical to those from SM $\ttbar$
dilepton and single lepton decays (see section \ref{secttbar}).
 However, the transverse momentum and angular distributions
differ. The CDF experiment 
has searched for events of this type at the Tevatron, with 
no evidence found \cite{Affolder:2000bp}. 

The signal (\ref{eqtt2chi0}) in the alternative scenario
for ${\tilde t}_1$ decay consists of an isolated high-$p_T$ lepton 
(if $W\to \ell \nu_\ell)$, a $b$ and $c$ jet, and substantial
missing transverse energy/momentum. Background is mainly due to
the production of $W$ plus jet events. The conclusion of \cite{Hosch:1997vf}
is that experiments at the Tevatron should eventually either
discover the decay  $t \to {\tilde t}_1  {\tilde \chi}^0_1$
 in the channel (\ref{eqtt2chi0}) or place an upper bound  of about
 $1\%$ on its
branching ratio. Analyses should account for
the possibility that hadronic production of supersymmetric particles
can increase the number of top and stop events. For instance, a
gluino $\tilde g$ being lighter than the non-top squarks would decay
into 
${\tilde g}\to t {\tilde t}_1^*, {\bar t}{\tilde t}_1$ if
$m_{\tilde g}> m_t +  m_{{\tilde t}_1}$ \cite{Baer:1991cb}.  

In addition, experiments at the Tevatron have explored the possibility
of top-squark pair production in $q \bar q$ annihilation and gluon
fusion, $q {\bar q}, gg \to {\tilde t}_1 {\tilde
  t}_1^*$. The search in \cite{Acosta:2003ys,:2007im} was conducted
under the assumption that  both ${\tilde t}_1$
decay to ${\tilde t}_1 \to b \ell^+ {\tilde\nu}_\ell$.
Searches for signatures of three- and four-body decays of top squarks
and of  both top squarks 
decaying to $c {\tilde \chi}^0_1$
were made in \cite{Abazov:2003wt} and
\cite{Abazov:2006wb,Aaltonen:2007sw,Duperrin:2007uy}, 
respectively.
All these searches were negative and
have excluded  substantial parts of  the kinematically accessible regions in the
$m_{{\tilde t}_1}, m_{\tilde\nu}$ and $m_{{\tilde t}_1}, m_{{\tilde \chi}^0_1}$
planes. 

Experiments at the Tevatron should be able to answer 
whether or not i) a top squark lighter than the top quark exists
and ii) the decay (\ref{eq-susytdec})
exists with a branching fraction $\gtrsim 1\%$. 
In any case, these questions will eventually be clarified at the LHC.

The answer to question i) is of great interest to cosmology.
Among the scenarios that try to explain the observed matter-antimatter
asymmetry of the universe, there is an attractive (viz. testable)
class that relates the generation of this asymmetry to the
electroweak phase transition in the early universe (which happens
at a temperature $T\sim 100$ GeV). These electroweak baryogenesis
scenarios require this phase transition to cause a thermal
non-equilibrium situation, i.e., to be of first order. 
This is excluded within the SM, but is possible for
instance within the MSSM if $m_{{\tilde t}_1}< m_t$ 
(see, e.g., \cite{Carena:1997ki}).

\subsubsection{Flavour-changing neutral current (FCNC) decays:}  
\label{subfcnc} \quad 
In the SM FCNC decays are induced by quantum corrections (at the one-loop level)
and are governed by the Glashow-Iliopoulos-Maiani (GIM) mechanism \cite{Glashow:1970gm}.
The transitions $t\to c$ and $t\to u$ are severely suppressed, because
the rates are determined, apart from the CKM matrix elements, by the differences
of the squared masses of the $b,s,d$ quarks which are very small compared to
$m_t^2$. This leads to tiny branching ratios 
\cite{DiazCruz:1989ub,Eilam:1990zc,Mele:1998ag}
that are many orders of magnitude below the detection
limits of the LHC  \cite{Ball:2007zza,Carvalho:2007yi}. 

In a number of SM extensions 
the branching ratios of these decays can be be significantly enhanced.
Supersymmetric extensions contain new flavour-violating interactions
involving supersymmetric particles.
As a consequence, GIM suppression of the $t\to c, \, u$ transitions may
be  overcome to a large extent in these models. Type-III 2-Higgs
doublet extensions  of the SM (see section
\ref{subchH}) allow for tree-level FCNC 
couplings of neutral Higgs bosons to quarks which are, as far as $t\to c$
transitions are concerned, not yet severely constrained.
Tree-level FCNC quark couplings of the $Z$ boson are present in models  with
exotic quarks, for instance extra heavy quarks being $SU(2)_L$
singlets. Sizeable $t\to c$ transitions 
naturally appear in models of dynamical electroweak symmetry breaking, where the
top quark plays a special role \cite{Hill:2002ap,Valenzuela:2005xr}.

For the decay modes $t\to c X^0$ ($X^0 =\gamma,\, Z,\, g,\, h$)
the order of magnitude of the maximally possible values 
of the  branching ratios $B$ are collected 
in  table \ref{tab-FCNC}, where $h$
denotes either
the SM Higgs boson or a light Higgs boson in one of the
above-mentioned
 SM  extensions
(assuming $m_h \sim 120$ GeV). Constraints from low-energy data were
taken into account. The SM values are from
\cite{AguilarSaavedra:2004wm}, for predictions 
in type-III 2HDM, in the MSSM and in  $R$-parity violating 
($\rpv$)  SUSY models
see \cite{Luke:1993cy,Atwood:1996vj},
 \cite{Yang:1993rb,Guasch:1999jp,Delepine:2004hr,Liu:2004qw,Cao:2007dk}, and 
\cite{Eilam:2001dh,Abraham:2000kx}, respectively. 
Topcolour-assisted technicolour models allow for $B(t\to c Z) \sim
10^{-5}$ \cite{Lu:2003yr}.
In models with
exotic quarks, for instance extra heavy quarks of charge $Q=2/3$ being $SU(2)_L$ singlets,
$B( t\to q Z)$ may be as large as $10^{-4}$ \cite{AguilarSaavedra:2002kr}.  
As to the expectations for the transitions $t\to u$:
in the SM they are   even smaller than $t\to c$, as they are 
CKM-suppressed with respect to this mode. This is not the case in
the MSSM and in  $\rpv$ SUSY: in these models 
$B(t\to c X^0) \simeq B(t\to c X^0)$.  
 If kinematically
possible, the decay mode $t\to c$ + sneutrino can have
 a branching $B\sim  10^{-5}$  in  $\rpv$ 
SUSY models \cite{BarShalom:1998uq}.

\begin{table}[h!]
\caption{\label{tab-FCNC} Branching ratios $B$ of some FCNC 
top quark decay modes, $t\to c \, X^0$, in the SM
and several of its extensions. Expected sensitivities at the LHC are
for a signal at the $5 \sigma$ level and $100{\rm fb}^{-1}$ integrated
luminosity. }
\begin{center}
\renewcommand{\arraystretch}{1.2}
\begin{tabular}{llllll}
\hline
Decay mode & SM & type-III 2HDM & MSSM &  $\rpv$ SUSY &
LHC sensitivity \\
\hline
$ \quad t\to c \; \gamma$ & $\sim 5 \times 10^{-14}$ &  $\sim 10^{-6}$ &
$\sim 5 \times 10^{-7}$ & $\sim 10^{-6}$ & $2 \times 10^{-4}$ \\
 $\quad t\to c\; Z$ & $\sim 10^{-14}$  &  $\sim 10^{-7}$  & 
$\sim 10^{-6}$ & $\sim 3 \times 10^{-5}$ & $3 \times 10^{-4}$ \\
$ \quad t\to c \; g$ & $\sim 5 \times 10^{-12}$ & $\sim  10^{-4}$ &
$\sim 3 \times 10^{-5}$ & $ \sim 10^{-4}$ & $2 \times  10^{-3}$ \\
$\quad t\to c \; h$ & $\sim  3 \times 10^{-15}$ & $\sim  10^{-3}$ &
$\sim 6 \times 10^{-5}$ & $ \sim 10^{-5}$ &  \\
\hline
\end{tabular}
\end{center}
\end{table}

So far, the experimental limits on the strength
of possible FCNC transitions of the top quark are not extremely tight
\cite{Abe:1997fz,Achard:2002vv,Chekanov:2003yt,:2008be,Yao:2006px}.
The presently most precise experimental upper bound on FCNC top decays
to a $Z$ boson was recently
 obtained by the CDF experiment, $B(t\to Z q)<3.7\%$ \cite{cdf-fcnc-note}.
 At the LHC FCNC transitions
can be searched for in (single) top production, which is
sensitive to anomalous $gqt$ couplings  (see
section~\ref{sub-singtnewphys}), and in top-decays, e.g., in $t\bar t$ events
where one top quark decays via the $W b$ decay mode and the other one
into $c \, X^0$ or $u \, X^0$. The best identification
will be reached for $t\to q \gamma$ and $t\to qZ\to q \ell^+ \ell^-.$
Simulation studies were
made by the D0  and the ATLAS
 collaboration  \cite{Ball:2007zza,Carvalho:2007yi}.
Resulting  estimates are given in table~\ref{tab-FCNC}
for the minimum size of the respective branching ratio allowing a discovery
at the LHC with $100 \, {\rm fb}^{-1}$ integrated
luminosity. 
\section{Top-quark pair production}
\label{secttbar}

The main physics goal associated with 
$\ttbar$ events at the
Tevatron, and even more so at the LHC, is the detailed investigation
of the top-quark pair production and top-quark decay dynamics.
Besides the determination of the top-mass, 
key measurements at the LHC will include the total cross section and 
differential distributions.  Specifically, top-spin effects, which offer
additional means to explore the interactions of these quarks, will also
be measurable.  
The LHC provides also the opportunity to dramatically
extend the  search for heavy resonances (which may or may not be associated with electroweak
symmetry breaking) that strongly couple to $\ttbar$ pairs.

At the Tevatron, and according to the SM also at the LHC, 
 $\ttbar$ pairs are produced by the strong
interactions. 
The main partonic subprocesses
are quark-antiquark annihilation, $q {\bar q} \to \ttbar$, which
dominates at the Tevatron,  and
gluon-gluon fusion, $gg \to \ttbar$, which makes up most of the
$\ttbar$ cross section at the LHC. 
As discussed
in section~\ref{sectdec} the $t$ and $\bar t$ quarks decay almost
exclusively into a $W$ boson and a $b$-jet.
The $\ttbar$ signals are then classified according to the decays
of the $W^+W^-$ bosons from $\ttbar \to bW^+ {\bar b} W^-$ as
dilepton, lepton +  jets, and
fully hadronic  decay channels.
\bea
{{p\bar{p},pp\to t\bar{t}\: X \to}}
\left\{\begin{array}{c}
{\mbox{${\ell^+ + \ell'^- + j_b +j_{\bar b}+p_T^{miss}}+ n  \ge 0\; {\rm jets}$}}, \\ 
{\mbox{${\ell^\pm +  j_b+ j_{\bar b} +  p_T^{miss}} + n  \ge 2\; {\rm jets}$}}, 
\label{eqttbarreac}\\
{\mbox{${j_b+j_{\bar b} +  n  \ge 4\; {\rm jets}}$}}. 
 \end{array} \right.  
\eea
From (\ref{eqbrtallow}) we obtain a branching fraction of 
$10.5\%$, $43.5\%$, and $45.5\%$ for the dilepton, lepton + jets, and
all jets modes, respectively. Many
top-physics analyses (will) use dilepton and lepton +
 jets final states with $\ell =e, \mu$ only. In this case the 
respective branching ratios are $4.7\%$, $29\%$, and  $45.5\%$.

All the channels (\ref{eqttbarreac}) were observed and analyzed at
the Tevatron (see \cite{Chakraborty:2003iw,Wagner:2005jh,Quadt:2006jk}
for reviews). Detailed simulation studies were made for the LHC
by the ATLAS \cite{:1999fr,Borjanovic:2004ce} and, more recently, by the CMS 
collaboration \cite{Ball:2007zza}. 

The cleanest signals for $\ttbar$
 production are provided by the dilepton and the lepton + jets
 channels. The signature for the $\ell \ell$ channel consists of 
two high $p_T$, isolated, and oppositely charged leptons, large
missing transverse energy/momentum, and at least two jets which
originate from $b$ quarks. The main background reactions with 
final states that can mimic the signal are the production
of $Z$ + jets, $VV$ + jets ($V=W,Z$), $\ell+4j$,  and fully
hadronic $\ttbar$ channels. After setting  appropriate selection criteria
including the requirement of $b$-tagging, a signal-to-background ratio
$S/B=12$ and a selection efficiency
$\epsilon \simeq 5\%$ was estimated  \cite{Ball:2007zza} for the $\ell
\ell$ channel at the LHC. 

The signature for the $\ell$ + jets channel consists of an isolated,
high $p_T$ charged lepton, large
missing transverse energy/momentum, and at least four jets, 
two of which originate from $b$ quarks. The main backgrounds come from the
production of $Wb{\bar b}$ + 2 jets, $W$ + 4 jets, and from the 
other $\ttbar$ channels.
For the $\ell$ + jets channel very pure samples should
be obtained at the LHC ($S/B\simeq 27$ and $\epsilon \simeq 6.3\%$ \cite{Ball:2007zza}). 
In this channel the complete final state can be reconstructed
(up to a two-fold ambiguity which results from the solution of a
quadratic equation) by solving kinematic equations,
assuming that $E_T^{\rm miss} = E_T^\nu$.

The fully hadronic channel has at least 6 jets in the final state.
It has the largest branching ratio of the three modes
(\ref{eqttbarreac}), and the event kinematics can be fully
reconstructed. However, this channel is polluted by a large QCD
multijet background. Requiring
$b$-tagging and requiring the signal events to contain rather 
high $p_T$, the  simulation study  \cite{Borjanovic:2004ce} obtained
a ratio $S/B=1/9$ and a selection efficiency
 $\epsilon \simeq 2.7\%$   for the low-luminosity phase of the LHC.
A similar result was reached in \cite{Ball:2007zza}.

In view of the rather large $\ttbar$ cross section and the 
distinct signatures of the dilepton and lepton +  jets channels,
these modes are perfect physics events to analyze during the first
data taking phase of the LHC detectors (with a few hundred ${\rm
  pb}^{-1}$ of integrated luminosity). In this phase
the focus will be  on 
detector issues such as measuring the  factors that determine the
calibration of the jet-energy scale, and measuring the $b$-tagging
efficiency. As far as top quark physics is concerned,
first measurements will include the $\ttbar$ cross section in the
various channels  and the 
determination of the top mass \cite{D'hondt:2007aj}.

In the ``discovery phase'' of the LHC
  millions of $\ttbar$ pairs will be produced already with
$10\, {\rm fb}^{-1}$ of integrated luminosity
(c.f. table~\ref{tab-topxsec}).  For most
top-quark observables, statistical uncertainties will then be below
the percent level; i.e., the measurements will
eventually be systematics dominated. 

In the following subsections we shall discuss $\ttbar$ production
mostly from the perspective of considering the top quark to be a signal.
Production of $\ttbar$ pairs is, on the other hand, also an important
background to the search for new particles, including the 
searches for the SM and/or non-standard Higgs bosons and for signals
of supersymmetry.  Obviously, both roles the top quark plays at the
Tevatron and at the LHC require accurate predictions of $\ttbar$
production and decay. 

\begin{figure}[ht]
\begin{center}
\includegraphics[width=12cm]{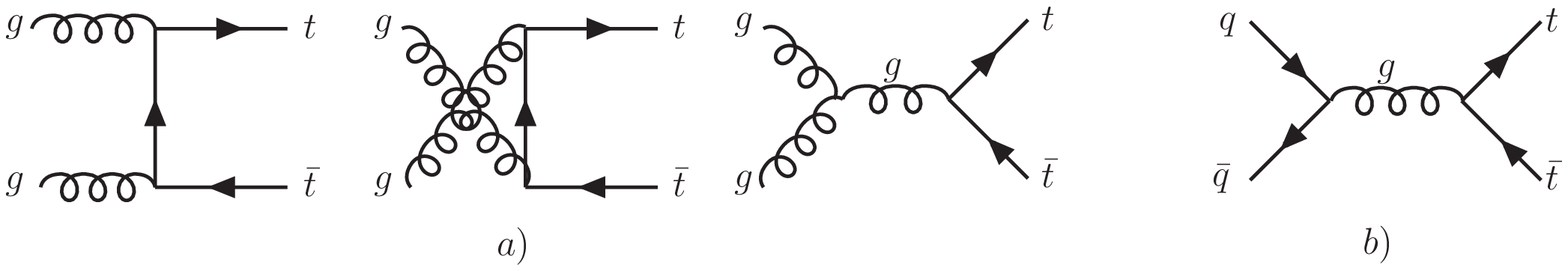}
\end{center}
\caption{\label{fig-ttborn} Lowest order Feynman diagrams for
$\ttbar$ production by the strong interactions: $gg\to \ttbar$ (a)
and $q {\bar q} \to \ttbar$ (b).}
\end{figure}

\subsection{Status of theory}
\label{sub-SMttbar}

Because $m_t\gg \Lambda_{QCD}$,  top-quark production and decay
processes are hard scattering reactions which can be computed in (QCD)
perturbation theory. The $\ttbar$ production processes are
depicted to lowest-order QCD  in figure~\ref{fig-ttborn}. At
next-to-leading order (NLO) in the QCD coupling $\as$, also
$qg$ and ${\bar q} g$ scatterings produce $\ttbar$ pairs.
To arbitrary order in QCD perturbation theory, the total $\ttbar$
cross section for
\be
 p {\bar p}, \, p p \, \to \, \ttbar \, + \, X 
\label{eq-hadrtt}
\ee
is  given as a convolution of the cross sections for the partonic subprocesses
and the parton distribution functions (PDF) -- up to terms which are
suppressed with some power of the hadronic center-of-mass energy
$\sqrt{s}$ (so-called higher twist terms):
\be
\sigma_{h_1 h_2}^{\ttbar}(s, m_t)   = 
\sum\limits_{i,j}\int_0^1 {\rmd x_1} {\rmd x_2} \: f_{i}^{h_1}(x_1,\mu_F) 
f_{j}^{h_2}(x_2,\mu_F) \: {\hat\sigma}_{ij}(\shat, m_t,\as(\mu_R), \mu_R, \mu_F) 
\,  .
\label{eqtotcs}
\ee
Here $i,j=g,q, {\bar q}$, and  $h_1,h_2 = p, {\bar p}$.
The PDF  $f_{i}^{h}(x, \mu_F)$ is  the probability density
 of finding
parton $i$ with longitudinal momentum fraction $x$ in hadron $h$ at
the factorization scale $\mu_F$. This scale, which is arbitrary in principle,
  is usually set equal to a typical scale of the problem,
  e.g. $m_t$, in order to avoid large logarithms in perturbation 
  theory. The cross section of a partonic subprocess is denoted
by ${\hat\sigma}_{ij}$, ${\hat s} =x_1x_2 s$ is the square of the 
partonic center-of-mass energy, and $\mu_R$ denotes the
renormalization scale. This scale is also arbitrary and need not be
the same as $\mu_F$; but again one should take it to be of the order
of a typical energy scale of the partonic processes in order not to
generate large logarithms. The hadronic cross section
 must not -- as an
observable -- depend on the choice of $\mu_R$ and $\mu_F$ -- but
computed in fixed-order perturbation theory, it does. 
The dependence of a hadronic observable $O_H$ on $\mu_R$ and $\mu_F$
decreases with increasing order of the perturbation series.
 Often one varies  $\mu_R$ and $\mu_F$ in some range and takes the
 resulting spread in the value of $O_H$ -- for no deeper reason -- as
 an estimate of the theoretical error, i.e.,  of the size of the
 uncalculated higher-order perturbative contributions. In higher-order
QCD the total cross section should be calculated in terms of a
short-distance mass parameter, e.g. the 
 $\overline{\rm MS}$ mass $\overline{m}_t$ rather than the pole mass $m_t$.
(Recall that in lowest-order perturbation theory the mass
parameter is not yet specified.)

Formula~\ref{eqtotcs} applies also when electroweak (and non-SM)
interactions are taken into account. Analogous formulae hold for
differential distributions, like the $p_T$ distributions of the $t$ and
$\bar t$ quarks or the $\ttbar$ invariant-mass distribution. Unlike
the total cross section, differential distributions often involve more
than one typical scale.  For instance, if $p_T,\mtt \gg m_t$, the choice of
$\mu_R$ and $\mu_F$ requires more scrutiny. Moreover, for adequate
predictions, top-quark decay must also be taken into account in the
computation of $\rmd{\hat\sigma}_{ij}$. 

\subsubsection{SM results:} \label{sub-smttresu} \quad
Quite a number of theoretical investigations have been made on
hadronic $\ttbar$ production. The results which were
 obtained to
higher orders in the SM couplings may be classified as follows.\\
1) Predictions for  on-shell $\ttbar$  states, summed over their spins: 
\begin{itemize}
\item NLO QCD corrections ($\Or(\as^3)$) for the total cross section
 \cite{Nason:1987xz,Beenakker:1988bq},  
 $p_T$ and rapidity distributions 
\cite{Nason:1989zy,Beenakker:1990maa}, and double-differential spectra
including the $\ttbar$ invariant-mass distribution and azimuthal
correlations \cite{Mangano:1991jk,Frixione:1995fj}. The cross section
and single particle distributions such as $p_T$ and rapidity
distributions 
are insensitive to the $t$ and $\bar t$ spin degrees of freedom, and
it is acceptable to calculate these quantities for on-shell stable top
quarks.
\item Threshold resummations: The NLO QCD 
differential cross sections 
$\rmd {\hat\sigma}_{ij}$ of the hard scattering subprocesses contain
logarithms that become large near threshold. Here `threshold'
refers not only
to the $\ttbar$ production threshold, but more generally to the
boundary of phase space where $\mtt/\shat$ becomes equal to 1.
These threshold logarithms can be summed \cite{Sterman:1986aj,Catani:1989ne}. 
This has been done, in various approaches, for the total cross section
 \cite{Kidonakis:1997gm,Bonciani:1998vc,Kidonakis:2001nj,Kidonakis:2003qe}
and for the $p_T$ distributions \cite{Kidonakis:2003qe,Banfi:2004xa}.
\item Mixed electroweak-QCD corrections ($\Or(\as^2\alpha)$):
The weak corrections ($W$, $Z$, and Higgs boson exchange)
 were first computed in \cite{Beenakker:1993yr}, and later  in a more
complete fashion for $q {\bar q} \to
\ttbar$ \cite{Bernreuther:2005is,Kuhn:2005it,Bernreuther:2008md}, 
$gg \to \ttbar$ \cite{Bernreuther:2006vg,Kuhn:2006vh,Moretti:2006nf},
and
$gq \, ({\bar q}) \to \ttbar q \, ({\bar q})$ \cite{Bernreuther:2006vg,Bernreuther:2008md}.
The photonic corrections were determined by
\cite{Hollik:2007sw}. Here partonic
$\ttbar$ production processes involving an initial-state photon appear, the most 
important one being $\gamma g \to \ttbar$.  The $\Or(\as^2\alpha)$
contributions  to the $\ttbar$ cross section are smaller than the
present QCD uncertainties, but these corrections are relevant for
 distributions, especially in the high-energy regime 
(see section~\ref{sub-topdist}).
\item $\ttbar$ + jet production: The NLO QCD corrections ($\Or(\as^4)$) 
to the  cross section of this  process were computed in 
 \cite{Dittmaier:2007wz}. It is important to know these corrections, as 
$\ttbar$ plus one hard jet constitutes a significant fraction of the
inclusive $\ttbar$ sample. Moreover, this process is
an important background for Higgs boson searches at the LHC. 
\end{itemize}
2) For  on-shell $\ttbar$ states, with $t$ and $\bar t$ spin degrees
of freedom fully taken into account,  the differential cross sections for
$\ttbar$ production by $gg, q {\bar q}, g q \, ({\bar q})$ initial
states were determined to NLO QCD \cite{Bernreuther:2001rq,Bernreuther:2004jv}.
The  mixed weak-QCD corrections were also computed 
\cite{Bernreuther:2005is,Bernreuther:2006vg}. These results allow
 for predictions of distributions induced by top-spin effects 
(see section~\ref{sub-tspcorr}). \\
3) For off-shell $t$ and $\bar t$ intermediate states, the
non-factorizable QCD corrections of $\Or(\as^3)$ are known
\cite{Beenakker:1999ya,Meyer} -- see below.

The physics effects of these corrections will be discussed in the
following subsections.   In addition to the above list, results
were published \cite{Czakon:2007ej,Czakon:2007wk}
for the two-loop virtual QCD corrections  for $q {\bar q}, g g\to
\ttbar$: these corrections were determined
by \cite{Czakon:2007ej,Czakon:2007wk}  in the
kinematic limit where all Mandelstam invariants are much larger than
$m_t^2$. Recently, the two-loop QCD corrections to  $q {\bar q} \to
 \ttbar$ were computed in the whole kinematic regime \cite{Czakon:2008zu}.
   These are ingredients required for
a computation of the
inclusive $\ttbar$ cross section to $\Or(\as^4)$ (c.f. also
\cite{Korner:2008bn}).
Such a computation requires, in addition, an efficient method to 
handle the soft and collinear singularities to NNLO QCD associated
with real radiation. 

\begin{figure}[ht]
\begin{center}
\includegraphics[width=10cm]{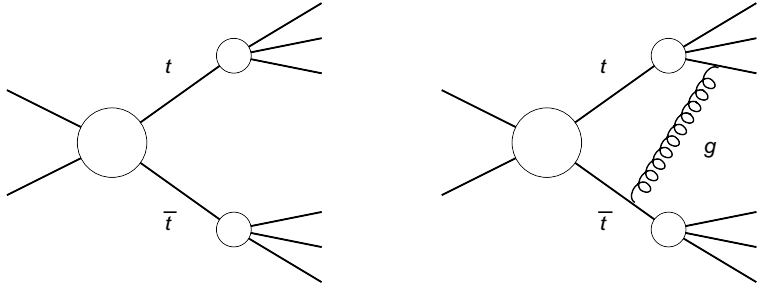}
\end{center}
\caption{\label{fig-nonfac} Illustration of factorizable (left) and
nonfactorizable (right) virtual QCD corrections.}
\end{figure}

In the remainder of this subsection we have a closer look at $\ttbar$
production and decay at NLO QCD. This involves the 
following $2 \to 6$ and $2\to 7$ parton reactions: 
\bea
{q\bar{q} \: {\stackrel{t\bar{t}}{\longrightarrow}} \:
b+\bar{b}+4\ f \, }  \; {(+\,g)}, \label{eqqqtt} \\
{gg  \: {\stackrel{t\bar{t}}{\longrightarrow}} \:
b+\bar{b}+4\ f \, }  \; {(+\,g)}, \label{eqggtt} \\
{g \, q  \: {\stackrel{t\bar{t}}{\longrightarrow}} \:
b+\bar{b}+4\ f \; +q },   \label{eqqgtt} \\
g \, \bar{q}  \: {\stackrel{t\bar{t}}{\longrightarrow}} \:
b+\bar{b}+4\ f \; + \;\bar{q},   \label{eqbqgtt}
\eea
where $f=q,\ell,\nu_\ell$. 
Because  ${\Gamma_{t}\ll m_{t}}$,  the  $t, \, \bar{t}$ quarks are narrow
resonances. 
Thus the double pole approximation is appropriate (we consider here
top as signal, not as background); 
i.e., the S matrix elements of the 
reactions (\ref{eqqqtt}) - (\ref{eqbqgtt}) 
(which can proceed through many intermediate
states other than $t \bar t$) are
expanded around their poles in the complex $t, \bar t$ energy planes,
and only the  term $\propto (D_t D_{\bar t})^{-1}$ of each 
matrix element is kept. ($D_t=p_t^2-m_t^2+\rmi m_{t}\Gamma_{t}$.)\\ 
In the double pole approximation the  radiative
corrections -- both the real and virtual ones --
can be classified into  factorizable and nonfactorizable corrections. 
In figure~\ref{fig-nonfac} this classification is illustrated for virtual corrections. 
While in  figure~\ref{fig-nonfac} (left) the radiative corrections are confined to
the $t \bar t$ production and/or the $t$ and/or $\bar t$ decay parts
of the amplitude, the gluon exchange depicted in figure~\ref{fig-nonfac} (right) connects
the production and decay parts. This classification applies also
to the squared matrix elements $|{\cal M}|^2$ of real gluon
radiation.

The differential  cross sections
for the above parton reactions  (\ref{eqqqtt}) - (\ref{eqbqgtt}) 
are to $\Or(\as^3)$:
\be
\rmd{\hat\sigma}_{ij} =\rmd{\hat\sigma}_{ij, B} + \rmd{\hat\sigma}_{ij,
  fact}  + 
\rmd{\hat\sigma}_{ij, nf} \, , 
\label{eq-bfac-nf}
\ee
where  $i,j = q,{\bar{q}}, g$ 
and $\rmd{\hat\sigma}_{ij, B}$ is the lowest order differential cross
section. For $ij=gq, g{\bar{q}}$ there is only  
$\rmd{\hat\sigma}_{ij, B}$ to this
order in $\as$. 
In computing $\rmd{\hat\sigma}_{ij, fact}$ we may apply the narrow width 
approximation,  $\Gamma_t/m_t \to 0$, for $t$ and $\bar t$. 
This means one neglects terms of
order $\as \Gamma_t/m_t$ with respect to the
Born term, which are parametrically smaller than the uncalculated NNLO
QCD corrections.
Then 
\be
{ \rmd{\hat\sigma}_{ij, fact} \propto {\rm Tr} \:
\left ( R^{(ij)}\rho_{f_1} {\bar \rho}_{{\bar f}_2}\right ) } \, .
\label{trrrr}
\ee
Here $(R^{(ij)})_{\alpha \alpha',\beta \beta'}$ are 
the  $t \bar t$ production density matrices, where $\alpha \alpha'$
and $\beta \beta'$ are the spin labels of the $t$ and $\bar t$ quarks,
respectively.
The density matrices that
describe the decays $t\to f_1$ and ${\bar t} \to {\bar f}_2$ are
  denoted by 
$\rho_{f_1}$ and  ${\bar \rho}_{{\bar f}_2}$, respectively.
The trace in (\ref{trrrr}) refers to the $t$ and $\bar t$ spin labels.
The  $R^{(ij)}$ are known to $\Or(\as^3)$ and $\Or(\as^2\alpha)$
 for the  initial states 
$ij=q {\bar{q}}, gg, gq, g {\bar{q}}$ and
intermediate states $t {\bar{t}}$ and $t {\bar{t}} g$
\cite{Bernreuther:2004jv,Bernreuther:2005is,Bernreuther:2006vg}.
The NLO decay density matrices can be extracted from the results of
\cite{Czarnecki:1990pe,Brandenburg:2002xr}; for details, 
see \cite{Bernreuther:2004jv}.
The $R^{(ij)}$, $\rho_{f_1}$, and ${\bar \rho}_{{\bar f}_2}$ serve also as
building blocks in the computation of $\rmd{\hat\sigma}_{ij, fact}$ when 
the intermediate  $t, {\bar t}$ quarks are taken to be 
off-shell (in the double pole approximation).

To order  $\as^3$, nonfactorizable QCD corrections
contribute to the differential cross sections of the reactions 
(\ref{eqqqtt}) and (\ref{eqggtt}). Figure~\ref{fig-nonfac} (right)
shows an example of a virtual nonfactorizable correction. The real
corrections can be grouped into diagrams where a gluon is radiated 
from an initial, intermediate, or final state, and
 the nonfactorizable real corrections arise from the interference of
these different classes.
The  nonfactorizable QCD corrections are dominated by 
gluon exchange/radiation with energy 
$E_g  \lesssim \Or(\Gamma_t)$ \cite{Beenakker:1999ya}. 
(Gluons with energies $E_g \gg \Gamma_t$
drive the intermediate $t$ and/or $\bar t$ quark far off-shell, and
these contributions to $\rmd{\hat\sigma}_{ij}$ can be neglected.)
They  contribute, e.g., to  $t, \, \bar t$ momentum distributions,   
and the   $t, \, \bar t$, and $t \bar t$ invariant mass distributions.
However, when computing observables which are inclusive in both the
$t$ and  $\bar t$  invariant masses, the 
nonfactorizable  QCD corrections of order $\as^3$
cancel \cite{Fadin:1993dz,Melnikov:1993np,Beenakker:1999ya}.

Studies where the
intermediate $t$ and $\bar t$ quarks are non-resonant were made at
leading-order by \cite{Kauer:2001sp,Kauer:2002sn}. This is relevant
for $\ttbar$ as background to new physics searches.

\subsubsection{Simulation tools:} \label{sub-simttool} \quad
A crucial ingredient in analyzing and interpreting experimental data on
top-quark production and decay is the accurate modeling of the signal
and background events. It must provide a description
of events for colour-singlet hadrons, while the perturbative results
listed in section~\ref{sub-smttresu} make only predictions
at the level of coloured final-state partons. 
This  modeling is 
done -- on the theoretical side -- 
with computer simulation programs,  
 so-called Monte Carlo (MC) generators. 
These are very complex and sophisticated tools, 
which we can mention here only
in passing. (For an introductory review, see \cite{Mangano:2005dj}.)
General purpose programs like PYTHIA \cite{Sjostrand:2003wg,Sjostrand:2007gs}
or HERWIG \cite{Corcella:2002jc,Bahr:2008pv} not only contain
(lowest order) hard scattering parton
matrix elements, but simulate also the emission of additional partons
from the initial and final-state partons in the hard process
 (parton showering), and model the formation of hadrons from 
 partons.
A number of programs exist that simulate hard hadronic collisions
a the level of partonic final states; for a detailed list, see \cite{cedar}.
Tools that include $\ttbar$ and single-top production and decay
and background processes at tree-level are the program packages  
TopRex \cite{Slabospitsky:2002ag}, ALPGEN \cite{Mangano:2002ea}, 
MadGraph/MadEvent \cite{Maltoni:2002qb}, and GR@PPA \cite{Tsuno:2006cu}. MC codes which incorporate
$\ttbar$ and single-top production at NLO QCD are MC@NLO \cite{Frixione:2006gn},
 MCFM \cite{CambEll,Ellis:2006ar}, and POWHEG 
\cite{Frixione:2007nw,Frixione:2007nu}.   For the simulation
of single top events at NLO QCD, see also \cite{Boos:2006af}. An important
aspect of  these codes is the use of a consistent
method of interfacing  the hard scattering part 
with a parton shower algorithm  in order to avoid double counting.
This problem arises when both  
the hard-scattering and the shower algorithm generate 
the same final state.

A further important issue in interpreting top-physics
data, which is least understood in terms of ab initio
calculations, is how the fragments of the (anti)proton that do not take part
in the hard scattering process evolve and affect the final states from
single top or $\ttbar$ decay. Some (long range) interaction between this
so-called underlying event and the partons involved in the
 hard scattering must occur in order to maintain the over-all colour
neutrality and the conservation of baryon number. Empirical models are tuned
with data on multiple interactions in hadronic collisions \cite{Mangano:2005dj}.

\subsection{The total cross section}
\label{sub-tottt}

The basic observable for the reactions (\ref{eq-hadrtt})
is the inclusive $\ttbar$ cross section. Predictions
involve the cross sections ${\hat\sigma}_{ij}$
for the parton processes
$ij\to \ttbar X$ and the parton luminosities
  $L_{ij}^{h_1 h_2}$. Formula
 (\ref{eqtotcs}) can be cast into the form  \cite{MochUw08}:
\be
\sigma_{h_1 h_2}^{\ttbar}(s, m_t) = \sum\limits_{i,j}\int_{4m_t^2}^s
\rmd\shat \, L_{ij}^{h_1 h_2}(s,\shat,\mu_F) \,
{\hat\sigma}_{ij}(\shat,m_t,\mu_R,\mu_F)\, ,
\label{eq-totcs2}
\ee
where 
\be
L_{ij}^{h_1 h_2}(s,\shat,\mu_F)= \frac{1}{s}\int_{\shat}^s
\frac{\rmd s'}{s'}
f_{i}^{h_1}\left(\frac{s'}{s},\mu_F\right)f_{j}^{h_2}
\left(\frac{\shat}{s'},\mu_F\right)\,,
\label{eq-lumi}
\ee
and $\shat = x_1x_2 s$. 

At the Tevatron the largest parton flux is
$L_{q \bar q}$, followed by $L_{qg}=L_{g\bar q}$, while  $L_{gg}$ is the smallest
one. On the other hand, ${\hat\sigma}_{qg}$ is much smaller
 than ${\hat\sigma}_{q \bar q}$ and ${\hat\sigma}_{gg}$, as the former is
a NLO correction of order $\as^3$.  Using the 
PDF sets from the CTEQ \cite{Pumplin:2002vw} or 
MRST \cite{Martin:2002aw,Martin:2003sk} collaborations
one arrives, to NLO QCD,
at the following relative contributions to 
the $\ttbar$ cross section: $\sim 85\%$ $(q {\bar q})$ and $\sim
15\%$ $(gg)$, while the contribution from $qg$ and $g {\bar q}$ initial 
states is only at the percent level. 

At the LHC the  $qg$ scattering occurs with the highest
parton luminosity, but the overall contribution to the hadronic
cross section is again small because of the small partonic cross section
compared with ${\hat\sigma}_{gg}$ and ${\hat\sigma}_{q \bar q}$. The
${\bar q} g$ contribution, which involves a smaller
parton flux than  $qg$ scattering,  is even further suppressed. 
At the LHC  the most important contribution to  $\sigma_{pp}^{\ttbar}$
comes from gluon-gluon fusion, where a large partonic cross section combines
with the second-largest flux, followed by the 
one from $q \bar q$ annihilation. At NLO QCD,  
the $gg$ and $q {\bar q}$ processes contribute
 about $90\%$ and $10\%$  to  $\sigma_{pp}^{\ttbar}$, respectively, 
 while that of the $qg$ channel is at the percent level.
One should note that the  $\ttbar$ cross section calculated 
to leading-order QCD has large uncertainties and
is significantly smaller than $\sigma_{\rm NLO}$:
the ratio $\sigma_{\rm NLO}/\sigma_{\rm LO}$ (computed with NLO and LO
PDF sets, respectively) is $\sim 1.25$ for the Tevatron 
and $\sim 1.5$ for the  LHC.

Detailed updates of the $\ttbar$ cross section at the Tevatron  
and the LHC at NLO QCD -- with threshold resummations included --
 have recently been made by \cite{MochUw08} and by \cite{Cacciari:2008zb}. Uncertainties were
 taken into account which arise from 
the choice of $\mu_R$, $\mu_F$ and from uncertainties
in the PDF sets (which are due to the uncertainties of the experimental
data used in the PDF fits). Previous predictions include 
\cite{Bonciani:1998vc,Kidonakis:2001nj,Kidonakis:2003qe,Cacciari:2003fi}.

In \cite{MochUw08} also an approximate
next-to-next-to-leading order (NNLO) result for the $\ttbar$ cross
section was derived. The argumentation of  \cite{MochUw08} yields
  all powers in $\ln\beta_t$ at
two loops (where $\beta_t$ is the top-quark velocity), the exact NNLO
scale dependence, and also the two-loop Coulomb corrections at the
$\ttbar$ threshold -- up to some constants. They can be obtained only by an explicit
NNLO computation of the $q {\bar q}$ and $gg$ initiated $\ttbar$ cross
sections.

Replacing the upper endpoint $s$ in the integral (\ref{eq-totcs2})
by $\shat_{max}$, it is interesting to determine the value
 of $\shat_{max}$ of the squared
parton center-of-mass energy for which the resulting $\sigma$ saturates the total
$\ttbar$ cross section to a large fraction, say $80\%$ ($95\%$). At the Tevatron
this happens at $\sqrt{\shat}\lesssim 470$ (600) GeV.
At the LHC the available phase space is significantly
 larger and a $80\%$ ($95\%$) saturation occurs at 
$\sqrt{\shat_{max}}\simeq 600$ (1000) GeV \cite{MochUw08}.

This means that at the Tevatron 
most of the $\ttbar$ production occurs near threshold.
Putting the momentum fractions of the initial-state
partons $x_i \simeq x_j= x_{thr}$,  one gets
$x_{thr} \simeq 2m_t/\sqrt{s} \simeq 0.17$ for the Tevatron. At these
values the valence quark PDFs are considerably larger than the gluon density.
The latter is rather poorly known in this regime. 
The near-threshold domination of the cross section at the Tevatron
had motivated the application of the threshold resummation methods
 (see section~\ref{sub-smttresu}). The computations
 of the cross section to fixed order ($\Or(\as^3)$)
and with threshold resummation included (properly incorporated in order to
avoid double counting) differ only $\sim 5\%$; but the inclusion of 
threshold resummation reduces the dependence of the prediction on 
variations of $\mu_{F}$ and $\mu_R$, which is reassuring. Several approximation
schemes have been
employed in the evaluation of 
$\sigma^{\ttbar}_{p \bar p}(\sqrt{s}=1.96 \, {\rm TeV})$. 
In  \cite{MochUw08,Cacciari:2008zb} analyses were made  based on the NLO
cross section including the resummed leading and
next-to-leading threshold logarithms (NLO+NLL approximation). Spreads of
 $\sigma^{\ttbar}_{p \bar p}$ are computed which arise from
scale variations  in the range
$m_t/2 \leq \mu_{F}, \mu_R \leq 2 m_t$, and from the use of
two recent PDF sets, CTEQ6.5 \cite{Tung:2006tb} and MRST-2006 
\cite{Martin:2007bv},
 including their uncertainties.  The resulting values of the
cross section, obtained as a function of $m_t$, have an uncertainty
of about $\pm  12\%$. For instance, for $m_t=171$ GeV,  \cite{Cacciari:2008zb}
obtains 
$\sigma^{\ttbar}_{p \bar p} =7.35^{+0.30}_{-0.53} \, (scales) \,
^{+0.53}_{-0.36} \, 
(PDF)$ pb using the CTEQ6.5 set. This is in accord with
\cite{MochUw08}. Employing the MRST-2006 set results in smaller PDF
uncertainties. The use of the approximate NNLO cross section of  \cite{MochUw08}
drastically reduces the uncertainty due to scale variations,
and results in an estimated total uncertainty for $\sigma^{\ttbar}_{p
  \bar p}$ of $\pm 8 \%$ (CTEQ6.5) and $\pm 6\%$ (MRST-2006) \cite{MochUw08}.

 The cross section decreases with increasing
$m_t$; the change is approximately given by 
$\Delta\sigma/\sigma \approx -5 \Delta m_t/m_t$ \cite{Catani:1996dj}.

At the Tevatron the CDF and D0 collaborations have measured the $\ttbar$
cross section in the channels (\ref{eqttbarreac}) with several
methods (see \cite{Abulencia:2006in,:2007qf,Abazov:2008yn} and references
therein, and \cite{Leone:2007mk} for a recent overview).
 The combination of the CDF results as of
summer 2007 yields $\sigma^{\ttbar}_{p \bar p} = 7.3\pm 0.5 ({stat.}) \pm 0.6
({syst.}) \pm 0.4 ({lumi.})$ pb 
\cite{Leone:2007mk}.
From a recent combined measurement of the
 cross section and the ratio $R$ defined in 
(\ref{eqcdfr}), D0 obtained $\sigma^{\ttbar}_{p \bar p} =
8.18^{+0.90}_{-0.84} ({stat. + syst.}) \pm 0.50 ({lumi.})$ pb \cite{Abazov:2008yn}.
The SM predictions cited above are in agreement with
these values.

\begin{figure}[ht]
\begin{center}
\includegraphics[width=10cm]{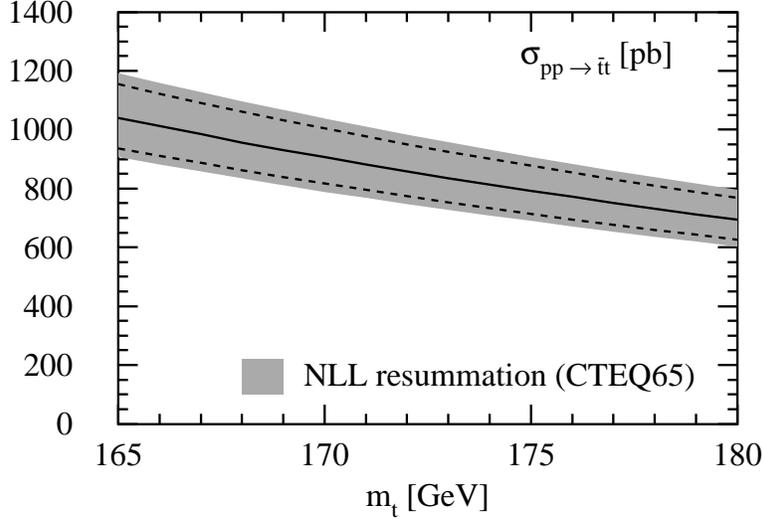}
\end{center}
\caption{\label{fig-totcrLHC} 
The total $t{\bar t}$ cross section at the LHC $(\sqrt{s}= 14$ TeV) resummed to 
NLL accuracy in QCD as a function of $m_t$ \cite{MochUw08}.
The solid line is the central value for $\mu=m_t$, the dashed lower and upper
lines correspond to $\mu=2m_t$ and $\mu=m_t/2$, respectively.
The shaded band denotes the range with the PDF uncertainty of the CTEQ6.5
set~\cite{Tung:2006tb} included.}
\end{figure}

As discussed above, most of the $\ttbar$ cross section at the LHC
comes from $gg$ fusion, and a simple kinematic consideration  as done above
shows that $\sigma^{\ttbar}_{\rm LHC}$ probes the gluon density in 
a regime where it is quite well known ($x \sim x_{thr}\sim 0.025$). 
The 
cross section shown in  figure~\ref{fig-totcrLHC}  exhibits
 the computation of \cite{MochUw08} based on 
the NLO+NLL approximation. The 
range of $\sigma^{\ttbar}_{\rm LHC}$ is plotted as a function of
$m_t$, where scale and PDF uncertainties are added linearly.
For a given $m_t$ 
the uncertainty of  $\sigma^{\ttbar}_{\rm LHC}$ with respect to its
central value is about $\pm 15\%$ and is dominated by the
scale uncertainty. The figure shows that
$\Delta\sigma/\sigma \approx -5 \Delta m_t/m_t$ holds also at the LHC.
The results obtained in  \cite{Cacciari:2008zb} have slightly smaller errors.
Again, the approximate NNLO calculation of $\sigma^{\ttbar}_{\rm LHC}$ 
by \cite{MochUw08} drastically reduces the scale uncertainites, 
and results in an estimated total uncertainty for 
$\sigma^{\ttbar}_{LHC}$ of $\pm 6 \%$ (CTEQ6.5) and $\pm 4\%$
(MRST-2006). While it is gratifying that this calculation is in accord
with the NLO+NLL approximation of figure~\ref{fig-totcrLHC}, it is --
needless to say -- no substitute\footnote{The
  recent suggestion \cite{Nadolsky:2008zw} to use top-quark pair production at the LHC as
an additional calibration process for the parton luminosities requires
a rather precise prediction of $\sigma^{\ttbar}$.} for an complete NNLO computation of
$\sigma^{\ttbar}$. 

The weak \cite{Bernreuther:2006vg,Kuhn:2006vh} 
and electromagnetic \cite{Hollik:2007sw} corrections of
$\Or(\as^2\alpha)$
are, as far as the total cross section is concerned, smaller
than the uncertainties of the present QCD predictions.
With respect to the  NLO QCD cross section, the weak and photonic
contributions amount to a correction of about $0.5\%$ and $-1.4\%$ for the Tevatron
and about $-1\%$ and $0.5\%$ for the LHC, respectively. 

With which precision will the $\ttbar$ cross section be measurable
at the LHC? Simulation studies
were made by the ATLAS \cite{:1999fr} and the CMS 
collaboration \cite{Ball:2007zza}. The lepton + jets  and the
dilepton channels (\ref{eqttbarreac}) offer
measurements with the smallest systematic uncertainties.
The analysis \cite{Ball:2007zza} concludes that
$\Delta\sigma/\sigma = 10.5\% \,(syst.) \pm 0.6\% \,(stat.) \pm 3\% \, (lumi.)$ 
and $\Delta\sigma/\sigma = 11\% \,(syst.) \pm 0.9\% \,(stat.) \pm 3\%
\, (lumi.)$ are possible for 
the lepton + jets  (with 5 ${\rm
  fb}^{-1}$ of integrated luminosity)
 and dilepton channels (with 10 ${\rm fb}^{-1}$), respectively $(\ell =e,\mu)$. The cross
 section measurement in the fully hadronic channel has an estimated
 systematic uncertainty of $20\%$.

One may optimistically expect that a measurement of
$\sigma^{\ttbar}_{\rm LHC}$ will eventually be achievable with 
a total relative error between $5\%$ and $10\%$.
This goal necessitates and motivates  improvements in the theory 
of the inclusive hadronic $\ttbar$ cross section. 

\subsection{Determination of the top-quark mass}
\label{sub-topmass}

As discussed in section~\ref{sub-secmass} the top-quark mass is
a key parameter of the SM, but also of many of its extensions, and
should therefore be determined as precisely as possible. It is, like
any other quark mass, a convention-dependent quantity. Distinguishing
between differently defined top-mass parameters requires 
the computation of observables (which depend on
this parameter) beyond the tree-level in a specific renormalization
scheme. 
The $\ttbar$ cross section discussed above is such an observable.
As we have seen, its sensitivity to the top-quark mass is rather good;
we have $\Delta\sigma/\sigma \simeq -5 \Delta m_t/m_t$ both
for the  the Tevatron and the LHC. One may eventually
compute the hadronic $\ttbar$ cross section to higher-order QCD in terms
of a well-defined short-distance mass, for instance the running 
${\overline{\rm MS}}$ mass ${\overline m}_t(\mu_R)$, rather than in
terms of the (ambiguous) pole mass $m_t$. 
Comparison of  the measured cross section  with 
$\sigma^{\ttbar}_{h_1 h_2}(s,\as,{\overline m}_t)$ for a specified value of $\mu_R$, 
the running mass ${\overline m}_t(\mu_R)$. 
The present and expected experimental uncertainties at the Tevatron
and the LHC, respectively, and the ones on the theory side preclude
for the time being a precise determination of
${\overline m}_t(\mu_R)$ in this way.
In any case, it allows, at some level of precision, to cross-check 
mass determinations that use other techniques (see below). In fact,
in a recent determination of $m_t$ in the dilepton channel by the CDF
 collaboration at the Tevatron,  the dependence of
$\sigma^{\ttbar}$ on $m_t$ was combined with the top mass
determination from event kinematics \cite{:2007jw}.

As an aside we recall the well-known result, based on many studies 
\cite{AguilarSaavedra:2001rg}, that the top-quark mass 
could be determined with unprecedented precision by a counting experiment
at a future linear $e^+ e^-$ collider. 
A scan of the cross section through the $\ttbar$ production threshold
and a fit of the resulting line-shape to the
theoretical predictions for $\sigma_{e^+ e^- \to \ttbar}(s)$, which can 
be made in terms of well-defined short-distance 
masses\footnote{For $e^+e^-\to \ttbar$ far above threshold, an interesting
theoretical analysis of a top-mass determination
from the invariant mass distribution was made in
\cite{Fleming:2007qr}.} (c.f., e.g., \cite{Hoang:2000yr}), would yield 
the ${\overline{\rm MS}}$ mass with an overall error of 
$\delta {\overline m}_t \simeq 150$ MeV.

Determination of the top mass at hadron colliders use the kinematic 
reconstruction of the events. A measure of the top mass is the
invariant mass $M_t=[(\sum_i p_i)^2]^{1/2}$ of the top-quark
 decay products. However, the peak 
$M_t^*$ of
the invariant mass distribution cannot be equal
to $m_t$, as this
parameter is associated with a coloured object, while the measured
$M_t$ distribution involves leptons and hadrons which are colour
 singlets. In the ``partonic phase'' of the
$\ttbar$ event, many additional partons are present
and at least one antiquark
$\bar q$ must combine with the partons from top-decay to form
colour-singlet states. Colour exchange occurs between the various stages
 of the hard scattering event  and between the final-state partons
 from this event and the underlying event (see sections~\ref{sub-smttresu}
and~\ref{sub-simttool}). When computing $\rmd\sigma/\rmd M_t$ in
perturbation theory one gets the following: As long as one considers
only factorizable corrections, figure~\ref{fig-nonfac} (left), the 
peak value $M_t^*$  of this distribution is 
equal to $m_t$, to any order in the 
perturbation expansion. Nonfactorizable (semisoft) gluon exchange, 
figure~\ref{fig-nonfac} (right),
shifts the  peak of $\rmd\sigma/\rmd M_t$ away from $m_t$, but the
effect, computed to $\Or(\as^3)$,
 is very small. Near the production threshold, 
$\shat \gtrsim 4 m_t^2$, a shift $\Delta M_t$ of $\sim -15$ MeV and
$\sim +10$ MeV occurs for $gg$ and $q \bar q$ initial states,
respectively, \cite{Beenakker:1999ya} and the sizes of the shifts, which are of
opposite sign for the two $\ttbar$ production channels, decrease with
increasing parton center-of-mass energy $\sqrt{\shat}$. When folded with
the PDF the overall effect on $\rmd\sigma/\rmd M_t$ is negligibly small.

However, no conclusion can be drawn from this on the non-perturbative
aspects of colour reconnection, which includes
the effect of long-wavelength colour fields from the underlying event
on the formation of hadrons in top decay. 
A study based on a heuristic model of non-perturbative colour
reconnection concludes that in top-mass determinations an uncertainty
$\delta m_t \sim \pm 0.5$ GeV is associated with 
this phenomenon \cite{Skands:2007zg}.

The experiments at the Tevatron use complex modeling techniques in order
to extract a value of  the  top-quark mass from the raw data. 
The modeling of the events involves simulations that use lowest-order parton
matrix elements for $\ttbar$ production and decay, various 
parton showering algorithms, and modeling of the hadronization and of
the underlying event. The methods are
reviewed in \cite{Wagner:2005jh,Quadt:2006jk}.
Recent results on the top-mass from the fully hadronic, dileptonic, 
and lepton + jets channels were obtained in 
\cite{:2007qf,:2007jw,Aaltonen:2007xx,Abazov:2007rk}. 
The average of all  measurements
was recently determined to be  
$m_t^{exp} = 172.6 \pm 0.8 \, (stat.) \, \pm 1.1 \, (syst.)$ GeV
\cite{:2007bxa}. Adding the uncertainties in quadrature yields
a total uncertainty of $1.4$ GeV. 
As the top-mass determinations at the Tevatron are based on 
kinematic reconstructions it seems natural to identify this value
with the pole mass $m_t$. One should, however, be careful with 
this interpretation. For instance, the value
of the upper bound on the SM Higgs boson mass,  derived 
by assuming that $m_t$
is known with this precision, may perhaps be premature.

For the LHC, simulation studies have been made by the 
ATLAS and the 
CMS collaborations in order to 
estimate the precision with which the top mass can be determined in
the three channels (\ref{eqttbarreac}). Based on kinematic
reconstructions and 
modeling similar to that used by the Tevatron experiments,
the CMS study \cite{Ball:2007zza} concludes that the top mass can be determined
with an uncertainty of $\Delta m_t^{exp} \simeq 1.2 \, (1.9)$ GeV in the
dilepton (lepton + jets) channel with $10\, {\rm fb}^{-1}$
of integrated luminosity, if the goal of a precise
 determination of the $b$-jet energy scale can be achieved.
The ATLAS study  \cite{Borjanovic:2004ce}
 arrived at a total error of 2 GeV for the same integrated luminosity.

In the high luminosity phase of the LHC the top mass can also be
determined
precisely from $\ttbar$ events with a $J/\psi$ from exclusive
$b$ decay in the final state
 \cite{Kharchilava:1999yj}.
Consider $\ttbar \to (b\to J/\psi) \ell \nu_\ell \: b qq'$ with
$J/\psi \to \ell^+\ell^-$. The top quark mass is correlated
 with the invariant mass $M_{J/\psi \ell}$  of the $J/\psi$ and the
lepton from the $W$ decay which comes from the same top quark
as the $b$ that decays into $J/\psi$. This correlation allows a
determination of the top mass, and this method considerably reduces the
uncertainty related to the knowledge of the jet energies. 
There are, however, other uncertainties, in particular
 theoretical ones associated with colour reconnection and 
quark fragmentation.  When combining the top-mass measurement with
the direct measurement methods mentioned in the previous paragraph,
an uncertainty on $m_t^{exp}$ of 1 GeV is feasible 
\cite{Borjanovic:2004ce,Ball:2007zza}.

The mean distance that $b$ hadrons from $\ttbar$ events travel before
they decay is also correlated with the top-quark mass, and this correlation
provides another method for determining  $m_t^{exp}$  \cite{Hill:2005zy}.
Here, systematic uncertainties are
associated with Monte Carlo modeling, $b$ fragmentation, and the
average $b$ hadron lifetime. 

Another observable that may be useful for determining the top-quark mass at the
LHC is the $\ttbar$ invariant mass distribution. 
The average $\langle\mtt\rangle$ and higher moments have some
sensitivity to $m_t$ \cite{Frederix:2007gi}. This distribution is also
affected by colour reconnection.

Thus, in view of the goal of determining at the LHC a well-defined 
top-quark mass with an overall uncertainty of $\sim 1$ GeV, there is
still some effort required, especially on the theoretical
side. This includes the
investigation of the theoretical
uncertainties of 
observables sensitive to the top-mass.

\subsection{Distributions}
\label{sub-topdist}

Besides  $\sigma^\ttbar$,
kinematic distributions are also important probes 
of the  dynamics of $\ttbar$ production. 
Investigations including higher-order SM corrections were listed
in section~\ref{sub-smttresu}, and  most of the higher-order
QCD results were incorporated in the NLO Monte Carlo codes mentioned
 in section~\ref{sub-simttool}.
Predictions 
for the $p_T$ or (pseudo)rapidity distribution of the $t$ or $\bar t$
quark can be made to good approximation for on-shell, stable top quarks
averaged over their spins. In general (multi-particle) distributions
should be computed at the level of the $t$ and $\bar t$  decay
products, and a number of distributions depend also on the 
spin configuration of the intermediate $\ttbar$
state. Some of them  will be discussed in the
next subsection.  Some variables, for instance $p^{\ttbar}_T =|{\bf
   p}_T^t +{\bf p}_T^{\bar t}|$ or the difference  
$\Delta\phi$ of azimuthal angles of the $t$ and $\bar t$
quark, have a non-trivial distribution only beyond the leading-order,
due to real radiation. This makes them sensitive to multiple gluon emission.

The precise measurement of the $\mtt$
distribution at the $\ttbar$ production threshold would allow for
interesting studies of $J=0$ colour 
singlet $\ttbar$ resonance effects \cite{Hagiwara:2008df}.
On the other side of the energy spectrum, 
 the measurement of the $p_T$ and $\ttbar$ invariant mass distributions
up to the highest possible values is crucial in the search for new
(TeV-scale) physics, such as heavy $s$-channel resonances (c.f. 
section~\ref{subheavH}). Therefore they should be known as precisely
as possible within the SM. 
The electroweak corrections, computed to
$\Or(\as^2\alpha)$, contribute to these distributions. 
The weak corrections 
grow in ``exclusive'' $t \bar t$ production (i.e., no real radiation
of $W$ and $Z$ bosons) due to the appearance of weak-interaction  Sudakov
logarithms.
The  electroweak corrections are
negative, both for the Tevatron and the LHC, above some minimum $p_T$
or $\mtt$ and increase in magnitude relative to the LO QCD corrections.
At the LHC, for example, the weak corrections to the LO $p_T$ and $\mtt$
distribution at $p_T=1$ TeV and $\mtt=2$ TeV are $-11\%$ and $-6\%$
(for a Higgs boson mass $m_H=120$ GeV),
respectively, \cite{Bernreuther:2006vg,Kuhn:2006vh}, 
while the purely photonic corrections  at these values of
$p_T$ and $\mtt$ are $-2\%$ and $+0.5\%$ \cite{Hollik:2007sw}.
A complete assessment of the significance of these corrections 
requires inclusion of the NLO QCD corrections and a discussion 
of the uncertainties resulting from PDF errors and scale variations.
On the experimental side, a measurement of these distributions
in the TeV range requires special $\ttbar$ identification criteria
\cite{Baur:2007ck}. In this energy region  
the $t$ and $\bar t$ quarks are highly Lorentz-boosted, which leads to
overlapping and merged jets in the case of hadronic top decays. 

For $p {\bar p} \to \ttbar + X$ kinematic distributions  in QCD 
need 
not be symmetric with respect to the interchange of the $t$- and $\bar
t$-quark  charges. This is
because the initial state, $|p({\bf q}){\bar p}(-{\bf q})\rangle$, is
not an eigenstate of charge conjugation.  
A charge asymmetry is generated at NLO
QCD by the interference of $C$-even and  -odd terms of the amplitudes
for $q {\bar q}$ annihilation and, likewise, for
$g q$ and $g {\bar q}$ fusion  
\cite{Halzen:1987xd,Nason:1989zy,Beenakker:1990maa,Kuhn:1998kw,Bowen:2005ap}.
(The contribution of the latter two processes to the asymmetry is small.)
One may define a differential and a total 
charge asymmetry
\be
A(y) = \frac{N_t(y_t) - N_{\bar t}(y_{\bar t})}{N_t(y_t) + N_{\bar
    t}(y_{\bar t})}\, , \quad A = \frac{N_t(y_t\geq 0) - N_{\bar
    t}(y_{\bar t}\geq 0)}{N_t(y_t\geq 0) + N_{\bar
    t}(y_{\bar t}\geq 0)} \, ,
\label{eq-diffasy1}
\ee
where $y_t$ and $y_{\bar t}$ are the  rapidities of 
the $t$ and $\bar t$ quark, and $N(y)=\rmd\sigma/\rmd y$.
Notice that $A(y)$ and $A$  are of order $\as$. 
A recent LO analysis \cite{Antunano:2007da} obtained 
$A=0.051\pm 0.006$ for the integrated
asymmetry, where the given uncertainty is due to scale variations and
variations of $m_t$ within the 
error given in (\ref{eq-mtdocdf}). The electroweak QCD interferences,  
which increase the LO QCD result by a factor 1.09, are included. 
If one takes only the strong interactions into
account, then one can invoke $C$ invariance which implies 
$N_{\bar t}(y_{\bar t}) = N_t(-y_t)$. With this proviso, $A$
is equal to the forward-backward asymmetry 
$A_{FB}^t = [N(y_t>0)-N(y_t<0)]/[N(y_t>0)+N(y_t<0)]$
of the $t$ quark. As the initial $p \bar p$ state is a $CP$
eigenstate in the laboratory frame, $CP$ invariance implies
$A_{FB}^{\bar t}= - A_{FB}^{t}$. Non-standard $CP$-violating
interactions, if existent, can lead to small deviations from 
this relation  \cite{Atwood:2000tu}.

For the production of $\ttbar$ pairs in association with a hard jet, 
the integrated charge asymmetry was  calculated to NLO QCD
by \cite{Dittmaier:2007wz}. Here $A=A_{FB}^t$, which to leading-order
is $-7\%$, is drastically reduced at NLO to $-1.5 \pm 1.5 \%$.
From this result, no conclusion can however be drawn concerning the size
of the unknown NLO QCD corrections to the inclusive $\ttbar$ asymmetry
discussed in the previous paragraph. 

Recent measurements at the Tevatron by the CDF and D0 collaborations
use an asymmetry different from (\ref{eq-diffasy1}).
Defining the difference $\Delta y= y_t  - y_{\bar t}$ of the rapidities
of the $t$ and $\bar t$ quark, these experiments use
$A'= (N_>-N_<)/(N_>+N_<)$, where $N_> \, (N_<)$ is the number of
events with positive (negative) $\Delta y$. The leading-order
SM prediction is $A'=0.078\pm 0.009$ \cite{Antunano:2007da}.
The lepton + jets final
states of the $\ttbar$ pair is well suited for this measurement, as
the charge of the lepton tags the charge of the top quark.
The size of this  asymmetry (and of (\ref{eq-diffasy1})) is strongly dependent on
acceptance cuts, as shown by a calculation of $A'$ \cite{:2007qb} with the MC@NLO event
generator. The D0 collaboration measured 
$A'_{D0}=0.12 \pm 0.08 \,({\rm stat.}) \pm 0.01\, ({\rm syst.})$ \cite{:2007qb}.
This result must be unfolded for detector efficiencies and migration
effects \cite{:2007qb} before it can be compared with predictions. 
The CDF experiment also measured this asymmetry and obtained, after
performing the unfoldings just mentioned:  
$A'=0.24 \pm 0.13 \,(stat.) \pm  0.04\, (syst.)$ \cite{cdf-afb-note}.
The central value of this result is higher than the SM expectation,
but the result  is still consistent with the SM
Monte Carlo prediction.

The asymmetries $A$, $A'$, and $A_{FB}^t$ are useful tools
in the search for new physics effects that involve axial vector
couplings to quarks. This has been
demonstrated for some hypothetical heavy $s$-channel resonances,
namely ``leptophobic'' $Z'$ vector bosons  (with vector-
and axial-vector couplings to quarks) \cite{:2007qb}
and axigluons (with axial vector couplings to quarks) 
\cite{Antunano:2007da,Sehgal:1987wi}.

At the LHC, the
initial state  $|p({\bf q}) p(-{\bf q})\rangle$ is an eigenstate of parity.
Thus, $A_{FB}^t=A_{FB}^{\bar t}=0$ in the laboratory frame, 
as long as only parity-invariant interactions -- more general, only
parity-even terms in the scattering operator -- are taken into account.
The charge asymmetries $A, A'$ induced by the SM interactions
are very small. They result from $\ttbar$ production by $q \bar q$
annihilation and $gq$ and $g{\bar q}$ fusion, which are subdominant
processes at the LHC. The size of the effect was investigated
 in \cite{Kuhn:1998kw,Antunano:2007da},
within the SM and for an axigluon with a mass in the TeV range,
in appropriately chosen kinematic regions.

\subsection{Top quark polarization and spin correlations}
\label{sub-tspcorr}

As emphasized in section~\ref{sub-secli}, the top quark is unique
among quarks in that polarization and spin
correlation phenomena provide important tools in exploring the dynamics of
these quarks. 
The SM predicts only a small
 polarization of the $t$- and $\bar t$-quark ensembles  when pair-produced in hadronic
collisions. Strong interactions lead to a  polarization  of
$t$ and $\bar t$ quarks orthogonal to the scattering plane, due to
absorptive parts 
of the scattering amplitudes of $q {\bar{q}} \to t {\bar{t}}$ and $g g
\to t \bar t$,  which are  of $\Or(\as^3)$. This polarization of $\Or(\as)$,
the size of which is dependent
 on the parton center-of-mass energy and on the
scattering angle,  does
not exceed $\sim 2 \%$ in magnitude 
\cite{Bernreuther:1995cx,Dharmaratna:1996xd}.
Parity-violating weak interactions, which affect  
both $q {\bar{q}} \to t {\bar{t}}$ and $g g
\to t \bar t$, induce a top quark polarization in the scattering plane (more
general, along some polar vector), which is, however, also small --
see below.

\begin{figure}[ht]
\begin{center}
\includegraphics[height=4cm,width=8cm]{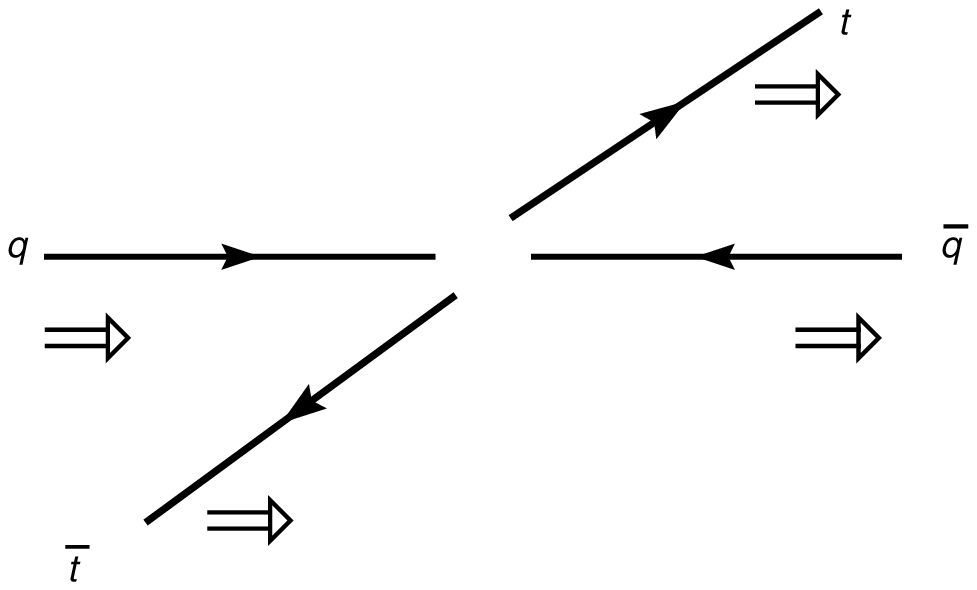}
\includegraphics[height=4cm,width=8cm]{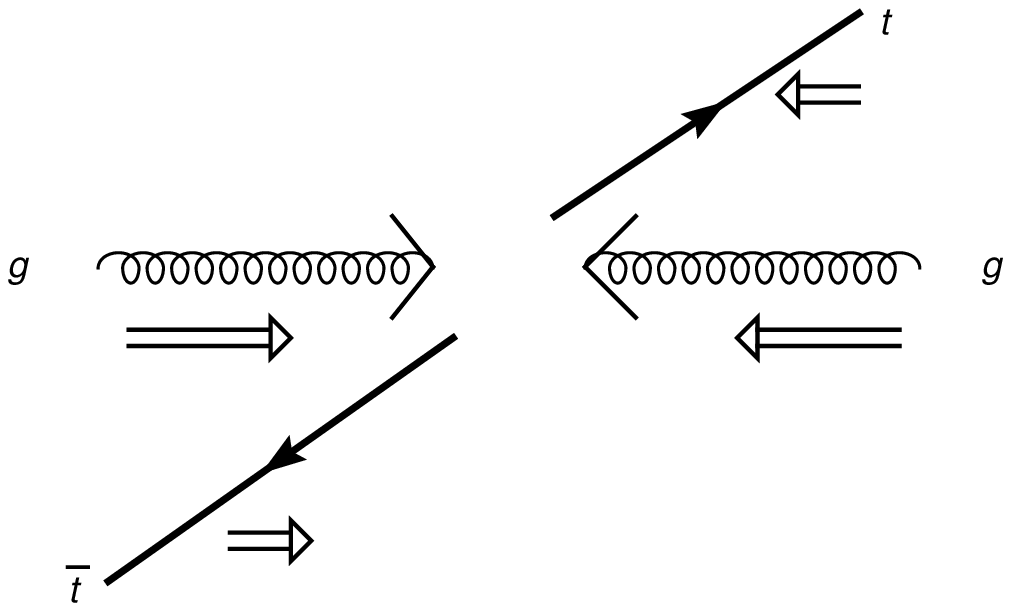}
\end{center}
\caption{\label{fig-ttspinc} Illustration of the correlation
of the $t$ and $\bar t$ spins
in the $q \bar q$ and
$gg$ production channels near threshold.}
\end{figure}

On the other hand the  correlation of the $t$ and $ \bar t$ spins
in the QCD-induced production reactions  is known to be
sizeable
\cite{Barger:1988jj,Arens:1992fg,Stelzer:1995gc,Brandenburg:1996df,Chang:1995ay,Mahlon:1995zn}.
 In fact, the strength of this correlation depends, like the
$t$ and $\bar t$  polarization, on the choice of suitable reference
axes, which can be interpreted as $t$ and $\bar t$ spin quantization
axes (i.e. as  $t$ and $\bar t$ spin bases) in the approximation of on-shell $t \bar t$ 
production and decay.
The correlation
of the  $t$ and $\bar t$ spins with respect to arbitrary reference axes
{$\hat{\bf a}$, $\hat{\bf b}$} is given by the expectation value
${\cal A} = \langle 4 ({\bf \hat{a}}\cdot \st)({\bf \hat{b}}\cdot
\stb) \rangle $. This correlation is equal to  the $t\bar{t}$ double spin  asymmetry
\be
{\cal A} \; = \; {\rm
N(\uparrow \uparrow)+\rm N(\downarrow \downarrow)
  - \rm N(\uparrow \downarrow)- \rm N(\downarrow \uparrow)\over
  \rm N(\uparrow \uparrow)+\rm N(\downarrow \downarrow)
  + \rm N(\uparrow \downarrow)+ \rm N(\downarrow \uparrow) } \, ,
\label{dspina}
\ee
where the first (second) arrow refers to the $t$ $({\bar t})$ 
spin projection onto  {$\hat{\bf a}$ ($\hat{\bf b}$}). 
Useful choices are the helicity basis, 
$\hat {\bf a} = {\bf{\hat{k}_t}},\;  \hat {\bf b} = {\bf{\hat{k}_{\bar
      t}}}$, the so-called beam basis $\hat {\bf a} = \hat{\bf b} =
\hat{\bf p}$, and off-diagonal basis $\hat {\bf a} = \hat{\bf b} = \bf{\hat{d}}$.
 Here  $\hat{\bf p}$ denotes the direction of one of the hadron beams
(i.e., the z axis in the laboratory frame), and $\bf{\hat{d}}$ is
given by 
\be
{\bf{\hat{d}}} =\frac{- \hat{\bf p} +(1-\gamma)(\hat{\bf p}\cdot
{\bf\hat{k}_{ t}}){\bf\hat{k}_{ t}}}{\sqrt{1-(\hat{\bf p}\cdot
{\bf\hat{k}_{t}})^2(1-\gamma^2)}}\, , \quad \gamma = E_t/m_t \, .
\label{eq-offdb}
\ee

Let us briefly discuss the significance of these reference axes.
At the Tevatron most of the $\ttbar$ pairs are produced by $q \bar q$
 annihilation not too far away from threshold. At threshold the
 $\ttbar$ pair, being produced in an $s$ wave, is in a $^3S_1$ state
 -- c.f. the illustration in figure~\ref{fig-ttspinc}.  Thus, when
 the top-quark velocity $\beta_t \to 0$, angular momentum conservation
 implies that  the $t$ and $\bar t$ spins are $100\%$ correlated with
 respect to the beam axis, while this correlation is smaller for any
 other axis. On the other hand, in the ultra-relativistic regime,  $\beta_t \to 1$,
 there is  $100\%$ correlation of the $t$ and $\bar t$ spins with
 respect to the $t$ and $\bar t$ helicity axes.  
This follows from the helicity conservation of quark-gluon interaction.
The vector (\ref{eq-offdb}) defines the so-called
 off-diagonal basis  \cite{Mahlon:1997uc} with respect to which the  
$t$ and $\bar t$ spins are 
$100\%$ correlated for any $\beta_t$. This holds for  $q \bar q \to \ttbar$ to Born
approximation. 

For $gg\to \ttbar$ no spin basis with this property exists.
This may be seen as follows. For $\beta_t = 0$ the $\ttbar$ is in
a $^1S_0$ state (recall the Landau-Yang theorem for 
ortho/para-positronium decay). Thus close to threshold  
mostly  $\ttbar$ pairs with like helicities are produced in $gg$
fusion, while the opposite is the case for $\beta_t \to 1$.

These considerations
imply that for the Tevatron the off-diagonal basis \cite{Mahlon:1997uc}  and the  beam basis 
\cite{Bernreuther:2001rq,Bernreuther:2004jv} yield the strongest
correlations, 
while for the LHC, where the top quarks have on average larger
velocities,  the helicity basis is a good choice. 
A prescription was given by \cite{Uwer:2004vp} to obtain at the LHC a 
correlation which is somewhat stronger than in the helicity basis. 

Notice that  the number of $\ttbar$ events with like and unlike 
spin projections  in (\ref{dspina}) is determined by the diagonal terms 
$(R^{(ij)})_{\alpha \alpha,\beta \beta}$
of the production density matrices  (c.f. section~\ref{sub-smttresu}).

At Born level the vectors involved in these three bases may be defined
in the center-of-mass frame of the colliding partons. However, this
frame is of no use,  once QCD corrections are taken
into account \cite{Bernreuther:2004jv}. 
The reconstruction of this frame requires the measurement of the
four-momenta of all final state particles/jets; but 
for real gluon radiation being collinear to one of the
initial partons this is not possible. Thus in this frame the 
correlation (\ref{dspina})
is not collinear-safe when using the helicity and the off-diagonal basis.
A suitable frame is the zero-momentum frame of the $t \bar t$ pair.
In the following the three bases above are defined with respect to
that frame. 

The correlations of the $t$ and  $\bar t$ spins manifest themselves in
decay  angular correlations which are to be measured with respect to
the chosen reference axes. If the $t$ ($\bar t$) decays
semileptonically then, as discussed in section~\ref{subang}, 
the charged lepton
is the best spin analyzer, while for non-leptonic   $t$ ($\bar t$)
decays the least-energetic non-$b$ jet will be the best choice, at least
from a theoretical point of view. This choice will be made in the
following.

\begin{figure}[ht]
\begin{center}
\includegraphics[height=8cm,width=8cm]{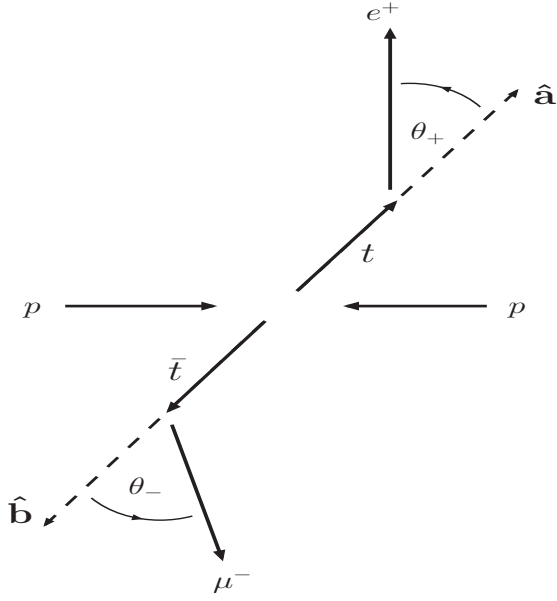}
\end{center}
\caption{\label{fig-helaxab} Illustration of the helicity axes in
  $\ttbar \to \ell^+ \ell'^- X$.}
\end{figure}

The dilepton and the lepton + jets channels 
\be
pp, \; p {\bar{p}} \longrightarrow  t{\bar{t}} \; X\to a\, b \; X \, ,
\label{reacab}
\ee
where $a \, b$ = $\ell^+ \, \ell'^-$, $\ell\, j_<$, and  $j_< \, \ell$ 
($\ell =e, \mu$),
are best suited for 
measurements of top-spin effects.  A useful observable is the
 following double distribution
\be
 \frac{1}{\sigma} \frac{\rmd^2\sigma}{\rmd\cos \theta_a \rmd\cos \theta_b}
\;  =  \; \frac{1}{4} \left[ 1+B_1 \cos\theta_a + B_2\cos\theta_b -
 {C} \cos\theta_a\cos\theta_b \right]  \, , \label{ddist}
\ee
where 
${\theta_a \; (\theta_b)}$ is the angle defined between the direction
of flight of  the particle/jet 
$a \, (b)$ in the  $t \; (\bar{t})$ rest frame and the reference axis  
$\bf{\hat{a} \; (\hat{b})}$. For the helicity
 axes these angles are illustrated in figure~\ref{fig-helaxab}.
The coefficients  $B_{1,2}$
reflect the polarization of $t$ and $\bar t$ with respect
to the axes $\bf\hat{a}$ and $\bf\hat{b}$, respectively. 
When choosing the beam, off-diagonal, or helicity basis,  then QCD absorptive
parts cannot  generate a  $t$ and $\bar t$ polarization along these
axes. Within the SM only weak interaction corrections lead to
non-zero  coefficients $B_i$, which are however small: $|B_1|,|B_2| <
1\%$ (see below). The coefficient $C$ reflects the correlation
of the $t$ and $ \bar t$ spins. Because (\ref{ddist}) is inclusive in
the $t$ and $\bar t$ invariant masses, the $\Or(\as^3)$
nonfactorizable QCD corrections, discussed
in section~\ref{sub-smttresu}, do not contribute. For factorizable
corrections, the 
following formula holds  to
all orders in $\as$ \cite{Bernreuther:2001rq}:
\be
 C\; = \; -  c_a c_b \: {\cal A} \, ,
\label{Cfor}
\ee
where ${\cal A}$ is the double spin asymmetry (\ref{dspina})
and $c_a$, $c_b$ are the $t$, $\bar t$ spin-analyzing powers of $a$,
$b$ given in table~\ref{tab-spinpo}. This formula tells us that the
double distribution   (\ref{ddist}) picks out the
diagonal terms $(R^{(ij)})_{\alpha \alpha,\beta \beta}$ in the 
production density matrices. (In the helicity basis $\alpha,\beta =R,
L$, where $R \, (L)$ refers to positive (negative) helicity of the
$t$ or $\bar t$ quark.) Off-diagonal terms can be probed with
higher-dimensional distributions 
\cite{Nelson:1997xd,Nelson:2005jp,Nelson:2005wh}.

Another useful observable for investigating  $t \bar t$ spin
correlations is the  opening angle distribution 
\cite{Bernreuther:1997gs,Bernreuther:2004jv}:
\be
\frac{1}{\sigma} \frac{\rmd\sigma}{\rmd\cos \varphi} \;  =  \;
\frac{1}{2} \left( 1 - D \, \cos\varphi \right) \, , 
\label{oadist}
\ee
where $\varphi$ =  $\angle ({\bf p_a},{\bf p_b})$,
with the directions of flight of $a,$ $b$  being defined 
in the respective  $t$, $\bar t$ rest frames. This distribution
reflects the correlation of $t$ and $\bar t$ spins when projected onto
each other, $\langle\st \cdot \stb \rangle$. 

The structures displayed on the  right hand sides
of (\ref{ddist}) and (\ref{oadist}) apply if no phase space cuts are
made. For estimators of $C$ and $D$ in the presence of cuts, 
see  \cite{Bernreuther:2004jv}.

The measurement of  (\ref{ddist})  and (\ref{oadist}) requires 
the reconstruction of the $t$ and $\bar t$ rest frames.
 For the $\ell$ + $j$ channel this can be done, as was discussed
in section~\ref{secttbar}. Here, $m_t$ should be considered to be known.
In the case of $\ttbar$ dilepton events, the equations which result
from the kinematics for the unknown
$\nu$ and ${\bar\nu}$ momenta can be solved up to a four-fold
ambiguity, if besides the $W$ mass also the
value of the top-quark mass is put in.  A weight can be assigned to
the different solutions by Monte Carlo simulation. The fully hadronic
channels can be completely reconstructed, but the
spin-correlation effects in the pertinent distributions are small.
Moreover, the signal-to-background ratio is unfavourable for these
events (c.f.  section~\ref{secttbar}). 

Table~\ref{tab-Corr} contains the predictions at LO and NLO in the
 QCD coupling for
the distributions  (\ref{ddist}) and (\ref{oadist}) for the dilepton and
lepton + jets channels. Predictions
 for the fully hadronic channel can be found in \cite{Bernreuther:2004jv}.
 For the
results of table~\ref{tab-Corr}, the LO and NLO PDFs
of \cite{Pumplin:2002vw} were used, and $\mu_F = \mu_R = m_t =175$ GeV.

%%%%%%%%%%%%%%%%%%%%%%%%%%%%%%%%
\begin{table}[t]
\caption{\label{tab-Corr}Coefficient $C$ of the
angular distribution (\ref{ddist})
 that reflects $t{\bar{t}}$
  spin correlations, at LO and NLO in $\as$, for the dilepton and lepton +
jets channels \cite{Bernreuther:2004jv}. For the LHC $C_{beam}$ and
$C_{off}$ are not given, as they are very small.}
\begin{center}
\begin{tabular}{ccc|cc}
 &\multicolumn{2}{c}{\footnotesize{Tevatron, $\sqrt{s}=1.96 $~TeV} }
&\multicolumn{2}{c}{\footnotesize{LHC, $\sqrt{s}=14$~TeV }}\\
$\ell \ell$ & LO & NLO & LO & NLO\\  \hline
${\rm C}_{\rm hel}$ & $-0.471$& $-0.352$ & $\hphantom{-}0.319$  &
$\hphantom{-}0.326$\\
${\rm C}_{\rm beam}$ & $\hphantom{-}0.928$&  $\hphantom{-}0.777$ &
$-0.005$ & $-0.072$\\
${\rm C}_{\rm off}$ & $\hphantom{-}0.937$ & $\hphantom{-}0.782$ &
$-0.027$ & $-0.089$\\ 
$D$ & $\hphantom{-}0.297$ & $\hphantom{-}0.213$ & $-0.217$ & $-0.237$ \\
\hline
$\ell+j$ & & & & \\ \hline
${\rm C}_{\rm hel}$ & $-0.240$ & $-0.168$ & $\hphantom{-}0.163$ & $\hphantom{-}0.158$ \\
${\rm C}_{\rm beam}$ & $\hphantom{-}0.474$ & $\hphantom{-}0.370$ & & \\
${\rm C}_{\rm off}$ & $\hphantom{-}0.478$ & $\hphantom{-}0.372$ & & \\
$D$ & $\hphantom{-}0.151$ & $\hphantom{-}0.10$1 & $-0.111$ & $-0.115$ \\ \hline
\end{tabular}
\end{center}
\end{table}
%%%%%%%%%%%%%%%%%%%%%%%%%%%%
We add the following remarks:
(i) The NLO predictions given in
table~\ref{tab-Corr}  remain basically unchanged when using the PDF set 
\cite{Martin:2003sk}.
(ii) Table~\ref{tab-Corr}   shows that for the Tevatron 
the beam basis is practically as good as the
off-diagonal basis for detecting the $t \bar t$ spin
correlations. From the experimental point of view the beam basis is
perhaps the best choice. 
(iii) For the LHC good choices are the  double
distribution (\ref{ddist}) in the helicity basis 
 and the opening angle distribution
(\ref{oadist}). The correlation coefficients  $C_{hel}$ and $D$ can be
enhanced by cutting away events with large $t \bar t$
invariant mass. 
(iv) Based on  a Monte Carlo
analysis of dilepton and lepton + jets events, the ATLAS study
\cite{Hubaut:2005er} concludes
that these correlations
can be measured at the LHC with
a relative error  of
$\delta D/D \simeq 4 \%$  and $\delta C_{hel}/C_{hel}  \simeq 6 \%$, 
including systematics and detector effects. The CMS study
of the distribution (\ref{ddist}) in the $\ell$ + $j$ channel
draws a more pessimistic conclusion: they estimate the
relative measurement uncertainty of  $C_{hel}$ to be $17\%$. 
(v) $q {\bar{q}}$ annihilation and $gg$
fusion contribute with opposite sign 
to  the above distributions. This makes them
quite sensitive to the  quark and gluon content of the
(anti)proton. (vi) The spin correlations are sensitive to new
interactions in $\ttbar$ production, especially when their chiral
structure differs from the dominant QCD vector coupling (see 
section~\ref{sub-BSM}). 

For a complete discussion of SM top-spin effects 
(electro)weak corrections must also be taken into account. 
The mixed weak QCD corrections 
of $\Or(\as^2\alpha)$ to (\ref{ddist})  and (\ref{oadist}) are also
known \cite{Bernreuther:2005is,Bernreuther:2006vg}. They contribute 
to the coefficients  $C_{hel}$ and $D$ only at the percent level.
For large $\ttbar$ invariant masses, $\mtt\gtrsim 1$ TeV, where the
event numbers become small, these corrections amount
to about $-10\%$ of the LO QCD values.

In addition, the weak
interaction corrections generate $P$-violating spin
effects, in particular a polarization $\langle\st\cdot
{\bf\hat{a}}\rangle$,  
$\langle\stb\cdot {\bf\hat{b}}\rangle$
of the $t$ and ${\bar t}$ quarks
along a polar vector, e.g., along the beam direction or along the 
$t$ and ${\bar t}$ directions of flight. 
$P$-violating (single) spin effects are again best analyzed in the 
$\ell \ell$ and $\ell+j$ channels. If one considers
$pp, p {\bar{p}} \to t{\bar{t}} X \to \ell^+ + X$,
information on the $t$ polarization
may be  obtained from the  angular distribution
\be
 \frac{1}{\sigma} \frac{\rmd\sigma}{\rmd\cos \theta_+} \;  = \; \frac{1}{2}
 \left( 1 + B \cos\theta_+ \right) \, ,
\label{Pviol}
\ee
where $\theta_+ =\angle (\ell^+,{\bf\hat{a}})$, and
 $\bf{\hat{a}}$ may be chosen to be the
beam axis (Tevatron)  or the helicity axis (LHC). (Of course,
 (\ref{Pviol}) follows from (\ref{ddist}).) The distribution
leads to the $P$-violating spin asymmetry
$A_{PV}= (N_>-N_<)/(N_>+N_<)=B/2$, where $N_>$ $(N_<)$ is the number
of events with $\cos\theta_+$ larger (smaller) than zero.
The spin asymmetry, considered as a function of
$\mtt$, increases with increasing $\mtt$. Eventually one should
measure this asymmetry bin-wise at the LHC, especially at large
$\mtt$, 
provided that sufficiently large data samples have been collected.
In the helicity basis  it becomes larger than $1\%$
at the LHC only for $\mtt \gtrsim 2$ TeV. However, the integrated asymmetry
$A_{PV}$ stays well below $1\%$ \cite{Bernreuther:2006vg,Bernreuther:2008md}.
Such a small asymmetry will hardly be measurable -- yet this makes
(\ref{Pviol})  a sensitive tool in the
search for non-standard $P$-violating interactions of top quarks.

\subsection{Associated production of $\ttbar X^0$, $X^0=\gamma,Z, H$}
\label{sub-assprod}

The couplings of the top quark to a photon and a $Z$ boson have not
yet been directly measured\footnote{The $ttZ$ couplings are, however,
 constrained by data from LEP, especially by  $Z\to b{\bar b}$.}. 
 At hadron colliders this can be done in
the associated production of a $\ttbar$ pair with a hard photon or a
$Z$ boson. In the high luminosity phase of the LHC the $\ttbar\gamma$
and $\ttbar Z$ event rates are large enough to allow for rather
sensitive tests of the top-quark couplings to the neutral gauge
bosons. Model-independent phenomenological analyses may be done using
a general form-factor decomposition of the $ttV$ vertices ($V=\gamma,
Z$) or equivalently, an effective, gauge-invariant Lagrangian
which describes possible $ttV$ interactions
in terms of anomalous couplings. In the approximation where apart
from the $V$ boson both top quarks in the  $ttV$ vertex are put
on-shell, this vertex can be parameterized, for each boson,
in terms of four anomalous couplings: a vector, an axial vector,
a  magnetic dipole moment and an electric dipole moment coupling (for a
 parameterization see, 
for instance, \cite{Baur:2004uw}). 

In the SM at tree level, the photonic vector coupling is equal to the top-quark
charge $Q_t e$,  while the axial vector coupling is zero, and the
$ttZ$ vector and axial vector couplings are given by the well-known
neutral current couplings $v_t^Z e $ and $a_t^Z e$. 
The magnetic and weak magnetic dipole moments are induced by quantum
fluctuations: to $\Or(\as^2)$
the anomalous (weak) magnetic moment of the top quark is $0.02$
$(0.007)$ \cite{Bernreuther:2005gq}. The SM radiative
corrections to the vector and axial vector couplings involving the
$Z$ boson are of
$\Or(10^{-2})$ \cite{Hollik:1988ii,Bernreuther:2005gq}. One should
note that the correct description of the $\ttbar V$ events
involves the $S$ matrix element of the respective process and not, in
general, these anomalous couplings or form factors. (See e.g. \cite{Bernreuther:2005gq}
for a discussion.) The primary use of anomalous couplings is to
parameterize new physics effects, in the same fashion as described in 
section~\ref{subnaom}.

Some of these effective couplings
may be modified significantly by new interactions. For instance in several
models of non-standard electroweak symmetry breaking, such as
technicolour \cite{Hill:2002ap}
 or Little Higgs models \cite{Berger:2005ht},
deviations may reach the level of $10\%$. 
If the top quark were a composite object, 
one might expect an anomalously large $\ttbar\gamma$ event
rate, due to de-excitation of higher-energetic top states.
The electric dipole moment and weak electric dipole moment of the top
quark in the $tt\gamma$ and $ttZ$ vertex, respectively, which are generated
by $CP$-violating interactions, are tiny within the SM. If non-standard $CP$ violation,
 in particular Higgs sector $CP$ violation exists, a sizeable top-quark
(weak) electric dipole moment can be induced \cite{Bernreuther:1992dz}.

How precisely can these couplings be measured in 
$\ttbar\gamma$ and $\ttbar Z$ events at the LHC? Rather detailed
studies have been made in \cite{Baur:2004uw,Baur:2001si,Baur:2005wi}.
As to $\ttbar\gamma$ events, one is interested in photon emission
 from top quarks. Cuts have been identified which suppress the
contributions of photon radiation from the other particles involved in
$\ttbar$ production and decay \cite{Baur:2004uw}. Assuming
$300 \, {\rm fb}^{-1}$ of integrated luminosity at the LHC, 
\cite{Baur:2004uw} concludes that the  $\ttbar\gamma$ vector and axial
vector couplings could be measured with a 1 s.d. error of
$\sim \pm 0.07$, while the magnetic and electric moments 
 may be determined with an accuracy of aobut $\pm 0.20$.
The event rate for  $\ttbar Z$ with subsequent leptonic $Z$ boson
decay is considerably smaller than the $\ttbar\gamma$ rate. Therefore,
one expects that the  
 $\ttbar Z$ couplings cannot be determined as precisely as the 
 $\ttbar\gamma$ couplings. Yet, for the axial vector coupling 
an accuracy of $\sim \pm 0.10$ could be reached \cite{Baur:2005wi}.
As far as the axial vector couplings are concerned this sensitivity
is competitive with the measurement expectations in $e^+ e^- \to
\ttbar$ at a future linear collider \cite{AguilarSaavedra:2001rg}.

An important objective at the LHC is the search for the associated
production of a $\ttbar$ pair with a Higgs boson, $p p \to \ttbar H$.
For a SM Higgs boson $H$, this channel may be observable if $H$ is
light, $m_H \lesssim 150$ GeV. 
In view of its cross section being significantly smaller
than those of other $H$ production reactions, this
 is very probably not the channel where $H$ will be
discovered (if it exists).  Yet, once a Higgs boson is found, it
will  be important to
measure its couplings to other particles in order to check whether
these couplings are really those of a standard-model Higgs boson.
The $\ttbar H$ channel provides a direct way to explore the
top Yukawa coupling $\lambda_t$, which the SM predicts to be $\lambda_t^{SM}=m_t/v
=m_t/(246{\rm GeV})$.
The cross section and distributions for this process are known to NLO QCD 
\cite{Beenakker:2001rj,Beenakker:2002nc,Dawson:2002tg,Dawson:2003zu}.
The QCD corrections increase the total $\ttbar H$ cross section
at the LHC by about $20\%$. 

A measurement of the absolute value  of the top Yukawa
coupling by counting the events $\ttbar H, \, H \to f$ is not possible
without further input. The determination requires, apart from a
precise prediction of the production cross section $\sigma(\ttbar H)
\propto |\lambda_t|^2$, the knowledge of the branching ratio $B(H\to f)$.
One can, however, measure ratios of couplings in a model-independent
way. For instance, a measurement of the ratio between the 
rate of $ttH$ production and  that of $WH$ production, where in
both cases $H$ decays to the same final state, would yield the ratio of the Higgs
couplings to the top quark and to the W boson. These coupling ratios
would also allow to discriminate between SM and non-SM Higgs bosons.

A light Higgs boson decays dominantly into $b \bar b$
pairs. Observation of the signal $\ttbar H$ followed by $H\to b \bar
b$ is difficult in view of the large background  \cite{:1999fr,Ball:2007zza}. 
The decay channel $H\to W^+W^-$ (where one or both $W$ bosons are
virtual in the case of  a light Higgs boson) is
also an option, because $B(H\to W^+W^-)\sim 10\%$.  The reaction
$\ttbar H, \, H\to \gamma \gamma$ has a rather clean signature
and allows to reconstruct the Higgs mass peak, but has a much
smaller rate than the $b {\bar b}$ and $W^+W^-$ channels. For
the $\gamma \gamma$ decay mode, a signal in excess of 3 s.d. should be
observable in the high luminosity phase of the LHC \cite{Ball:2007zza}.

In SM extensions with more than one Higgs (doublet) field, the
couplings of the Higgs bosons to the other particles depend on additional
parameters which are unknown. In two-Higgs doublet extensions like the
MSSM, a key parameter is $\tan\beta$, introduced in
section~\ref{subchH}.  For the MSSM to be phenomenologically viable,
$\tan\beta$ is required to be larger than one. This implies, for a
large parameter range, that the
Yukawa coupling of the lightest MSSM Higgs boson $h$ to top quarks is
smaller  than the corresponding SM coupling, while its coupling to $b$
quarks is enhanced. In the MSSM one expects that Higgs-boson
radiation off $b$ quarks, $b{\bar b} h$, dominates over $\ttbar h$.

\subsection{BSM effects in $\ttbar$ production}
\label{sub-BSM}

The existence of physics beyond the standard model (BSM) could
affect $\ttbar$ production in several ways. New particles which
strongly 
couple to top quarks
could show up as resonance bumps in the $\ttbar$ invariant-mass
spectrum, or may be produced in association with $\ttbar$ pairs.
Virtual new  particle
exchanges may significantly modify the total cross section 
 and/or kinematic 
distributions. Some effects of this type were already discussed
 in section~\ref{sub-assprod}. 
 Of particular interest in the search for BSM effects
in (future) high-statistics data is the measurement of
distributions/observables that signify $P$ or $CP$ violation
 in $\ttbar$ production, because such
effects are small to tiny according to the SM.

\subsubsection{Effects of virtual particle exchanges:}
\label{subvirpaex} \quad
The effects of virtual new particle exchanges on the $\ttbar$
cross section and distributions has been extensively studied 
in the literature
 for a number of SM extensions. If resonance effects are
absent, one expects  significant deviations from SM predictions
only if the new particles $X$ that couple to top quarks are not too
heavy, $m_X \sim$ a few hundred GeV, assuming that the associated
 couplings are not much stronger  than $g_{QCD}$.

Popular new physics  
models with such particles include non-supersymmetric two-Higgs
 doublet extensions and the MSSM. For a 2HDM  of type II
 the one-loop corrections to hadronic $\ttbar$ production were computed
in \cite{Stange:1993td,Zhou:1996dx,Hollik:1997hm,Kao:1999kj}. Within the MSSM the
full supersymmetric QCD (SQCD) corrections (squark and gluino
exchanges) were determined by \cite{Berge:2007dz},
 both for the cross section and for a number of
distributions. (In this paper
earlier discrepancies in the literature
\cite{Alam:1996mh,Sullivan:1996ry} were resolved.)
The corrections due to the exchange of colour-singlet supersymmetric
particles were calculated by 
\cite{Kim:1996nza,Yang:1996dma,Hollik:1997hm,Kao:1999kj}, and more
recently by \cite{Ross:2007ez} which incorporated also the SQCD
corrections.

As to the size of these corrections: in a 2HDM they arise from the
exchange of one charged and three neutral Higgs bosons. In a large
region  of the
 parameter space of the model, the corrections to the cross section
 are negative and amount to not more than 
a few percent of the Born cross section,
 even if one or several of the Higgs bosons are relatively light 
($m_{Higgs}=\Or(100\, {\rm GeV}))$ \cite{Hollik:1997hm}. 
However, if one of the neutral Higgs bosons
is heavier than $2m_t$ it can be resonantly produced in the gluon
fusion subprocess, $gg \to \ttbar$, and produce a 
peak in the $\mtt$ spectrum. This scenario is relevant for the LHC
-- see section~\ref{subheavH}. Of special interest are
 $P$-violating asymmetries, for instance the 
spin asymmetry $A_{PV}$ defined below (\ref{Pviol}), to which charged
Higgs boson exchanges can contribute, assuming that the neutral Higgs
states have a definite $CP$ parity. 
The resulting effect was found to be small: even for a rather 
light charged Higgs boson, $|A_{PV}|\lesssim 1\%$ \cite{Kao:1999kj}. 
In general, the Higgs
self-interactions, i.e., the (effective) Higgs potential of a 2HDM -- but also
of the MSSM  \cite{Pilaftsis:1998dd} -- may break $CP$ 
invariance. As a consequence, the pseudoscalar state, $A$, having $CP$
parity $-1$, mixes with the two scalar, $CP=+1$ states $h$ and $H$.
The resulting mass eigenstates $\phi_i \, (i=1,2,3)$ no longer
 have a definite $CP$ parity -- they couple to both  scalar and
 pseudoscalar lepton and quark currents. As the Yukawa couplings of 
the $\phi_i$ are proportional to the mass of the respective fermion, the 
top-quark scalar and pseudoscalar Yukawa couplings can be
of order one. Resulting $P$-
and $CP$-violating effects in top-quark reactions may reach observable
levels.
CP-violating phenomena were investigated for
non-resonant\cite{Schmidt:1992et}
 and resonant $\phi_i$
exchanges \cite{Bernreuther:1993df,Bernreuther:1993hq} 
in hadronic $\ttbar$ production;  c.f. 
also \cite{Zhou:1998wz,Khater:2003wq}. 
For non-resonant $\phi_i$ exchanges the $CP$-violating effects 
in $\ttbar$ production at the LHC are small, they are
 below the percent level 
\cite{Schmidt:1992et,Bernreuther:1993df,Bernreuther:1993hq}.
We shall discuss appropriate observables
 in somewhat more detail in
section~\ref{subheavH} for the  case of resonant Higgs production and 
decay into $\ttbar$ pairs where the effects become larger.  

The one-loop MSSM corrections to hadronic $\ttbar$ production
 comprise, apart from the
exchange of the Higgs bosons $h,$ $H$, $A$, and $H^\pm$, the contributions
of the non-coloured and coloured SUSY particles. In general the SQCD 
corrections are the dominant ones. The largest corrections occur
if the mass of the gluino is close to its present experimental lower
bound, $m_{\tilde g} \gtrsim 230$ GeV, and the splitting between the
masses of the two top squarks ${\tilde t}_1$, ${\tilde t}_2$ is large.
In this case the contribution to the $\ttbar$ cross section at the
Tevatron and at the LHC can reach $\pm 5\%$ of the Born cross
section, where the sign of the correction depends on the stop mixing angle
\cite{Berge:2007dz}. In a large portion of the
parameter space of the model the SQCD corrections to the $\mtt$ and $p_T$ 
distributions are  of moderate size, of the order of a few percent, both for the
Tevatron and the LHC. However, the  gluino pair which appears
 in the intermediate state
can become resonant, $q {\bar q}, gg\to {\tilde g} {\tilde g} \to \ttbar$. For
rather light gluinos, $m_{\tilde g} \gtrsim 230$ GeV, this would show
up as a bump in the $\mtt$ spectrum and would cause also a significant
distortion of the $p_T$ distribution \cite{Berge:2007dz}.
As the  gluino-quark-squark interactions violate parity, the SQCD
corrections contribute also to parity-odd asymmetries such as
$A_{PV}$. For the LHC, effects can reach $2\%$ at the differential
level, i.e. in some $\mtt$ bins, but the corrections to the integrated
asymmetry are below $1\%$ \cite{Berge:2007dz}.

The gluino-quark-squark interactions not only violate parity, but
can also break $CP$. This effect can be parameterized by a
phase in the interaction Lagrangian. While for light quarks this possibility is
strongly constrained by the experimental upper bound on the electric
dipole moment of the neutron, no such strong constraint exists for
a possible $CP$ phase in the ${\tilde g}t{\tilde t}_{1,2}$ interaction.
The observables with which this
interaction was investigated in \cite{Schmidt:1992kt,Zhou:1998wz}
yield, however, effects below the percent level.

BSM effects in $\ttbar$ production may be parameterized in a
(rather) model-independent way by effective quark-gluon interactions
(induced by heavy particle exchanges), the strength of which 
is described anomalous couplings. There is a vast  literature on
this topic, which we cannot cover in detail here (see, for instance,
\cite{Beneke:2000hk}).
Hadronic $\ttbar$ production is obviously the right place to search
for anomalous $\ttbar g$ interactions, in particular for 
chromomagnetic and chromoelectric top-quark couplings, which are
described by the effective Lagrangian
\be
{\cal L}_{eff}= -\frac{\mu_t^c}{2} \, 
{\bar t}\sigma_{\mu\nu} G^{\mu\nu} t 
 -  i \frac{d_t^c}{2} \, {\bar t} \sigma_{\mu\nu}\gamma_5 G^{\mu\nu} t \,
 ,
\label{eq-chromedm}
\ee
where  $G^{\mu\nu} =G^{a\mu\nu}T^a$ is the gluon field strength tensor,  
$\mu_t^c={\hat\mu}_t^c/m_t $ and $d_t^c={\hat d}_t^c/m_t$ denote 
the dimensionful chromomagnetic and
-electric dipole moments of the top quark, and 
${\hat\mu}_t^c$ and ${\hat d}_t^c$ their dimensionless analogues. 
In renormalizable theories
they are induced at the loop level;  $d_t^c\neq 0$ requires
$CP$-violating interactions. The chromoelectric moment generated by
the CKM phase is tiny. Therefore, any observable effect would signify a new
$CP$-violating interaction.

The contribution of the dimension-five interactions to the partonic
differential cross sections $q {\bar q}, gg \to \ttbar$ grows with $\shat$ relative
to the Born terms. If the
dipole moments are nearly constant or vary only weakly with $\shat$,
then the anomalous contributions (\ref{eq-chromedm}) would
considerably distort the
high-end tail of the $p_T$ and $\mtt$ spectra. A number of other
top-quark distributions would also be affected by (\ref{eq-chromedm}), in
particular the $\ttbar$ spin correlations discussed in
section~\ref{sub-tspcorr}. Effects of an anomalous chromomagnetic
moment were investigated in
\cite{Atwood:1994vm,Haberl:1995ek,Cheung:1996kc,Beneke:2000hk}.
A non-zero chromoelectric top-quark dipole moment would leave its mark
 in suitably constructed $CP$-odd and/or $T$-odd triple correlations or
 energy asymmetries\footnote{A general kinematic analysis of
observables
in  hadronic $\ttbar$ production obtained the result \cite{Bernreuther:1993hq} 
that interactions which violate both $P$ and $CP$ induce a $CP$-odd transverse spin-spin
correlation and/or a $CP$-odd longitudinal polarization asymmetry (see
section~\ref{subheavH}).}. These observables
  are best measured in the dilepton and lepton + jets channels.
Such effects were studied in
\cite{Atwood:1992vj,brama,Haberl:1995ek,Choi:1997ie,Grzadkowski:1997yi}.
Assuming these form factors to be real, one finds that 
anomalous couplings as small as $|{\hat\mu}_t^c|, |{\hat d}_t^c| \sim
0.04$ can be detected at the LHC. As these form factors are probed in
the time-like region, they can have real and imaginary parts, which
can be measured separately with appropriate correlations. 

Decay angular correlations can also be used to search for new physics
effects in top quark decay. 
The exclusion limits on the mass  and couplings
of a charged Higgs boson implied by the negative searches  of the D0 and CDF
experiments \cite{Abazov:2001md,Abulencia:2005jd,Grenier:2007xj} do not yet preclude
the existence of the decay mode $t\to b H^+$. If it exists one would expect a preference of
$t\to b \tau^+\nu_\tau$ over the other semileptonic decay modes (c.f.
 section~\ref{subchH}). 
For the  $\ttbar \to \tau + 2 j_b + 2 j + E_T^{miss}$ channels one can
construct appropriate azimuthal angle correlations that are sensitive
to the Lorentz structure of a charged Higgs-boson coupling \cite{Eriksson:2007fx}.

\subsubsection{Heavy resonances:}
\label{subheavH} \quad

Many BSM physics scenarios predict heavy, electrically neutral
bosons  $X^0$, with masses $m_{X^0}$ up to a few TeV, 
that (strongly) couple to $\ttbar$ pairs. Thus these resonances
  would show up as
bumps  in the $\mtt$ invariant mass distribution.
 2HDM or supersymmetric extensions predict a spectrum of neutral
Higgs bosons, some of which can be heavy enough to decay into
$\ttbar$. Models that aim to explain the mechanism of electroweak
gauge symmetry breaking ``dynamically'' by a new
strong force, like technicolour models and their descendants
\cite{Hill:2002ap}, contain new spin-zero and spin-one states.
In top-colour \cite{Hill:1991at,Hill:1993hs} and Little Higgs models
 \cite{ArkaniHamed:2001nc} (c.f. \cite{Schmaltz:2005ky} for a review) 
the top quark
plays a special role. New vector resonances
appear in these models with reasonably strong couplings to top quarks.
Models with extra  dimensions
\cite{Antoniadis:1990ew,ArkaniHamed:1998rs,Randall:1999vf} 
have  massive spin-one and
spin-two Kaluza-Klein (KK) excitations. In some of these
models the couplings of the new states to light quarks and gluons
is suppressed 
 \cite{Agashe:2003zs,Fitzpatrick:2007qr,Lillie:2007yh,Djouadi:2007eg,Lillie:2007ve}, 
and their decay into $\ttbar$ is expected to be their main discovery channel. 

The prime observable in the search for such objects is of course the
$\mtt$ spectrum, but also  the $p_T$ distribution and distributions
due to top-quark polarization and spin correlation effects are useful
tools. Phenomenological studies on resonance production in the
$\ttbar$ channel were made 
for Higgs bosons (more general, spin-zero resonances) 
\cite{Gaemers:1984sj,Dicus:1994bm,Bernreuther:1997gs}, for
spin-one bosons from technicolour \cite{Eichten:1994nc},
topcolour \cite{Hill:1993hs}, and Little Higgs
models \cite{Schmaltz:2005ky}, axigluons \cite{Choudhury:2007ux},
for KK excitations of the graviton
\cite{Fitzpatrick:2007qr,Arai:2004yd,Arai:2007ts},
and  KK excitations of the weak  \cite{Antoniadis:1999bq,Agashe:2007ki} 
and strong 
\cite{Agashe:2003zs,Fitzpatrick:2007qr,Djouadi:2007eg,Lillie:2007ve,Burdman:2006gy,Baur:2008uv}
gauge bosons. Recent model-independent resonance 
studies include \cite{Frederix:2007gi,Barger:2006hm}.
The expected discovery reach in the $\ttbar$ channel at the LHC for such states is typically 
 $m_{X^0} \sim$ a few TeV, depending on the couplings and widths of
 these particles. 

These new states may also be produced in association with a top-quark
pair, $\ttbar X^0$. If $X^0$  couples predominantly to quarks of the
third generation, one expects an enhanced $\ttbar \ttbar$ production
cross section \cite{Han:2004zh,Schwinn:2005qa}. An enhanced production
rate of this four top state could also be a sign of a top-quark
substructure \cite{Lillie:2007hd}.

The experiments at the Tevatron have searched for (colour) neutral, narrow spin-one
resonances, $Z'$, that decay into $\ttbar$, with  no evidence found
for such objects so far. Exclusion limits depends on the mass and
couplings of $Z'$. Assuming that such a particle has the same
couplings as the $Z$ boson, masses $m_{Z'} \lesssim 700$ GeV can be
excluded \cite{Schwanenberger:2006tv,Cabrera:2007ad}. 
However, some new resonances with lower masses may be produced dominantly in gluon
fusion (see  the next paragraph), so a precise measurement of the $\mtt$
spectrum over the whole accessible energy range is indispensable at the
LHC.

\begin{figure}[ht]
\begin{center}
\includegraphics[width=10cm]{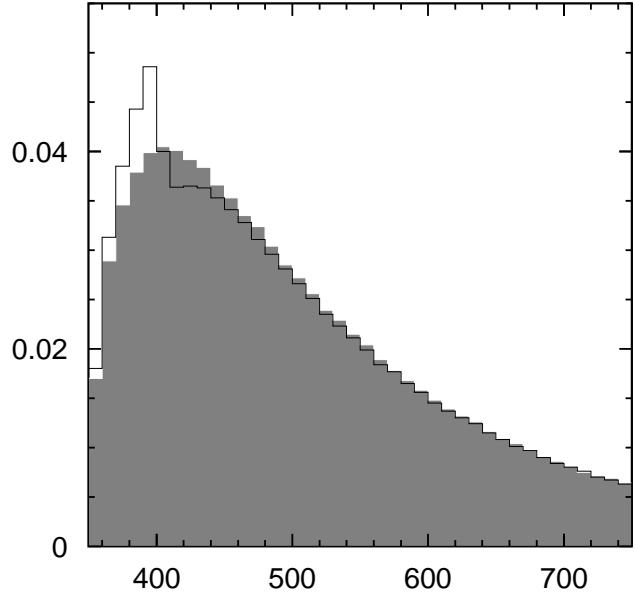}
\end{center}
\caption{\label{fig-heavyH} Signal of a heavy pseudoscalar Higgs boson
  resonance $A$ in the $\mtt$ spectrum of
 the $\ttbar$ lepton + jets channel \cite{Bernreuther:1997gs}. ($m_A=400$ GeV, $\Gamma_A=10
 GeV$, $\tan\beta=1$.) The solid line represents
signal plus background, and the shaded area is the non-resonant
$\ttbar$ background only.}
\end{figure}

For  brevity we discuss here  only one scenario which is of interest
for the LHC, namely heavy non-standard Higgs bosons which strongly couple to top
quarks, as predicted by many SM extensions. 
Consider, for instance,  a  2HDM  or the MSSM where the spectrum of physical
Higgs particles contains three neutral states:
two scalars $h, H$ with $J^{PC}= 0^{++}$
and a pseudoscalar $A$  with $J^{PC}= 0^{-+}$.
 Depending on the parameters of the
respective model some of these states may be heavy, e.g., $H$
and $A$, with masses of the
order of 300 GeV or  larger. Of particular interest here is the
case of a pseudoscalar, as 
$A {\rightarrow\!\!\!\!\!\!\!/}$ $W^+ W^- , ZZ$ in lowest order.
If the ratio of the vacuum expectation values of the two
Higgs doublet fields, $\tan\beta$, is of order 1, these states will
strongly couple to top quarks.
Consider the production of a heavy Higgs
 boson $\phi$  via gluon fusion at
the LHC.  If $\phi = A$ this boson will dominantly decay into
$t \bar t$ pairs, $g g  \to \phi \to  \ttbar$, but also for $\phi = H$
this channel will be significant.
The amplitude of this reaction interferes with
the amplitude of the QCD-induced  non-resonant $t \bar t$ background,
$g g \to \ttbar$, and this interference is not negligible, even
in the vicinity of the resonance, $\sqrt{\shat} \sim m_\phi$, because the
width $\Gamma_\phi$ of the Higgs particle is not necessarily
 very small compared to $m_\phi$; i.e., the resonance need not be narrow.
In the case of a non-narrow resonance the interference generates a  peak-dip structure  in the 
$\ttbar$ invariant mass distribution $\mtt$. For a scalar Higgs boson 
his was investigated
 first in \cite{Gaemers:1984sj},
  and for scalar and
pseudoscalar states in \cite{Dicus:1994bm,Bernreuther:1997gs}.
 The ``golden'' channel to search for resonances
in $\ttbar$ production is  the $\ell$ + jets final state. 
An example of the signature of a pseudoscalar Higgs boson with
parameters $m_A=400$ GeV, $\Gamma_A=10$ GeV, and $\tan\beta=1$ is shown in
figure~\ref{fig-heavyH}. In the MSSM one expects $\tan\beta>1$. In
this case 
the $A\ttbar$ coupling and therefore also the width
$\Gamma_A$ will be smaller. This results in a narrower and higher
peak in the $\mtt$ spectrum. Analogous studies were made for 
a scalar resonance  \cite{Bernreuther:1997gs,Frederix:2007gi}.
A scalar Higgs boson $H$ with mass $m_H \gtrsim 350$ GeV and
SM couplings to the other particles would, however, to be too broad to
be visible in the $\mtt$ distribution.
Statistically significant signals for scalar and pseudoscalar
resonances are possible
 in the mass range 350 GeV $\lesssim m_\phi \lesssim$ 500 GeV,
 depending on the strength of the Yukawa couplings and on the width of
 $\phi$ \cite{Bernreuther:1997gs,:1999fr}.
Needless to say, this is a difficult channel which requires a very
good resolution of the $\mtt$  distribution 
and a precise knowledge of the
SM background contribution.

Suppose experiments will be lucky and discover a heavy boson  in 
$t \bar t$ production. The spin of this resonance may be inferred from
the polar angle distribution of the top quarks. Another observable
which is sensitive to the spin of such an intermediate state is the
Collins-Soper angle \cite{Collins:1977iv,Frederix:2007gi}.
Let's assume the
outcome of such an analysis is 
that the resonance has spin zero. How to find out  whether
it is a scalar or a pseudoscalar? In \cite{Bernreuther:1997gs} it was
proposed to use spin correlations for answering this question, and it
was found that  $\langle\st \cdot \stb\rangle$ is the best choice, which is easy
to understand in simple quantum mechanical terms. 
Consider $g g \to \phi \to \ttbar$.
If $\phi$ is a scalar ($J^{PC}=0^{++}$) then
$\ttbar$ is in a $^3P_0$ state, and a straightforward calculation yields
$\langle\st \cdot \stb \rangle= 1/4$. 
If $\phi$ is a pseudoscalar ($J^{PC}=0^{-+}$)
then  $\ttbar$ is in a $^1S_0$ state 
and $\langle\st \cdot \stb \rangle = -3/4$.
The non-resonant $\ttbar$ background dilutes this striking difference in the  values of this 
correlation.
As discussed in 
section~\ref{sub-tspcorr} the correlation $\langle\st \cdot \stb \rangle$
induces the opening angle distribution  (\ref{oadist}) which is best
studied in the dilepton channel. Depending on the couplings and on the 
width of $\phi$ a statistically significant effect may be found with
this distribution. In order to preserve the discriminating power of
the underlying spin correlation, this distribution should be measured only for $\ttbar$ 
events that lie in a suitably chosen $\mtt$ bin below $m_\phi$ \cite{Bernreuther:1997gs}.

However, things could be more complex. 
 If the (effective) Higgs potential does not conserve $CP$, the
 scalars $h,H$
and the pseudoscalar $A$ will mix, and the Higgs states $\phi_i$ of definite 
mass will no longer
have a definite $CP$ quantum number (as already mentioned
 in  section~\ref{subvirpaex}).
While in a non-supersymmetric 2HDM this can occur at tree level 
 it is a loop-induced effect in
the MSSM, but it can nevertheless be 
sizeable \cite{Pilaftsis:1998dd}.
If this is the case, the states $\phi_i$  will have both scalar and pseudoscalar couplings
to fermions; i.e., the $\ttbar \phi_i$ interactions
 violate both $P$ and $CP$. Then
the spin correlation $\langle\st \cdot \stb \rangle$ and the resulting
coefficient $D$ in (\ref{oadist}) 
 will lie between the pure scalar and pseudoscalar cases
 discussed above. Their values depend on the ratio of the scalar and
 pseudoscalar $\ttbar\phi_i$ couplings.
Yet (non-resonant and resonant) $\phi_i$ exchange
  induces in addition the  $CP$-odd and $T$-odd transverse spin-spin correlation
$\langle {\cal O}_1\rangle = \langle {\hat{\bf k}}_t \cdot(\st\times
\stb)\rangle$ 
and the $CP$-odd and $T$-even 
longitudinal polarization asymmetry $\langle {\cal O}_2\rangle = \langle{\hat{\bf k}}_t
\cdot(\st - \stb)\rangle$ \cite{Bernreuther:1993df,Bernreuther:1993hq}. The latter
asymmetry corresponds to a difference in the produced numbers of $\ttbar$
pairs with negative and positive helicities, $N(t_L{\bar t}_L) - N(t_R{\bar
  t}_R) \neq 0$. If one or several of the $\phi_i$ have
masses near or above the $\ttbar$ threshold, $s$-channel $\phi_i$
exchanges become resonant, and the induced $CP$-odd correlations can
become quite sizeable. 
They, in turn, generate corresponding angular correlations and
asymmetries in the various $\ttbar$ decay channels \cite{Bernreuther:1993hq,Bernreuther:1998qv}.
Again, the dilepton and  lepton + jets channels are most suited for
such investigations. 
For instance,  once sufficiently large
$\ttbar$ data sets have been collected, one may consider the samples
$\ttbar \to \ell^+ X$ and $\ttbar \to \ell^- X$. For each sample one
can measure the 
$\cos\theta_+$ $(\cos\theta_-)$ distribution, which was
defined in (\ref{Pviol}), where $\theta_+$ $(\theta_-)$ in the angle
between the $\ell^+$ $(\ell^-)$ and the $t$ $({\bar t})$ direction of
flight. The  $\ell^+$ $(\ell^-)$ direction is taken to be in the  $t$
$({\bar t})$ rest frame, while the $t$ $({\bar t})$ direction of
flight is to be determined in the $\ttbar$ ZMF of the respective sample.
A signal of $CP$ violation would be a difference in the two
distributions; that is, a difference in the expectation values,
$\Delta_{CP}=\langle\cos\theta_+\rangle - \langle\cos\theta_-\rangle
\neq 0$. This difference results from $\langle {\cal O}_2\rangle  \neq
0$. For resonant $\phi$ exchange, $\Delta_{CP} \sim 5\%$ is
possible if these expectation values are evaluated in suitably chosen
$\mtt$ bins in the vicinity of a resonance \cite{Bernreuther:1998qv}.
The spin correlation $\langle {\cal O}_1\rangle$ leads to triple
correlations among the momenta of the final state particles/jets.
For instance, for the dilepton channel $\ttbar \to \ell^+ \ell'^- X$
one may measure ${\hat{\bf k}}_t \cdot({\hat{\bf p}}_{\ell} \times 
{\hat{\bf p}}_{\ell'})$. Its expectation value, evaluated in a $\mtt$
bin near the resonance, can reach a value of 
several percent \cite{Bernreuther:1998qv}.

The search for non-standard $CP$ violation, in particular Higgs sector
 $CP$ violation, at the LHC is of great interest, not only to particle
 physics but also to cosmology.
Viable scenarios that try to explain the matter-antimatter asymmetry as a
dynamical effect require non-standard $CP$ violation. The discovery of
Higgs bosons with $CP$-violating couplings would support the
hypothesis that this asymmetry was generated at the 
electroweak phase transition \cite{Cohen:1993nk}.
There are several reactions where one can search for 
 Higgs sector  $CP$ violation (c.f. \cite{Accomando:2006ga} for a review), and
 $\ttbar$ production is among the more promising ones. One should note
 that the $CP$-odd observables referred to above probe non-standard $CP$
violation, irrespective of whether or not such effects are induced by
Higgs bosons. 
 
\section{Single-top-quark production}
\label{secsingletop}

\begin{figure}[ht]
\begin{center}
\includegraphics[width=12cm]{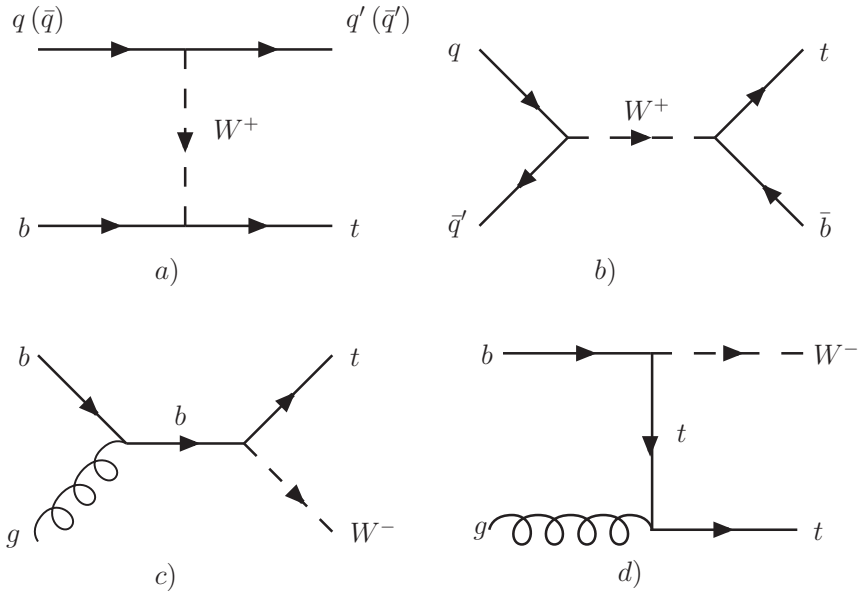}
\end{center}
\caption{\label{fig-singlet} Lowest order Feynman diagrams for 
single-top-quark production processes: $t$ channel (a), $s$ channel (b),
and associated $tW$ production (c,d).}
\end{figure}

In the hadronic production of single (anti)top quarks 
the weak interactions are involved 
in an essential way.
Therefore, these reactions
provide, besides top-quark decay, another important  opportunity
 to study the charged weak current interactions of this quark.
In the SM there are three main hadronic production modes,
namely  top-quark production via the exchange of a virtual $W$ boson
 in the $t$-channel and in the $s$-channel, 
and the associated production of a $t$ quark and  real $W$ boson:
\be
 q \, ({\bar q}) \,  b \to  q' \, ({\bar q}') \, t \, ,
\quad 
q \, {\bar q}' \to {\bar b} \, t \, ,
\quad 
b \, g \to W^- \, t \, .
\label{eq-streac} 
\ee
These reactions are depicted to lowest order in the gauge couplings
 in figure \ref{fig-singlet}. 
The cross sections of these processes are
proportional to $|V_{tb}|^2$. Thus, single-top-quark production
provides a means of directly measuring the strength of the $Wtb$ vertex.
Moreover, the reactions (\ref{eq-streac})  are a
source of highly polarized top quarks, which allow for dedicated
investigations of the 
structure of the charged weak current interactions of this quark.
Exotic $t$ and $\bar t$  production processes involving new
particles/interactions are also conceivable; for instance,  the
associated production of a top  quark and charged Higgs boson,
or enhanced production of single top quarks by
sizeable flavour-changing neutral currents.

Thus, there are interesting physics issues associated with the hadronic
production of single top quarks. However, their observation is much
more challenging than detecting $\ttbar$ pairs. This is partly due to
smaller cross sections, but mainly due to the fact that the
final-state signatures suffer from larger backgrounds (see below). 
Although a few thousand single $t$ and $\bar t$ have been
produced at the Tevatron (with 1 ${\rm fb}^{-1}$ of integrated
luminosity), is was only in 2006 that the D0 collaboration
\cite{Abazov:2006gd} -- and
recently also CDF \cite{CDFsinglet} -- reported evidence for the detection of these processes.
At the LHC prospects are brighter, as this collider will not only be a
factory  for the production of $\ttbar$
pairs, but also for single top quarks.

Single-top-quark production constitutes also an
important  background to  a number of possible new
physics processes. Prominent examples include the associated production of
a neutral (SM) Higgs boson and a $W$ boson, $ p {\bar p}, pp \to W^+H
X,$ $H\to b {\bar b}$, for which $ p {\bar p}, pp \to t {\bar b} X \to
W^+ b {\bar b} X$ is a significant background. The $tW$
mode is an important background to Higgs production and
decay into $W$ bosons, $gg \to H \to W^+W^-$. Needless to say that
also this aspect requires the theoretical description of single-top
processes to be as precisely and detailed as possible.

\subsection{The production cross sections: status of theory}
\label{sub-sintprod}

The three modes (\ref{eq-streac}) probe the charged-current
interactions of the top quark in different kinematical regions. 
Moreover, these reactions are sensitive to different
types of new particles/interactions. Each mode has a relatively
distinct event kinematics. Therefore they may  eventually be
observable separately.
In the following subsections, we review the theoretical results 
for the three modes  within
the SM. The reactions  (\ref{eq-streac}) are hard scattering
processes.
Thus, formulae
analogous to  (\ref{eqtotcs})
 apply for predictions of the cross sections and distributions.

\subsubsection{$t$-channel production:} 
\label{sub-tchanp} \quad 
This production process is the most important one for
the Tevatron and for the LHC, as it has the largest cross section of
the three channels (\ref{eq-streac}) for both colliders.
 
At the LHC the dominant $t$-channel parton processes are $u b \to d t$
(which makes up about $74\%$ of the cross section),
${\bar d} b \to {\bar u} t$ ($\sim 12\%$), ${\bar s} b \to {\bar c} t$
($\sim 8\%$), and $c b \to s t$ ($\sim 6\%$). In addition there are a
number of CKM-suppressed reactions. At the Tevatron the same reactions 
take place, but with with different percentage contributions to the cross
section, the most important ones being $u b \to d t$ ($\sim 70\%$)
and ${\bar d} b \to {\bar u} t$ ($\sim 21\%$).
 
The production of a  $\bar t$ quark requires initial-state
$\bar u$, $d$, $s$, $\bar c$ quarks. The
relative contributions of the CKM-allowed
subprocesses to the $t$-channel cross section at the LHC are 
$d {\bar b} \to u {\bar t}$ ($\sim 56\%$),  
$s {\bar b} \to c {\bar t}$ ($\sim 13\%$),
${\bar u} {\bar b} \to {\bar d} {\bar t}$ ($\sim 20\%$), and
${\bar c} {\bar b} \to {\bar s} {\bar t}$ ($\sim 11\%$).
As the proton contains more
valence $u$ quarks than $d$ quarks, the $t$-quark production cross section
 at the LHC, which is a  $pp$
 collider, is larger than the $\bar t$-quark cross section. 
The SM predicts these cross
sections to be equal at the Tevatron.

Figure~\ref{fig-singlet}a shows the Feynman diagrams for the
$t$-channel processes to leading-order QCD in the so-called 
5-flavour scheme, where the (anti)proton is considered to contain also
$b$ and $\bar b$ quarks in its partonic sea, with a $b$-quark PDF
 which is obtained from QCD evolution of the gluon
and light-quark PDFs. The $b$ and $\bar b$ quarks in the (anti)proton
sea arise from the splitting of virtual gluons into nearly collinear
$b\bar b$ pairs, $g\to b \bar b$. Thus, if one considers the
(anti)proton to be composed only of the four lightest quarks and
gluons, one would take into account the reactions 
$q g \to q' t {\bar b}$ instead of  $q  b \to  q' t$.  (That is why
the $t$-channel reactions  are often called
``$W$-gluon fusion processes'' in the literature.)
However, the cross section for the  $W$-gluon fusion process contains
terms proportional to $\as \ln(m_t/m_b)^2$ from the region where the
outgoing
$\bar b$ is parallel to the gluon in the initial state. These
logarithmically enhanced terms, which make perturbation theory
in $\as$ unreliable, can be summed up by introducing a $b$-quark PDF
\cite{Aivazis:1993pi}. In this approach, the process 
$q g \to q' t {\bar b}$ (with
the contribution from the collinear region subtracted) is one of the
next-to-leading order  QCD contributions to the LO cross section.

To NLO QCD the $t$-channel cross section was calculated  in 
\cite{Bordes:1994ki,Stelzer:1997ns,Stelzer:1998ni}. 
The corrections are relatively modest, they increase the LO  $t$-quark
 cross
section at the LHC (Tevatron)  by about about $5\%$ ($9\%$).
NLO QCD results for the fully
differential cross section were presented  in 
\cite{Harris:2002md,Sullivan:2004ie,Sullivan:2005ar}. 
NLO QCD analyses including the semileptonic
top decays  were made in \cite{Campbell:2004ch} and in
\cite{Cao:2004ky,Cao:2005pq}, where also   
top-spin effects were taken into account. Matching of the  next-to-leading order QCD results 
with parton shower Monte Carlo simulations, according
to the prescription used in the Monte Carlo  program  MC@NLO, 
was made in \cite{Frixione:2005vw}. 
The electroweak corrections
were computed  in
\cite{Beccaria:2008av,Beccaria:2006ir} within the SM and the MSSM. 
The corrections turn out to be
small in both models, at the percent level and below.

\begin{table}[h!]
\caption{\label{tab-singlet} Predictions for
single top-quark production cross
  sections at the Tevatron and 
the LHC  according to the recent update
 \cite{Kidonakis:2006bu,Kidonakis:2007ej}. The given errors
include scale uncertainties, PDF uncertainties, and uncertainties in
$m_t$. The value  $m_t=171.4 \pm 2.1$ GeV was used. At the
Tevatron
the SM prediction for ${\bar t}$ production is equal to 
$\sigma^t$.}
\begin{center}
\renewcommand{\arraystretch}{1.2}
\begin{tabular}{llll}
\hline
cross section  & $t$ channel & $s$ channel & $tW$ mode \\
\hline
 $\sigma_{\rm Tevatron}^t$   &
 $1.15 \pm 0.07$ pb  & $0.54 \pm 0.04$  pb  & $0.14 \pm 0.03$ pb \\
\hline
$\sigma_{\rm LHC}^t$  & $150 \pm 6$  pb & $7.8 \pm 0.7$ pb &
$44 \pm 5$  pb \\
$\sigma_{\rm LHC}^{\bar t}$  &
$92\pm 4$  pb & $4.3 \pm 0.3$ pb & $44\pm 5$ pb  \\
\hline
\end{tabular}
\end{center}
\end{table}

Table~\ref{tab-singlet} contains predictions for the $t$-channel cross
sections which were updated in
\cite{Kidonakis:2006bu,Kidonakis:2007ej}. The value given for the
Tevatron was obtained taking higher-order soft gluon corrections into
account. For the LHC the incorporation of these threshold corrections
is not meaningful  in the case of the $t$-channel processes. Therefore
the values for $\sigma_{\rm LHC}^t$ and $\sigma_{\rm LHC}^{\bar t}$
given  in table~\ref{tab-singlet} are
based on the fixed-order NLO QCD results. The predictions were made 
with the parton distribution functions MRST2004
\cite{Martin:2002aw,Martin:2003sk} and with $m_t=171.4\pm 2.1$
GeV. The given errors include the uncertainties in the PDF, in $m_t$,
and those due to variation of the factorization and renormalization
scales, $\mu_F$ and $\mu_R$, between $m_t/2$ and $2m_t$. 
For fixed-order NLO predictions, see \cite{Harris:2002md,Sullivan:2004ie}.

\subsubsection{$s$-channel production:} 
\label{sub-schanp} \quad At the Tevatron the $s$-channel processes
$q  {\bar q}' \to {\bar b} t$ have the second largest cross section
of the three single-top-production modes (\ref{eq-streac}), while at
the LHC this channel is very small compared to the other ones. 
The dominant process is $u {\bar d} \to {\bar b} t$. Additional
 $q  {\bar q}'$ annihilation channels are ${\bar s} c \to  {\bar b} t$ and CKM-suppressed
reactions.

The leading-order processes are depicted in figure~\ref{fig-singlet}b.
To NLO QCD the cross section was calculated  in \cite{Smith:1996ij},
and in fully differential form in \cite{Harris:2002md}. 
The corrections are rather large; they increase the LO  $t$-quark cross
section at the LHC (Tevatron)  by about $44\%$ ($47\%$).
NLO QCD results including the semileptonic
top decays  were presented in  \cite{Campbell:2004ch,Cao:2004ap}.

Effects on the cross section
due to resummation of higher-order soft gluon radiation are
quite sizable for this channel, both for the Tevatron and the LHC
\cite{Kidonakis:2006bu,Kidonakis:2007ej,Mrenna:1997wp}, so one may
take them into account. 
The values given in table~\ref{tab-singlet} from
\cite{Kidonakis:2006bu,Kidonakis:2007ej} are based on incorporating
such corrections. 
The errors arise from the uncertainties mentioned at the end
of section~\ref{sub-tchanp}.

The hadronic $s$-channel cross section has a rather small PDF uncertainty, because
it arises mostly from light quarks in the initial state. The largest
part of the errors  given in  table~\ref{tab-singlet} 
is due to   $\Delta m_t
= \pm 2.1$ GeV. If the uncertainty on $m_t$ will be significantly
reduced (and the precise meaning of the measured value $m_t^{exp}$
will be clarified), the predictions for the $s$-channel cross
section will become quite precise. Thus from a theoretical point of
view, the $s$-channel mode is well suited to determine the strength
of the $tbW$ vertex, i.e., to directly measure  $|V_{tb}|$. This is of
relevance for the Tevatron, where this mode makes up about $30\%$ of
the single $t$ cross section. 

\subsubsection{Associated $tW$ production:} 
\label{sub-asstwp} \quad
Associated $tW$ production proceeds via $g b \to t W^-$. The
CKM-suppressed contributions from $gs$ and $gd$ initial states are 
negligibly small. This mode plays no role at the Tevatron, but is of
interest for the LHC.

The leading-order 
Feynman diagrams are shown  in figure~\ref{fig-singlet}c, d.
A key issue for this mode is the separation of final states arising 
from $tW$ and $\ttbar$ intermediate states. This problem shows up when
one considers  
real radiation corrections to NLO QCD. The process
$g g \to t W^- {\bar b}$, where one gluon splits into a virtual $b$
and a real $\bar b$, is among the QCD corrections to  $g b \to t W^-$
(c.f. the analogous discussion in section~\ref{sub-tchanp}). However,
$g g \to t W^- {\bar b}$ can proceed also via a intermediate $\ttbar$ 
state, where not only the $t$, but also the $\bar t$ is on-shell
(resonant) and decays into $W^- {\bar b}$. When integrated over the
total available phase space the contribution from the $\ttbar$ amplitude
to the cross section is, at the LHC, 
about one order of magnitude larger than the lowest order $tW$ cross
section. Thus, if one wants to investigate the ``genuine'' $tW$ mode,
the $\ttbar$ terms must be suppressed by appropriate cuts.

Several methods were proposed and studied  in the literature. 
One approach is to make a cut on the invariant mass of the $W^- {\bar b}$
system to prevent the $\bar t$ propagator from becoming resonant
 \cite{Belyaev:1998dn,Belyaev:2000me}. Another method
 \cite{Tait:1999cf,Campbell:2005bb} is to use
a veto on the additional $\bar b$ jet, namely to accept only ${\bar b}$
with a transverse momentum below some value, typically
$p_T^{\bar b} < 50$ GeV at the LHC. This suppresses contributions from
$\ttbar$ intermediate states, too. In order to obtain the ``genuine'' $tW$
cross section to NLO QCD one should subtract from 
$\sigma(gg\to t W^- {\bar b})$ the leading-order $\ttbar$
contribution, i.e., $\sigma_{LO}(gg\to \ttbar) B({\bar t } \to W^-
{\bar b})$ \cite{Tait:1999cf}.

The NLO QCD corrections to the $tW$ cross section were investigated  in
\cite{Giele:1995kr,Zhu:2002uj}.
NLO QCD predictions taking into account the semileptonic top decays
and the leptonic $W$ decays are also available \cite{Campbell:2005bb}.
Applying a $b$-jet veto, the NLO QCD corrections are moderate:
$\sigma_{NLO}(p_T^{\bar b} < 50 {\rm GeV})/\sigma_{LO} \simeq 1.1$.
Table~\ref{tab-singlet}   contains the results of
 \cite{Kidonakis:2006bu,Kidonakis:2007ej} for the hadronic $tW$ cross
sections at the Tevatron and LHC, where higher-order soft gluon corrections were taken into
account, but no cut on $p_T^{\bar b}$ was applied. Input and error
estimates are as mentioned at the end of section~\ref{sub-tchanp}.

\subsection{Signals and backgrounds;  prospects for the LHC}
\label{sub-prospLHC}

Table~\ref{tab-singlet} and the results given in
section~\ref{secttbar}
imply that the total production rate for single $t$- and $\bar t$-quarks
  at the
Tevatron and LHC is about $48\%$ and $38\%$ of the $\ttbar$ production
rate, respectively. Thus, as already stated, the LHC will be also
a single-top factory, with $\sim 3\times 10^6$ $t$- and $\bar t$-quarks
  being
produced already with $10\,{\rm fb}^{-1}$ of integrated luminosity.
However, observing singly produced top quarks is much more difficult
than observing $\ttbar$ pairs, because the final states from 
single-top events are clouded by large backgrounds. Let us briefly
discuss this, first for the $t$-channel production 
mode. In order to suppress the  QCD background
 one is forced to search for semileptonic top-decays and to rely on $b$
tagging. Thus one searches for the final states $tq \to
Wbq \to \ell \nu_{\ell} b q$.
A fair fraction of the signal events contain also an additional
$\bar b$ jet, which has, however, most of the time a transverse
momentum too low to be observable. A characteristic signature is
provided by the light-quark jet. The $t$-channel $W$ boson by which
$tq$ is produced deflects 
the initial-state light quark only a bit.
At the Tevatron, the outgoing light-quark jet
is therefore emitted  most of time at large rapidity,
i.e., very forward in the detector. At the $pp$ collider LHC the light
jets are emitted preferentially  both into the forward and backward direction.
Thus the signal consists of an
isolated, high $p_T$ charged lepton, large missing transverse
energy/momentum, and two jets -- one of which is a $b$ jet and the
other, light-quark jet has large pseudorapidity $\eta$ ($|\eta|$) at the
Tevatron (LHC).  

For the $s$-channel production mode the signal is
 $t{\bar b} \to \ell \nu_{\ell} b {\bar b}$, i.e., an
isolated, high $p_T$ charged lepton, large missing transverse
energy/momentum, and two $b$-jets.

The background to these modes is huge and consists of a number of processes.
Irreducible background arises from gauge boson plus heavy-quark
production: $W db$, $Wb \bar b$, $WZ$, with $W\to \ell \nu_{\ell}$ and
$Z\to b {\bar b}$. Considerable background arises
also from $W c$ and $W c{\bar c}$ production. Moreover, a significant background is due to 
the production of $W$ + two light jets, where one of the  jets
fakes a $b$ quark. Severe background is due
to $\ttbar$  production with subsequent decay into dileptonic and lepton + jets
channels, when only one of the leptons  (in the $\ell\ell$ channels), 
respectively only 2 jets (in the $\ell+j$ channels) are observed.
QCD multijet events have a huge cross section, but such background can
be reduced substantially by considering only the above-mentioned
signals.

The D0 and CDF collaborations reported evidence
for single-top production at the Tevatron \cite{Abazov:2006gd,CDFsinglet}.
The analyses used a number of discriminating variables and 
sophisticated statistical methods. Using $0.9 \, {\rm fb}^{-1}$ of
data, D0 obtained for the sum of
the $t$- and $\bar t$-quark  production cross sections
$\sigma^t + \sigma^{\bar t} = 4.7 \pm 1.3$ pb \cite{Abazov:2006gd}.
Assuming SM physics -- i.e. putting the anomalous couplings
in the $tWb$ vetex to zero -- this cross-section measurement implies
the bound $0.68 < |V_{tb}|\leq 1$  ($95\%$ C.L.) on the 
CKM matrix element $V_{tb}$. The  CDF experiment reports
 the result $\sigma^t + \sigma^{\bar t} = 2.2 \pm 0.7$ pb
 \cite{CDFsinglet}, based on  $2.2 \, {\rm fb}^{-1}$ of
data. From this measurement the value $|V_{tb}|=0.88 \pm 0.14 \,(exp.) \,
\pm 0.07 \, (th.)$ was extracted. While the central value
of the D0 and CDF cross-section result is a bit on the high and low side,
respectively, compared with the $t$- plus $s$-channel predictions collected in 
table~\ref{tab-singlet}, the measurements are nevertheless in
 agreement with the SM.

For the LHC prospects are brighter to actually disentangle the
different single-top production modes and to use them for 
a detailed exploration of the physics involved. Simulation studies
were   performed by the ATLAS \cite{:1999fr}
and the CMS collaboration \cite{Ball:2007zza}.
 For the $t$-channel mode 
a signal-to-background ratio $N_S/N_B=1.34$ and a significance $S_{\rm
  stat} = N_S/\sqrt{N_S+N_B} =37$ can be reached \cite{Ball:2007zza}. 
The $t$-channel
cross section is expected to be measurable with a  total error of
$10\%$. This implies, assuming SM physics, that the
CKM matrix element $V_{tb}$ can be determined with $5\%$ accuracy.

The $tW$ mode, which has the second-highest yield at the LHC, has two $W$ bosons
and a $b$ quark in the final state. In the simulation 
\cite{Ball:2007zza} only leptonic $W$ decays were considered. Then the
signals for this mode are $\ell^+ \ell'^- b E_T^{\rm miss}$ and
 $\ell^\pm b j jE_T^{\rm miss}$ if the top quark decays semi- and
 non-leptonically, respectively. The dominant background arises from
 $\ttbar$ events. Other background comes from $Wb \bar b$, $W$ + jets,
$WW$ + jets, from QCD multijets, and from $t$-channel single-top
production. The study concludes that a statistical significance of
6.4 can be reached by combining the two channels. The estimated
uncertainties for the cross section measurements  in the dilepton and
single lepton channels is $26\%$ and $23\%$, respectively. 
 
For the cross section of the
$s$-channel top production mode  at the LHC a
measurement uncertainty of  $36\%$ was  estimated \cite{Ball:2007zza}.

\subsection{Top-quark polarization}
\label{sub-singtpol}

Because the weak interactions are involved in single-top production,
the produced sample of $t$ and $\bar t$ quarks is highly polarized,
and the decay products are correlated with the top spin 
-- c.f. the decay-angular 
distributions  (\ref{dcost}), (\ref{dcostbar}). These distributions 
depend not only on the structure of the 
top-decay vertex, but also on the  polarization degree $p_t$ 
 ($p_{\bar t}$)
of the (anti)top quark.  The polarization degree 
 characterizes the production dynamics of the top quark.
Sizeable contributions from the exchange of new particles, for
instance, heavy $W'^\pm$ bosons having also right-handed couplings, or
charged Higgs bosons $H^\pm$ would leave their mark in this observable.
Thus, these decay angular distributions contain important information
about top quark production and decay. (See, however, the remark at the
end of section~\ref{subnaom}.)

For a specific production 
dynamics the top polarization degree depends on the choice of the reference
axis, i.e., on which ``top-spin quantization axis'' is chosen. Within
the SM this issue was
investigated  exhaustively for the $s$- and $t$-channel modes
at the Tevatron \cite{Mahlon:1996pn} and for the $t$-channel mode at
the LHC \cite{Mahlon:1999gz}.
Here we review only the case of the  $t$-channel mode at
the LHC in some detail, for which 
the experimental investigation of top-spin effects
 may be feasible with reasonable precision. First we recall that 
about $80\%$ of the processes for $t$-quark production,
 $u b \to d t$ and $c b \to s t$,
have a $d$-type quark in the final state. Crossing symmetry
relates the lowest-order amplitude for these processes, 
figure~\ref{fig-singlet}a, to the amplitude for nonleptonic top-decay,
$t\to b u {\bar d}$. From the discussion in section~\ref{subang} we
know that the spin of the $t$ quark is  maximally correlated
with the direction of the $d$-type quark. For the
case at hand this implies that the  sample 
of $t$ quarks produced by these processes is $100\%$ polarized in the
direction of the $d$-type jet -- the spectator jet -- with the jet
direction determined in the top-quark rest frame. This direction is
called the spectator basis. Higher-order QCD
corrections (gluon radiation) will dilute this maximal correlation
somewhat.  Considering the subsequent semileptonic 
decay of the $t$ quark, i.e., $pp \to d t  \to d \ell^+ \nu_{\ell} b$,
we have the decay-angular distribution
\be
\frac{1}{\sigma^t}\frac{\rmd\sigma^t}{\rmd\cos\theta_{+}} =
  \frac{1}{2} \, (1 + p_t\, c_{+} \, \cos\theta_+) \, , \quad
p_t = \frac{N_\uparrow - N_\downarrow}{N_\uparrow + N_\downarrow} \, ,
\label{eq-decstdd}
\ee
where $\theta_{+}$ is the angle between the direction of
charged lepton $\ell^+$
and the chosen spin axis,  in the $t$ rest frame. The  coefficient
$c_{+}$ is the spin-analyzing power of $\ell^+$ which is $+1$ in the
SM (see table~\ref{tab-spinpo}). This distribution implies a lepton 
asymmetry with respect to the chosen axis,
$A_{\uparrow\downarrow}^t = p_t c_+$. From the above discussion -- and the remark
in brackets in the middle of the next paragraph --  we expect 
$A_{\uparrow\downarrow}^t 
\sim 100\%$ for the spectator basis. 

In the case of $\bar t$ production, the probability for the spectator
jet being a $d$-type jet is only about $31\%$, while in about
$69\%$ the $d$-type quarks are in the initial state, being supplied by
one of the proton beams (c.f. section~\ref{sub-tchanp}). 
At first sight it seems that using
the spectator direction as  spin basis for the $\bar t$ quark
 is not a good choice.
However, as was discussed in section~\ref{sub-prospLHC}, the direction of the
spectator jet nearly coincides with the direction of the incoming
light quark. Thus, even when
the $d$-type quark is in the initial state, 
the polarization degree $p_{\bar t}$ will not be degraded very much 
when choosing the spectator direction as spin basis.
(This applies also to the case of $t$-quark  production discussed above,
where about $20\%$ of the events arise from
$d$-type quarks in the initial state.) Considering  again the crossed
process, ${\bar t} \to {\bar b} d {\bar u}$, we know from 
section~\ref{subang} that the spin of the $\bar t$ quark is  maximally anti-correlated
with the direction of the $d$-type quark. Thus the  produced sample 
of $\bar t$ quarks will have a large negative
polarization $p_{\bar t}$, close to  $-100\%$, with respect to the 
direction of the spectator jet.
For the subsequent semileptonic 
decay of the $\bar t$ quark, 
$pp \to j {\bar t}  \to j \ell^- {\bar \nu}_{\ell} {\bar b}$,
we have, in analogy to (\ref{eq-decstdd}),
 the decay-angular distribution
\be
\frac{1}{\sigma^{\bar t}}\frac{\rmd\sigma^{\bar t}}{\rmd\cos\theta_{-}} =
  \frac{1}{2} \, (1 + p_{\bar t}\, c_{-} \, \cos\theta_-) \, , 
\label{eq-decstbard}
\ee
where $c_{-}=-1$ is the spin-analyzing power of $\ell^-$. Thus
the spin-asymmetries $A_{\uparrow\downarrow}^t$ and 
$A_{\uparrow\downarrow}^{\bar t}$ which appear in the 
distributions (\ref{eq-decstdd}) and 
(\ref{eq-decstbard}) are both positive. This implies that in 
experimental analyses the $t$- and $\bar t$-quark samples may be combined
without diluting the resulting decay-angular distribution. This
would be necessary if the sign of the charged lepton cannot be
determined.

This discussion suggests that another
good choice for the top-spin axis should be
 the direction of one of the proton beams, as
seen in the $\bar t$ (respectively $t$) quark rest frame. As the two beams are
not back-to-back in this frame, one should choose the direction of
that beam which is most closely aligned with the spectator jet
on an event-by-event basis  \cite{Mahlon:1996pn,Mahlon:1999gz}. This 
reference axis is called the beamline basis.

As the top quarks produced in the $t$-channel process at the LHC have 
relatively large velocities, one might envisage also 
the traditional helicity basis. In fact, inspection of the 
spin configurations allowed by the $V-A$ law and angular momentum
conservation in the $2\to 2$ processes figure~\ref{fig-singlet}a yields that
 the $t$ and $\bar t$ quarks have negative and positive helicities, 
respectively,
in the center-of-mass frame of the initial partons.
However, this frame is not collinear-safe, as discussed in
section~\ref{sub-tspcorr}.  One may determine the $t$- and $\bar t$-quark
directions of flight in the laboratory frame (lab. helicity basis),
which is unambiguous theoretically. With respect to this
frame a sizable fraction of $t$ ($\bar t$) quarks has however  
also positive (negative) helicities. Thus the polarization degrees
$p_t$ and $p_{\bar t}$ with respect to the laboratory helicity basis
of the  $t$ and $\bar t$ quarks are
expected to be smaller than in the above spin bases.

Table~\ref{tab-stspinas} contains the results of a leading-order
computation \cite{Mahlon:1999gz} of the spin fractions
and  polarization degrees $p_t$ and $p_{\bar t}$, for the case of the LHC
and the three spin bases discussed above. The $2\to 3$ processes
$q g \to q' t {\bar b}$ (see section~\ref{sub-tchanp})
were also taken into account. The applied selection cuts
for $p_T^\ell$, $p_T^{\rm miss}$, $p_T^b$, and the $p_T$ and
pseudorapidity range of the spectator jet increase the spin fractions.
The results corrobate the qualitative discussion above: the spectator
and the beamline bases yield the highest polarization degrees in the SM.
Nevertheless, the $t$- and $\bar t$-quark  polarizations should be measured for 
all spin bases in order to explore the
production dynamics in detail. If  single-top production is 
influenced by new interactions with a chiral structure different
from $V-A$ then this can affect the polarization degree for each  
spin basis in a different way.

An experimental issue is the precision with which
the top-quark rest frame can
be reconstructed. The unobserved momentum of
the neutrino from semileptonic
top decay leads to a two-fold 
ambiguity in the kinematic reconstruction
(c.f. section~\ref{sub-tspcorr}). 
More detailed studies must take into account also the effect of
background contributions to and the effects of energy smearing on the 
distributions (\ref{eq-decstdd}) and   
(\ref{eq-decstbard}). Studies made so far include
\cite{:1999fr,Beneke:2000hk}.

\begin{table}[h!]
\caption{\label{tab-stspinas} SM predictions for the dominant
spin fraction and polarization degree of $t$ and $\bar t$
quarks for several
$t$ and $\bar t$ spin bases 
in the $t$-channel mode at the LHC \cite{Mahlon:1999gz}, 
with selection cuts as specified in this reference. The acronym bml
denotes the beamline basis; L, R denote negative and positive helicity, 
respectively.}
\begin{center}
\renewcommand{\arraystretch}{1.2}
\begin{tabular}{llll}
\hline
  & basis & fraction  & polarization $p$ \\
\hline
$t$: &  lab. helicity &
 $74\%$ $\downarrow$ (L) &  -0.48  \\
 & spectator & $99\%$ $\uparrow$ & 0.99 \\
 & bml &  $98\%$ $\uparrow$ & 0.96\\
\hline
$\bar t$: &  lab. helicity &
 $70\%$ $\uparrow$ (R) &  0.41  \\
 & spectator & $98\%$ $\downarrow$ & -0.96 \\
 & bml &  $99\%$ $\uparrow$ & -0.97\\
\hline
\end{tabular}
\end{center}
\end{table}

As to the polarization of top quarks produced by the $tW$ mode,
which has the second-largest cross section at the LHC, the situation
 is more complicated. Consider the lowest-order amplitude, given by
the sum of the diagrams
figure~\ref{fig-singlet}c and \ref{fig-singlet}d. While in the 
contribution figure~\ref{fig-singlet}c the top-quark state has left-handed
chirality at the production vertex ($btW$ vertex), left- and
right-handed chiralities appear with equal strength in 
the diagram figure~\ref{fig-singlet}c ($g t{\bar t}$ vertex), which
will dilute the top-quark polarization degree. The issue was
analyzed in \cite{Boos:2002xw}.
For the final states 
$tW^- \to b \ell^+ \nu_\ell \ell'^- {\bar \nu}_{\ell'}$ it
 was found that one can select a sample of top quarks with
polarization vector preferentially close to the direction of
of the charged lepton $\ell'^-$ from $W^-$ decay by applying appropriate
cuts which reduce top-quark contributions with opposite polarization.
It remains to be demonstrated  whether 
polarization measurements with reasonable precision are feasible for
the $tW$ mode at the LHC.

\subsection{New physics effects}
\label{sub-singtnewphys}
New physics effects in single-top production 
could manifest themselves by modifications of the
strength and structure of the SM $tWb$ vertex 
or, more generally, by the effects of virtual new particle
exchanges. New particles might also appear as resonances
in $s$-channel single-top production or in association with a top quark.
New mechanisms like  FCNC interactions could lead to enhanced
production rates or exotic final states, e.g., ``like-sign'' $tt$ or
${\bar t} {\bar t}$ events.
 
First of all, single-top production probes the strength  of the 
charged weak interactions of the top quark. The expected  measurement
uncertainty of the $t$-channel cross section at the LHC (see
section~\ref{sub-prospLHC}) implies that this strength may be
determined eventually with an accuracy of about $5\%$. If one assumes
that the main effect of new physics is the modification of the SM
$tWb$ vertex, one may use the general parameterization 
 (\ref{eq-ffdecomp}) and compute the effects of the anomalous couplings
 on the single-top production cross sections.
(More generally, one may take additional couplings into account
\cite{AguilarSaavedra:2008gt}, which appear when the $t$ and $b$
quarks involved are off-shell.) 
 In combination with the
helicity fractions $F_{0,\pm}$ discussed in 
sections~\ref{subWbhel} and~\ref{subnaom} one can uniquely
determine the four anomalous couplings by fits to
(future) data \cite{Chen:2005vr}. For analyses taking into account also
effective $qq'tb$ production vertices, see for instance 
\cite{Yue:2006qx,Cao:2007ea}.

A number of investigations were made in specific extensions of the SM
to determine the effects  on the production rates of virtual new particle
exchanges, in particular in the MSSM,
the presently most popular paradigm for new physics. 
The supersymmetric QCD corrections to the three
single-top production modes were calculated in \cite{Zhang:2006cx}, 
and the electroweak MSSM corrections to the $t$-channel and $tW$ modes
were computed in \cite{Beccaria:2008av,Beccaria:2007tc}. The effects
on the cross sections and on distributions are modest to small, of the
order of a few percent at most. This complies with the smallness
of the supersymmetric quantum corrections in top decays $t\to b f f'$
-- c.f. section~\ref{subtext}.

New heavy charged boson resonances may exist that (strongly)
couple to top quarks. Heavy
vector bosons $W'^\pm$ are predicted by several SM extensions, such as 
left-right-symmetric 
models  \cite{Pati:1974yy,Mohapatra:1974hk,Mohapatra:1974gc},
or technicolour models and their descendants
\cite{Simmons:1996ws,Hill:2002ap}. Heavy charged Higgs bosons $H^\pm$
appear in many SM extensions, for instance in the MSSM. Topcolour
models predict top-pions $\pi_t^\pm$ \cite{Hill:2002ap} whose masses
might be as low as several hundred GeV. If such resonances exist they
can contribute significantly to the $s$-channel rate, 
$q {\bar q}' \to W'^+, W^+ \to t {\bar b}$ (with interference
effects which might be significant, depending on the mass and width of
$W'$) and 
$c{\bar b} \to \phi^+ \to  t {\bar b}$
$(\phi^+ = H^+, \pi_t^+)$
\cite{Simmons:1996ws,He:1998ie,Tait:2000sh,Boos:2006xe}. The LHC should
eventually be sensitive to resonances which preferentially
couple to heavy quarks,  with masses up to a few TeV.
The $t$-channel  exchange of heavy charged bosons $X^\pm$ 
is marginal to negligible, as their contribution to the cross section
is suppressed by $1/m_X^2$.
Heavy $W'$ bosons with SM couplings and masses below 1 TeV have been
excluded by the D0 collaboration
at the Tevatron from the search for their  leptonic decays
$W' \to \ell \nu_{\ell}$ \cite{Magass:2007zz}. The CDF experiment has
searched for  $W' \to t b$ and excludes a $W'$ with a mass below $790$
GeV having this decay mode \cite{Duperrin:2007uy}.
Lower bounds on the masses and couplings  of such resonances can also be derived 
from the measurement 
of the sum of the $s$- and $t$-channel single top-production cross
section at the Tevatron \cite{Ball:2007zza} which is in agreement, 
within errors, with the SM prediction. 

Charged bosons could also be produced in
association with top quarks. For charged Higgs bosons $H^\pm$ 
this was investigated for the
LHC in \cite{Maltoni:2001hu}, with the conclusion that 
some perspectives for a discovery
exist if the $H^\pm$ are rather light. 

Single-top production is also a good place to search for 
FCNC interactions involving the top quark. If sizeable 
$t\leftrightarrow c, u$ transitions exist, they would lead not only
to new top-decay modes (as discussed in section~\ref{subfcnc}),
but also to new production processes, e.g., $c g\to t$,
$gg\to t {\bar c}$,  $c g\to tX^0$ ($X^0 =g,Z,\gamma,h$).
Many theoretical investigations have been made  on this subject.
A rather general approach is to use effective Lagrangians, i.e., 
to classify the relevance of possible FCNC interactions of the top quark
in terms of inverse powers of some large scale $\Lambda \sim 1$ TeV, and
to parameterize them by dimensionless anomalous couplings 
\cite{Beneke:2000hk,Han:1995pk,Han:1998tp,Larios:2003jq,Liu:2005dp}.
Other studies have investigated the above FCNC production processes
within specific SM extensions, in particular in the MSSM
\cite{Liu:2004bb,Cao:2007dk,LopezVal:2007rc} and in 
topcolour-type models \cite{Cao:2007bx}.
In the MSSM the
supersymmetry-breaking terms provide a 
 new source of flavour violation, 
which leads to FCNC processes at one-loop order.
A systematic investigation of various FCNC single-top production and 
decay processes
in the MSSM was made in \cite{Cao:2007dk}. Taking into account 
 phenomenological 
constraints, this study and \cite{LopezVal:2007rc} conclude that the cross section
for $gg\to t {\bar c}$ is at most 0.7  to  1 pb. 
However, this mode has a huge irreducible SM background:
the $t{\bar c}$ production rate in the SM is more than 12 times larger. 
Moreover, it is unlikely that
this final state can be separated from the much more frequent
$t{\bar q}$ events  (see section~\ref{sub-tchanp}). A cleaner signature
arises from the FCNC process $c g \to t$, whose cross section in the MSSM 
is, however, also not larger than about 1 pb  \cite{Cao:2007dk}.
Topcolour-type models predict somewhat larger FCNC effects
\cite{Cao:2007bx}.

An interesting signature caused by FCNC $gqt$ couplings is the
production of like-sign top pairs,
$qq \to tt$, ${\bar q}{\bar q}\to {\bar t}{\bar t}$ $(q=u,c)$  
\cite{Gouz:1998rk,Larios:2003jq}
 whose signature is two like-sign high $p_T$ leptons plus two
hard $b$-jets. The ATLAS collaboration has investigated the potential
reach of the LHC to this mode in terms of sensitivity limits to
anomalous $gtc$ and $gtu$ couplings \cite{:1999fr}.

While the best place to search for
 FCNC transitions $t\leftrightarrow c,u$ involving the photon and
$Z$ boson are top decays, 
hadronic single-top production is mostly sensitive to FCNC couplings
involving the gluon. The D0 collaboration at the Tevatron
searched  for the production of single top quarks by FCNC
$gtc$ and $gtu$ couplings and set upper limits on respective
anomalous coupling parameters, as no significant deviation from the
SM prediction was observed \cite{:2008be}. At the LHC the sensitivity to 
FCNC gluon top-quark couplings should increase by one to two orders 
of magnitude \cite{Beneke:2000hk}. 

In summary, from the measurements of
the cross sections of the
three production modes (\ref{eq-streac}) and the $t$- and $\bar t$-quark
polarizations, and from the search for exotic final states, a detailed
exploration of the weak interactions -- or, more general, flavour-changing
interactions -- of the top quark should be possible at the LHC.  

\section{Summary and outlook}
\label{secsumout}
An impressive amount of insight  has been gained to date 
into the properties and interactions
of the top quark. Yet it seems fair to say that
the physics of this quark remains to be fully explored in the years to
come, both in $\ttbar$ and single top-quark production and decay. 
This particle  provides the unique opportunity to
explore the physics of a bare quark at distances below the  attometer scale.
LHC experiments will probe, in the
$\ttbar$ channel, the existence of heavy resonances with masses up to
several TeV. These searches will, in particular, help to clarify the
mechanism of electroweak gauge symmetry breaking. 
The search for non-standard Higgs bosons and/or non-standard $CP$
violation will  provide important laboratory tests of  electroweak
baryogenesis scenarios. 
 A  precise measurement  of the $\mtt$ and  $p_T$ distributions, especially 
of their high-energy tails, and of correlations and  asymmetries 
will be an important issue in the high-luminosity
phase of the LHC. Angular distributions and correlations due to
top-spin effects should become, at the LHC, important tools in 
the investigations of the dynamics of top quarks. 
At the LHC the sensitivity 
to FCNC interactions involving the top quark should reach a level
$\sim 10^{-4}$.

The standard model predictions for top quarks as a signal at the LHC
are, by and large, in reasonably good shape. Yet there are still a
number of challenges which include the following issues: As reviewed above,
the top mass has been determined already with a relative uncertainty of
$0.8\%$, and this error on $m_t^{exp}$ will be further reduced in future
measurements. An interpretation of $m_t^{exp}$ in terms of a Lagrangian mass parameter
is, however, presently not possible at this level of precision. Observables should be identified, which can
be both computed and measured with reasonable precision, that allow
the determination of a well-defined top mass parameter with an
uncertainty of about $2$ GeV or better. Furthermore, the 
$\ttbar$ cross section at the LHC may eventually be
measurable with an uncertainty of about $5\%$. 
This requires the SM prediction for $\sigma^{\ttbar}_{LHC}$ to reach the
same level of precision, 
which necessitates the complete computation of the NNLO QCD corrections. 
Moreover, a number of distributions, including
the $\mtt$ and  $p_T$ distribution, and their theory errors should be
determined  more precisely. Furthermore,
these improvements will have to be included into Monte Carlo
simulation programs. In view of the progress that hadron collider
phenomenology has experienced in the recent past, one can be quite
confident that at least some of these issues will be resolved in the
not too distant future.

%% Acknowledgements %%%%%%%%%%%
\subsubsection*{Acknowledgements}
I am grateful to Martin Beneke, Stefan Berge, Sven Moch, 
Peter Uwer, Zongguo Si, Wolfgang Wagner, and
Peter Zerwas for helpful discussions, to Sven Moch for providing figure
4,  and to Alan D. Martin for suggesting to
write this review. This work was supported by Deutsche
Forschungsgemeinschaft SFB/TR9.


\begin{thebibliography}{555}

%%%%%%%%%%%%%%%%%%%%%% reviews %%%%%%%%%%%%%%%%%%%%%%%%

%%%%%%%%%%%%%%%%%%%%%%% observation %%%%%%%%%%%%%%%%%%%

%\cite{Abe:1995hr}
\bibitem{Abe:1995hr}
  F.~Abe {\it et al.}  [CDF Collaboration],
  %``Observation of top quark production in $\bar{p}p$ collisions,''
  Phys.\ Rev.\ Lett.\  {\bf 74} (1995) 2626
  [arXiv:hep-ex/9503002].
  %%CITATION = PRLTA,74,2626;%%
%\cite{Abachi:1995iq}
\bibitem{Abachi:1995iq}
  S.~Abachi {\it et al.}  [D0 Collaboration],
  %``Observation of the top quark,''
  Phys.\ Rev.\ Lett.\  {\bf 74} (1995) 2632
  [arXiv:hep-ex/9503003].
  %%CITATION = PRLTA,74,2632;%%
%\cite{Abazov:2006gd}
\bibitem{Abazov:2006gd}
  V.~M.~Abazov {\it et al.}  [D0 Collaboration],
  %``Evidence for production of single top quarks and first direct  measurement
  %of |V(tb)|,''
  Phys.\ Rev.\ Lett.\  {\bf 98} (2007) 181802
  [arXiv:hep-ex/0612052]; \\
  %%CITATION = PRLTA,98,181802;%%
 V.~M.~Abazov {\it et al.}  [D0 Collaboration],
  ``Evidence for production of single top quarks,''
  arXiv:0803.0739 [hep-ex].
  %%CITATION = ARXIV:0803.0739;%%


\bibitem{CDFsinglet}
{\tt http://www-cdf.fnal.gov/physics/new/top/public\_singletop.html}
\\
CDF Collaboration,
``Combination of CDF Single Top Quark 
Searches with 2.2 ${\rm fb}^{-1}$ of Data,''
CDF Note 9251 (2008).


%%%%% reviews %%%%%%%%%%%%


%\cite{Beneke:2000hk}
\bibitem{Beneke:2000hk}
  M.~Beneke {\it et al.},
  ``Top quark physics,''
  arXiv:hep-ph/0003033.
  %%CITATION = HEP-PH/0003033;%%

%\cite{Chakraborty:2003iw}
\bibitem{Chakraborty:2003iw}
  D.~Chakraborty, J.~Konigsberg and D.~L.~Rainwater,
  %``Review of top quark physics,''
  Ann.\ Rev.\ Nucl.\ Part.\ Sci.\  {\bf 53} (2003) 301
  [arXiv:hep-ph/0303092].
  %%CITATION = ARNUA,53,301;%%

%\cite{Wagner:2005jh}
\bibitem{Wagner:2005jh}
  W.~Wagner,
  %``Top quark physics in hadron collisions,''
  Rept.\ Prog.\ Phys.\  {\bf 68} (2005) 2409
  [arXiv:hep-ph/0507207].
  %%CITATION = RPPHA,68,2409;%%

%\cite{Quadt:2006jk}
\bibitem{Quadt:2006jk}
  A.~Quadt,
  %``Top quark physics at hadron colliders,''
  Eur.\ Phys.\ J.\  C {\bf 48} (2006) 835.
  %%CITATION = EPHJA,C48,835;%%






%%%%%%%%%%%%%%%%%%%%%%%%%   %%%%%%%%%%%%%%%%%%%%%%%%%%%%%%%%%%%%%%%%%%%



%\cite{Beneke:1994sw}
\bibitem{Beneke:1994sw}
  M.~Beneke and V.~M.~Braun,
  %``Heavy Quark Effective Theory Beyond Perturbation Theory: Renormalons, The
  %Pole Mass And The Residual Mass Term,''
  Nucl.\ Phys.\  B {\bf 426} (1994) 301
  [arXiv:hep-ph/9402364].
  %%CITATION = NUPHA,B426,301;%%

%\cite{Bigi:1994em}
\bibitem{Bigi:1994em}
  I.~I.~Y.~Bigi, M.~A.~Shifman, N.~G.~Uraltsev and A.~I.~Vainshtein,
  %``The Pole mass of the heavy quark. Perturbation theory and beyond,''
  Phys.\ Rev.\  D {\bf 50} (1994) 2234
  [arXiv:hep-ph/9402360].
  %%CITATION = PHRVA,D50,2234;%%

%\cite{Ball:1995ni}
\bibitem{Ball:1995ni}
  P.~Ball, M.~Beneke and V.~M.~Braun,
  %``Resummation of (beta0 alpha-s)**n corrections in QCD: Techniques and
  %applications to the tau hadronic width and the heavy quark pole mass,''
  Nucl.\ Phys.\  B {\bf 452} (1995) 563
  [arXiv:hep-ph/9502300].
  %%CITATION = NUPHA,B452,563;%%

%\cite{Smith:1996xz}
\bibitem{Smith:1996xz}
  M.~C.~Smith and S.~S.~Willenbrock,
  %``Top-quark pole mass,''
  Phys.\ Rev.\ Lett.\  {\bf 79} (1997) 3825
  [arXiv:hep-ph/9612329].
  %%CITATION = PRLTA,79,3825;%%

%\cite{Gray:1990yh}
\bibitem{Gray:1990yh}
  N.~Gray, D.~J.~Broadhurst, W.~Grafe and K.~Schilcher,
  %``Three Loop Relation Of Quark (Modified) Ms And Pole Masses,''
  Z.\ Phys.\  C {\bf 48} (1990) 673.
  %%CITATION = ZEPYA,C48,673;%%

%\cite{Chetyrkin:1999ys}
\bibitem{Chetyrkin:1999ys}
  K.~G.~Chetyrkin and M.~Steinhauser,
  %``Short distance mass of a heavy quark at order alpha(s)**3,''
  Phys.\ Rev.\ Lett.\  {\bf 83} (1999) 4001
  [arXiv:hep-ph/9907509].
  %%CITATION = PRLTA,83,4001;%%

%\cite{Melnikov:2000qh}
\bibitem{Melnikov:2000qh}
  K.~Melnikov and T.~v.~Ritbergen,
  %``The three-loop relation between the MS-bar and the pole quark masses,''
  Phys.\ Lett.\  B {\bf 482} (2000) 99
  [arXiv:hep-ph/9912391].
  %%CITATION = PHLTA,B482,99;%%


%\cite{:2007bxa}
\bibitem{:2007bxa}
  T.~T.~E.~Group {\it et al.}  [CDF Collaboration],
``A Combination of CDF and D0 Results on the Mass of the Top Quark,''
  arXiv:0803.1683 [hep-ex].
  %%CITATION = ARXIV:0803.1683;%%

\bibitem{:2007tmass}
    [CDF Collaboration],
  ``A Combination of CDF and D0 results on the mass of the top quark,''
  arXiv:hep-ex/0703034.
  %%CITATION = HEP-EX/0703034;%%

%\cite{Alcaraz:2007ri}
\bibitem{Alcaraz:2007ri}
  J.~Alcaraz {\it et al.}  [LEP Collaborations],
  ``Precision Electroweak Measurements and Constraints on the
   Standard Model,'' arXiv:0712.0929 [hep-ex].
  %%CITATION = ARXIV:0712.0929;%%


%\cite{Yao:2006px}
\bibitem{Yao:2006px}
  W.~M.~Yao {\it et al.}  [Particle Data Group],
  %``Review of particle physics,''
  J.\ Phys.\ G {\bf 33} (2006) 1.
  %%CITATION = JPHGB,G33,1;%%



%\cite{Bigi:1986jk}
\bibitem{Bigi:1986jk}
  I.~I.~Y.~Bigi, Y.~L.~Dokshitzer, V.~A.~Khoze, J.~H.~K\"uhn and P.~M.~Zerwas,
  %``Production and Decay Properties of Ultraheavy Quarks,''
  Phys.\ Lett.\  B {\bf 181} (1986) 157.
  %%CITATION = PHLTA,B181,157;%%

%
\bibitem{Kuhn:1983ix}
  J.~H.~K\"uhn,
  %``How To Measure The Polarization Of Top Quarks,''
  Nucl.\ Phys.\  B {\bf 237} (1984) 77.
  %%CITATION = NUPHA,B237,77;%%

%\cite{Hubaut:2005er}
\bibitem{Hubaut:2005er}
  F.~Hubaut, E.~Monnier, P.~Pralavorio, K.~Smolek and V.~Simak,
  %``ATLAS sensitivity to top quark and W boson polarization in t anti-t
  %events,''
  Eur.\ Phys.\ J.\  C {\bf 44S2} (2005) 13
  [arXiv:hep-ex/0508061].
  %%CITATION = EPHJA,C44S2,13;%%


%\cite{Ball:2007zza}
\bibitem{Ball:2007zza}
  G.~L.~Bayatian {\it et al.}  [CMS Collaboration],
  %``CMS technical design report, volume II: Physics performance,''
  J.\ Phys.\ G {\bf 34}, 995 (2007).
  %%CITATION = JPHGB,G34,995;%%


%\cite{Berger:2000zu}
\bibitem{Berger:2000zu}
  E.~L.~Berger and T.~M.~P.~Tait,
  ``Top spin and experimental tests,''
  arXiv:hep-ph/0002305.
  %%CITATION = HEP-PH/0002305;%%

%%%%%%%%%%%%%%%%%%%%%%%%%%%% charge %%%%%%%%%%%%%%%

%\cite{Chang:1998pt}
\bibitem{Chang:1998pt}
  D.~Chang, W.~F.~Chang and E.~Ma,
  %``Alternative interpretation of the Tevatron top events,''
  Phys.\ Rev.\  D {\bf 59} (1999) 091503
  [arXiv:hep-ph/9810531].
  %%CITATION = PHRVA,D59,091503;%%

%\cite{Abazov:2006vd}
\bibitem{Abazov:2006vd}
  V.~M.~Abazov {\it et al.}  [D0 Collaboration],
  %``Experimental discrimination between 
% charge $2e/3$ top quark and charge 4e/3
  %exotic quark production scenarios,''
  Phys.\ Rev.\ Lett.\  {\bf 98} (2007) 041801
  [arXiv:hep-ex/0608044].
  %%CITATION = PRLTA,98,041801;%%
 
%\cite{Beretvas:2006sw}
\bibitem{Beretvas:2006sw}
  A.~Beretvas {\it et al.}  [CDF Collaboration],
  ``Finding the charge of the top quark in the dilepton channel,''
  arXiv:0707.1339 [hep-ex].
  %%CITATION = ARXIV:0707.1339;%%

%\cite{Leone:2007mk}
\bibitem{Leone:2007mk}
  S.~Leone  [CDF Collaboration],
  ``Electroweak and Top Physics at the Tevatron and Indirect Higgs Limits,''
  arXiv:0710.4983 [hep-ex].
  %%CITATION = ARXIV:0710.4983;%%

%\cite{Abazov:2008yn}
\bibitem{Abazov:2008yn}
  V.~M.~Abazov {\it et al.}  [D0 Collaboration],
  ``Simultaneous measurement of the ratio 
 $B(t \to Wb)/B(t\to Wq)$ and the top quark
  pair production cross section with the D0 detector at $\sqrt s$ =1.96 TeV,''
  arXiv:0801.1326 [hep-ex].
  %%CITATION = ARXIV:0801.1326;%%




%%%%%%%%%%%%%%%%%%%%%%%%%%%%%%%%%%%%%%%%%%%%
%%%%%%%%%%%%%%%%%%%%%%%%%%% top decay %%%%%%%%%%%%%%%%%%%%%%%%%%%%
%%%%%%%%%%%%%%%%%%%%%%%%%%%%%%%%%%%%%%%%%%%%

%%%%%%%%%%%%%%%%% width in SM %%%%%%%%%%%%%%%

\bibitem{JezKu89}
 M.~Jezabek and J.~H.~K\"uhn,
 Nucl.\ Phys.\  B {\bf 314} (1989) 1.

%\cite{Denner:1990ns}
\bibitem{Denner:1990ns}
  A.~Denner and T.~Sack,
  %``The Top width,''
  Nucl.\ Phys.\  B {\bf 358} (1991) 46.
  %%CITATION = NUPHA,B358,46;%%

%\cite{Eilam:1991iz}
\bibitem{Eilam:1991iz}
  G.~Eilam, R.~R.~Mendel, R.~Migneron and A.~Soni,
  %``Radiative corrections to top quark decay,''
  Phys.\ Rev.\ Lett.\  {\bf 66} (1991) 3105.
  %%CITATION = PRLTA,66,3105;%%


%\cite{Jezabek:1993wk}
\bibitem{Jezabek:1993wk}
  M.~Jezabek and J.~H.~K\"uhn,
  %``The Top width: Theoretical update,''
  Phys.\ Rev.\  D {\bf 48} (1993) 1910
  [Erratum-ibid.\  D {\bf 49} (1994) 4970]
  [arXiv:hep-ph/9302295].
  %%CITATION = PHRVA,D48,1910;%%

%\cite{Czarnecki:1998qc}
\bibitem{Czarnecki:1998qc}
  A.~Czarnecki and K.~Melnikov,
  %``Two-loop {QCD} corrections to top quark width,''
  Nucl.\ Phys.\  B {\bf 544} (1999) 520
  [arXiv:hep-ph/9806244].
  %%CITATION = NUPHA,B544,520;%%


\bibitem{Chet99}
  K.~G.~Chetyrkin, R.~Harlander, T.~Seidensticker and M.~Steinhauser,
  %``Second order {QCD} corrections to Gamma(t --> W b),''
  Phys.\ Rev.\  D {\bf 60} (1999) 114015
  [arXiv:hep-ph/9906273].
  %%CITATION = PHRVA,D60,114015;%%


%%%%%%%%%%%%%%%%%%%%%%%%%%%%%%%%% W helicity %%%%%%%%%%%%%%%


%\cite{Do:2002ky}
\bibitem{Do:2002ky}
  H.~S.~Do, S.~Groote, J.~G.~K\"orner and M.~C.~Mauser,
  %``Electroweak and finite width corrections to top quark decays into
  %transverse and longitudinal W-bosons,''
  Phys.\ Rev.\  D {\bf 67} (2003) 091501
  [arXiv:hep-ph/0209185].
  %%CITATION = PHRVA,D67,091501;%%

%\cite{Fischer:2001gp}
\bibitem{Fischer:2001gp}
  M.~Fischer, S.~Groote, J.~G.~K\"orner and M.~C.~Mauser,
  %``Complete angular analysis of polarized top decay at O(alpha(s)),''
  Phys.\ Rev.\  D {\bf 65} (2002) 054036
  [arXiv:hep-ph/0101322].
  %%CITATION = PHRVA,D65,054036;%%


%\cite{AguilarSaavedra:2006fy}
\bibitem{AguilarSaavedra:2006fy}
 J.~A.~Aguilar-Saavedra, J.~Carvalho, N.~Castro, F.~Veloso and A.~Onofre,
  %``Probing anomalous W t b couplings in top pair decays,''
  Eur.\ Phys.\ J.\  C {\bf 50} (2007) 519
  [arXiv:hep-ph/0605190].
  %%CITATION = EPHJA,C50,519;%%


%\cite{Abulencia:2006ei}
\bibitem{Abulencia:2006ei}
  A.~Abulencia {\it et al.}  [CDF II Collaboration],
  %``Measurement of the Helicity Fractions of W Bosons from Top Quark Decays
  %using Fully Reconstructed \boldmath${t\bar{t}}$ Events with CDF II,''
  Phys.\ Rev.\  D {\bf 75} (2007) 052001
  [arXiv:hep-ex/0612011].
  %%CITATION = PHRVA,D75,052001;%%

%\cite{Abulencia:2006iy}
\bibitem{Abulencia:2006iy}
  A.~Abulencia {\it et al.}  [CDF Collaboration],
  %``Search for V + A current in top quark decay in p anti-p collisions at
  %s**(1/2) = 1.96-TeV,''
  Phys.\ Rev.\ Lett.\  {\bf 98} (2007) 072001
  [arXiv:hep-ex/0608062].
  %%CITATION = PRLTA,98,072001;%%

%\cite{Abazov:2006hb}
\bibitem{Abazov:2006hb}
  V.~M.~Abazov {\it et al.}  [D0 Collaboration],
  %``Measurement of the W boson helicity in top quark decay at D0,''
  Phys.\ Rev.\  D {\bf 75} (2007) 031102
  [arXiv:hep-ex/0609045].
  %%CITATION = PHRVA,D75,031102;%%

%\cite{Abazov:2007ve}
\bibitem{Abazov:2007ve}
  V.~M.~Abazov {\it et al.}  [D0 Collaboration],
  ``Model-independent measurement of the W boson helicity in top quark
   decays,''
  arXiv:0711.0032 [hep-ex].
  %%CITATION = ARXIV:0711.0032;%%



%\cite{AguilarSaavedra:2007rs}
\bibitem{AguilarSaavedra:2007rs}
  J.~A.~Aguilar-Saavedra, J.~Carvalho, N.~Castro, A.~Onofre and F.~Veloso,
  %``ATLAS sensitivity to Wtb anomalous couplings in top quark decays,''
  arXiv:0705.3041 [hep-ph].
  %%CITATION = ARXIV:0705.3041;%%


%%%%%%%%%%%% distributions, helicity %%%%%%%%

%\cite{Czarnecki:1990pe}
\bibitem{Czarnecki:1990pe}
  A.~Czarnecki, M.~Jezabek and J.~H.~K\"uhn,
  %``LEPTON SPECTRA FROM DECAYS OF POLARIZED TOP QUARKS,''
  Nucl.\ Phys.\  B {\bf 351} (1991) 70.
  %%CITATION = NUPHA,B351,70;%%

%\cite{Brandenburg:2002xr}
\bibitem{Brandenburg:2002xr}
  A.~Brandenburg, Z.~G.~Si and P.~Uwer,
  %``QCD-corrected spin analysing power of jets in decays of polarized top
  %quarks,''
  Phys.\ Lett.\  B {\bf 539} (2002) 235
  [arXiv:hep-ph/0205023].
  %%CITATION = PHLTA,B539,235;%%

  

%\cite{Fischer:1998gsa}
\bibitem{Fischer:1998gsa}
  M.~Fischer, S.~Groote, J.~G.~K\"orner, M.~C.~Mauser and B.~Lampe,
  %``Polarized top decay into polarized W: t(pol.) --> W(pol.) + b at
  %O(alpha(s)),''
  Phys.\ Lett.\  B {\bf 451} (1999) 406
  [arXiv:hep-ph/9811482].
  %%CITATION = PHLTA,B451,406;%%
 
%\cite{Jezabek:1994zv}
\bibitem{Jezabek:1994zv}
  M.~Jezabek and J.~H.~K\"uhn,
  %``V-A tests through leptons from polarized top quarks,''
  Phys.\ Lett.\  B {\bf 329} (1994) 317
  [arXiv:hep-ph/9403366].
  %%CITATION = PHLTA,B329,317;%%

%\cite{Bernreuther:2003xj}
\bibitem{Bernreuther:2003xj}
  W.~Bernreuther, M.~Fuecker and Y.~Umeda,
  %``Semileptonic decays of polarised top quarks: V + A admixture and QCD
  %corrections,''
  Phys.\ Lett.\  B {\bf 582} (2004) 32
  [arXiv:hep-ph/0308296].
  %%CITATION = PHLTA,B582,32;%%


%%%%%%%%%%%%%%%%% NON-SM t decay %%%%%%%%%%%%%%%


%%%%%%%%%%%%%%% non-SM effects in 
%%%% distributions of SL top decay, anomalous couplings  %%%%%%%%%%%%
%%%%%%%%%%%%%%%% general form factor decomp. %%%%%%%%%%%%%%%%%

%\cite{Bernreuther:1992be}
\bibitem{Bernreuther:1992be}
  W.~Bernreuther, O.~Nachtmann, P.~Overmann and T.~Schr\"oder,
  %``Angular correlations and distributions for searches of CP violation in top
  %quark production and decay,''
   Nucl.\ Phys.\ B {\bf 388} (1992) 53
  [Erratum-ibid.\ B {\bf 406} (1993) 516].
  %%CITATION = NUPHA,B388,53;%%

%\cite{Ma:1991ry}
\bibitem{Ma:1991ry}
  J.~P.~Ma and A.~Brandenburg,
  %``CP violation and top quark decays,''
  Z.\ Phys.\  C {\bf 56} (1992) 97.
  %%CITATION = ZEPYA,C56,97;%%

%\cite{Kane:1991bg}
\bibitem{Kane:1991bg}
  G.~L.~Kane, G.~A.~Ladinsky and C.~P.~Yuan,
  %``Using the top quark for testing standard model polarization and CP
  %predictions,''
  Phys.\ Rev.\ D {\bf 45} (1992) 124.
  %%CITATION = PHRVA,D45,124;%%

%\cite{Fujikawa:1993zu}
\bibitem{Fujikawa:1993zu}
  K.~Fujikawa and A.~Yamada,
  %``Test of the chiral structure of the top - bottom charged current by the
  %process b $\to$ s gamma,''
  Phys.\ Rev.\  D {\bf 49} (1994) 5890.
  %%CITATION = PHRVA,D49,5890;%%

%\cite{Cho:1993zb}
\bibitem{Cho:1993zb}
  P.~L.~Cho and M.~Misiak,
  %``b $\to$ s gamma decay in SU(2)-L x SU(2)-r x U(1) extensions of the
  %Standard Model,''
  Phys.\ Rev.\  D {\bf 49} (1994) 5894
  [arXiv:hep-ph/9310332].
  %%CITATION = PHRVA,D49,5894;%%


%\cite{Grzadkowski:2008mf}
\bibitem{Grzadkowski:2008mf}
  B.~Grzadkowski and M.~Misiak,
  ``Anomalous Wtb coupling effects in the weak radiative B-meson decay,''
  arXiv:0802.1413 [hep-ph].
  %%CITATION = ARXIV:0802.1413;%%

%\cite{Larios:1999au}
\bibitem{Larios:1999au}
  F.~Larios, M.~A.~Perez and C.~P.~Yuan,
  %``Analysis of t b W and t t Z couplings from CLEO and LEP/SLC data,''
  Phys.\ Lett.\  B {\bf 457} (1999) 334
  [arXiv:hep-ph/9903394].
  %%CITATION = PHLTA,B457,334;%%

%\cite{Burdman:1999fw}
\bibitem{Burdman:1999fw}
  G.~Burdman, M.~C.~Gonzalez-Garcia and S.~F.~Novaes,
  %``Anomalous couplings of the third generation in rare B decays,''
  Phys.\ Rev.\  D {\bf 61} (2000) 114016
  [arXiv:hep-ph/9906329].
  %%CITATION = PHRVA,D61,114016;%%




%\cite{Brandenburg:2002xa}
\bibitem{Brandenburg:2002xa}
  A.~Brandenburg and M.~Maniatis,
  %``Impact of SUSY-QCD corrections on top quark decay distributions,''
  Phys.\ Lett.\  B {\bf 545} (2002) 139
  [arXiv:hep-ph/0207154].
  %%CITATION = PHLTA,B545,139;%%

%\cite{Cao:2003yk}
\bibitem{Cao:2003yk}
  J.~j.~Cao, R.~J.~Oakes, F.~Wang and J.~M.~Yang,
  %``Supersymmetric effects in top quark decay into polarized W-boson,''
  Phys.\ Rev.\  D {\bf 68} (2003) 054019
  [arXiv:hep-ph/0306278].
  %%CITATION = PHRVA,D68,054019;%%

%\cite{He:1999vp}
\bibitem{He:1999vp}
  H.~J.~He, T.~Tait and C.~P.~Yuan,
  %``New topflavor models with seesaw mechanism,''
  Phys.\ Rev.\  D {\bf 62} (2000) 011702
  [arXiv:hep-ph/9911266].
  %%CITATION = PHRVA,D62,011702;%%


%\cite{Hikasa:1998wx}
\bibitem{Hikasa:1998wx}
  K.~i.~Hikasa, K.~Whisnant, J.~M.~Yang and B.~L.~Young,
  %``Probing anomalous top quark interactions at the Fermilab Tevatron
  %collider,''
  Phys.\ Rev.\  D {\bf 58} (1998) 114003
  [arXiv:hep-ph/9806401].
  %%CITATION = PHRVA,D58,114003;%%

%\cite{Boos:1999dd}
\bibitem{Boos:1999dd}
  E.~Boos, L.~Dudko and T.~Ohl,
  %``Complete calculations of W b anti-b and W b anti-b + jet production at
  %Tevatron and LHC: Probing anomalous W t b couplings in single top
  %production,''
  Eur.\ Phys.\ J.\  C {\bf 11} (1999) 473
  [arXiv:hep-ph/9903215].
  %%CITATION = EPHJA,C11,473;%%

%\cite{Chen:2005vr}
\bibitem{Chen:2005vr}
  C.~R.~Chen, F.~Larios and C.~P.~Yuan,
  %``General analysis of single top production and W helicity in top decay,''
  Phys.\ Lett.\  B {\bf 631} (2005) 126
  [AIP Conf.\ Proc.\  {\bf 792} (2005) 591]
  [arXiv:hep-ph/0503040].
  %%CITATION = APCPC,792,591;%%




%\cite{Bernreuther:1993xp}
\bibitem{Bernreuther:1993xp}
  W.~Bernreuther and P.~Overmann,
  %``CP asymmetries in top quark pair production and decay: Contributions from
  %neutral Higgs boson and gluino exchange,''
  Z.\ Phys.\  C {\bf 61} (1994) 599.
  %%CITATION = ZEPYA,C61,599;%%




%%%%%%%%%%%%% T-odd final state




%\cite{Nelson:1997xd}
\bibitem{Nelson:1997xd}
  C.~A.~Nelson, B.~T.~Kress, M.~Lopes and T.~P.~McCauley,
  %``General tests for t --> W+ b couplings at hadron colliders,''
  Phys.\ Rev.\  D {\bf 56} (1997) 5928
  [arXiv:hep-ph/9707211].
  %%CITATION = PHRVA,D56,5928;%%

%\cite{Nelson:1998pu}
\bibitem{Nelson:1998pu}
  C.~A.~Nelson and A.~M.~Cohen,
  %``Measurement of helicity parameters in top quark decay,''
  Eur.\ Phys.\ J.\  C {\bf 8} (1999) 393
  [arXiv:hep-ph/9806373].
  %%CITATION = EPHJA,C8,393;%%

%\cite{Nelson:2000dn}
\bibitem{Nelson:2000dn}
  C.~A.~Nelson and L.~J.~.~Adler,
  %``On measurement of helicity parameters in top quark decay,''
  Eur.\ Phys.\ J.\  C {\bf 17} (2000) 399
  [arXiv:hep-ph/0007086].
  %%CITATION = EPHJA,C17,399;%%



%\cite{Grzadkowski:1999iq}
\bibitem{Grzadkowski:1999iq}
  B.~Grzadkowski and Z.~Hioki,
  %``New hints for testing anomalous top quark interactions at future linear
  %colliders,''
  Phys.\ Lett.\  B {\bf 476} (2000) 87
  [arXiv:hep-ph/9911505].
  %%CITATION = PHLTA,B476,87;%%

%\cite{Rindani:2000jg}
\bibitem{Rindani:2000jg}
  S.~D.~Rindani,
  %``Effect of anomalous t b W vertex on decay-lepton distributions in  e+ e-
  %--> t anti-t and CP-violating asymmetries,''
  Pramana {\bf 54} (2000) 791
  [arXiv:hep-ph/0002006].
  %%CITATION = PRAMC,54,791;%%


%\cite{Godbole:2006tq}
\bibitem{Godbole:2006tq}
  R.~M.~Godbole, S.~D.~Rindani and R.~K.~Singh,
  %``Lepton distribution as a probe of new physics in production and decay of
  %the t quark and its polarization,''
  JHEP {\bf 0612} (2006) 021
  [arXiv:hep-ph/0605100].
  %%CITATION = JHEPA,0612,021;%%

%%%%%%%%%%%% exp. sensitivity %%%%%%%%%%%%%%


%%% see the papers by Hubault and by Saavedra

%%%%%%%%%%%%%%%%%%%%%%% non-SM top decays %%%%%%%%%%%%%%%%%%%%%%
%%%%%%%%%%%%%%%%%%%% charged Higgs %%%%%%%%%%%%%%

%\cite{Gunion:1989we}
\bibitem{Gunion:1989we}
  J.~F.~Gunion, H.~E.~Haber, G.~L.~Kane and S.~Dawson,
  ``The Higgs Hunter's Guide,'' Westview Press, Cambridge, Mass. (1990).
  %%CITATION = BNL-41644;%%

%\cite{Gambino:2001ew}
\bibitem{Gambino:2001ew}
  P.~Gambino and M.~Misiak,
  %``Quark mass effects in anti-B --> X/s gamma,''
  Nucl.\ Phys.\  B {\bf 611} (2001) 338
  [arXiv:hep-ph/0104034].
  %%CITATION = NUPHA,B611,338;%%


%\cite{Li:1990cp}
\bibitem{Li:1990cp}
  C.~S.~Li and T.~C.~Yuan,
  %``QCD Correction To Charged Higgs Decay Of The Top Quark,''
  Phys.\ Rev.\  D {\bf 42} (1990) 3088
  [Erratum-ibid.\  D {\bf 47} (1993) 2156].
  %%CITATION = PHRVA,D42,3088;%%

%\cite{Czarnecki:1992zm}
\bibitem{Czarnecki:1992zm}
  A.~Czarnecki and S.~Davidson,
  %``QCD corrections to the charged Higgs decay of a heavy quark,''
  Phys.\ Rev.\  D {\bf 48} (1993) 4183
  [arXiv:hep-ph/9301237].
  %%CITATION = PHRVA,D48,4183;%%

%\cite{Coarasa:1996qa}
\bibitem{Coarasa:1996qa}
  J.~A.~Coarasa, D.~Garcia, J.~Guasch, R.~A.~Jimenez and J.~Sola,
  %``Quantum effects on t --> H+ b in the MSSM: A window to *virtual*
  %supersymmetry?,''
  Eur.\ Phys.\ J.\  C {\bf 2} (1998) 373
  [arXiv:hep-ph/9607485].
  %%CITATION = EPHJA,C2,373;%%


%\cite{Carena:1999py}
\bibitem{Carena:1999py}
  M.~S.~Carena, D.~Garcia, U.~Nierste and C.~E.~M.~Wagner,
  %``Effective Lagrangian for the anti-t b H+ interaction in the MSSM and
  %charged Higgs phenomenology,''
  Nucl.\ Phys.\  B {\bf 577} (2000) 88
  [arXiv:hep-ph/9912516].
  %%CITATION = NUPHA,B577,88;%%

%%%%%%%%%%%%%%%%%%%% exp. search for charged Higgs %%%%%%%%%%%%%%

%\cite{Abazov:2001md}
\bibitem{Abazov:2001md}
  V.~M.~Abazov {\it et al.}  [D0 Collaboration],
  %``Direct search for charged Higgs bosons in decays of top quarks,''
  Phys.\ Rev.\ Lett.\  {\bf 88} (2002) 151803
  [arXiv:hep-ex/0102039].
  %%CITATION = PRLTA,88,151803;%%

%\cite{Abulencia:2005jd}
\bibitem{Abulencia:2005jd}
  A.~Abulencia {\it et al.}  [CDF Collaboration],
  %``Search for charged Higgs bosons from top quark decays in $p\bar{p}$
  %collisions at $\sqrt{s} =$ 1.96-TeV,''
  Phys.\ Rev.\ Lett.\  {\bf 96} (2006) 042003
  [arXiv:hep-ex/0510065].
  %%CITATION = PRLTA,96,042003;%%

%\cite{Grenier:2007xj}
\bibitem{Grenier:2007xj}
  G.~Grenier,
  ``Search for supersymmetric charged Higgs bosons at the TeVatron,''
  arXiv:0710.0853 [hep-ex].
  %%CITATION = ARXIV:0710.0853;%%

%\cite{:1999fr}
\bibitem{:1999fr}
 ``ATLAS detector and physics performance. 
 Technical design report.  Vol. 2,'',
 report CERN-LHCC-99-15.
  %%CITATION = ATLAS-TDR-15;%%


%%%%%%%%%%% spin analyzer %%%%%%%%%%
%\cite{Korner:2002fx}
\bibitem{Korner:2002fx}
  J.~G.~K\"orner and M.~C.~Mauser,
  ``O(alpha(s)) radiative corrections to polarized top decay into a charged
  Higgs t(pol.) $\to$ $ H^+$ + b,''
  arXiv:hep-ph/0211098.
  %%CITATION = HEP-PH/0211098;%%

%%%%%%%%%%%%%%%%%%%%% t -> stop %%%%%%%%%%%%%%%%%%%%%%%

%\cite{Ellis:1983ed}
\bibitem{Ellis:1983ed}
  J.~R.~Ellis and S.~Rudaz,
  %``Search For Supersymmetry In Toponium Decays,''
  Phys.\ Lett.\  B {\bf 128} (1983) 248.
  %%CITATION = PHLTA,B128,248;%%

%\cite{Hikasa:1987db}
\bibitem{Hikasa:1987db}
  K.~I.~Hikasa and M.~Kobayashi,
  %``Light Scalar Top at e+ e- Colliders,''
  Phys.\ Rev.\  D {\bf 36} (1987) 724.
  %%CITATION = PHRVA,D36,724;%%

%\cite{Baer:1991cb}
\bibitem{Baer:1991cb}
  H.~Baer, M.~Drees, R.~Godbole, J.~F.~Gunion and X.~Tata,
  %``Phenomenology of light top squarks at the Fermilab Tevatron,''
  Phys.\ Rev.\  D {\bf 44} (1991) 725.
  %%CITATION = PHRVA,D44,725;%%

%\cite{Porod:1998yp}
\bibitem{Porod:1998yp}
  W.~Porod,
  %``More on higher order decays of the lighter top squark,''
  Phys.\ Rev.\  D {\bf 59} (1999) 095009
  [arXiv:hep-ph/9812230].
  %%CITATION = PHRVA,D59,095009;%%

%\cite{Boehm:1999tr}
\bibitem{Boehm:1999tr}
  C.~Boehm, A.~Djouadi and Y.~Mambrini,
  %``Decays of the lightest top squark,''
  Phys.\ Rev.\  D {\bf 61} (2000) 095006
  [arXiv:hep-ph/9907428].
  %%CITATION = PHRVA,D61,095006;%%

%\cite{Affolder:2000bp}
\bibitem{Affolder:2000bp}
  A.~A.~Affolder {\it et al.}  [CDF Collaboration],
  %``Search for the supersymmetric partner of the top quark in $p\bar{p}$
  %collisions at $\sqrt{s} = 1.8$ TeV,''
  Phys.\ Rev.\  D {\bf 63} (2001) 091101
  [arXiv:hep-ex/0011004].
  %%CITATION = PHRVA,D63,091101;%%

%\cite{Hosch:1997vf}
\bibitem{Hosch:1997vf}
  M.~Hosch, R.~J.~Oakes, K.~Whisnant, J.~M.~Yang, B.~l.~Young and X.~Zhang,
  %``Probing top quark decay into light stop in the supersymmetric standard
  %model at the upgraded Tevatron,''
  Phys.\ Rev.\  D {\bf 58} (1998) 034002
  [arXiv:hep-ph/9711234].
  %%CITATION = PHRVA,D58,034002;%%

%\cite{Acosta:2003ys}
\bibitem{Acosta:2003ys}
  D.~E.~Acosta {\it et al.}  [CDF Collaboration],
  %``Search for the supersymmetric partner of the top quark in dilepton events
  %from $p\bar{p}$ collisions at $\sqrt{s} = 1.8$ TeV,''
  Phys.\ Rev.\ Lett.\  {\bf 90} (2003) 251801
  [arXiv:hep-ex/0302009].
  %%CITATION = PRLTA,90,251801;%%

%\cite{:2007im}
\bibitem{:2007im}
  V.~M.~Abazov {\it et al.}  [D0 Collaboration],
  %``Search for the lightest scalar top quark in events with two leptons in $p
  %\bar{p}$ collisions at $\sqrt{s}$ = 1.96-TeV,''
  Phys.\ Lett.\  B {\bf 659} (2008) 500
  [arXiv:0707.2864 [hep-ex]].
  %%CITATION = PHLTA,B659,500;%%


%\cite{Abazov:2003wt}
\bibitem{Abazov:2003wt}
  V.~M.~Abazov {\it et al.}  [D0 collaboration],
  %``Search for 3- and 4-body decays of the scalar top quark in $p\bar{p}$
  %collisions at $\sqrt{s} = 1.8$ TeV,''
  Phys.\ Lett.\  B {\bf 581} (2004) 147.
  %%CITATION = PHLTA,B581,147;%%

%\cite{Abazov:2006wb}
\bibitem{Abazov:2006wb}
  V.~M.~Abazov {\it et al.}  [D0 Collaboration],
  %``Search for the pair production of scalar top quarks in 
  % the acoplanar  charm
  %jet topology in p anti-p collisions at s**(1/2) = 1.96-TeV,''
  Phys.\ Lett.\  B {\bf 645} (2007) 119
  [arXiv:hep-ex/0611003].
  %%CITATION = PHLTA,B645,119;%%

%\cite{Aaltonen:2007sw}
\bibitem{Aaltonen:2007sw}
  T.~Aaltonen {\it et al.}  [CDF Collaboration],
  %``Search for Direct Pair Production of Supersymmetric Top and Supersymmetric
  %Bottom Quarks in $p \bar{p}$ Collisions at $\sqrt{s}$ = 1.96-TeV,''
  Phys.\ Rev.\  D {\bf 76} (2007) 072010
  [arXiv:0707.2567 [hep-ex]].
  %%CITATION = PHRVA,D76,072010;%%

%\cite{Duperrin:2007uy}
\bibitem{Duperrin:2007uy}
  A.~Duperrin  [CDF Collaboration],
  ``Searches for Higgs and BSM at the Tevatron,''
  arXiv:0710.4265 [hep-ex].
  %%CITATION = ARXIV:0710.4265;%%


%\cite{Carena:1997ki}
\bibitem{Carena:1997ki}
  M.~S.~Carena, M.~Quiros and C.~E.~M.~Wagner,
  %``Electroweak baryogenesis and Higgs and stop searches at LEP and the
  %Tevatron,''
  Nucl.\ Phys.\  B {\bf 524} (1998) 3
  [arXiv:hep-ph/9710401].
  %%CITATION = NUPHA,B524,3;%%


%%%%%%%%%%%%%%%%%%% FCNC in SM and BSM %%%%%%%%%%%%%%%%%%
%%%%%%%%%%%%%%%%%%% in SM %%%%%%%%%%%%%%%%%%%
%\cite{Glashow:1970gm}
\bibitem{Glashow:1970gm}
  S.~L.~Glashow, J.~Iliopoulos and L.~Maiani,
  %``Weak Interactions with Lepton-Hadron Symmetry,''
  Phys.\ Rev.\  D {\bf 2} (1970) 1285.
  %%CITATION = PHRVA,D2,1285;%%

%\cite{DiazCruz:1989ub}
\bibitem{DiazCruz:1989ub}
  J.~L.~Diaz-Cruz, R.~Martinez, M.~A.~Perez and A.~Rosado,
  %``FLAVOR CHANGING RADIATIVE DECAY OF THF t QUARK,''
  Phys.\ Rev.\  D {\bf 41} (1990) 891.
  %%CITATION = PHRVA,D41,891;%%

%\cite{Eilam:1990zc}
\bibitem{Eilam:1990zc}
  G.~Eilam, J.~L.~Hewett and A.~Soni,
  %``Rare decays of the top quark in the standard and two Higgs doublet
  %models,''
  Phys.\ Rev.\  D {\bf 44} (1991) 1473
  [Erratum-ibid.\  D {\bf 59} (1999) 039901].
  %%CITATION = PHRVA,D44,1473;%%

%\cite{Mele:1998ag}
\bibitem{Mele:1998ag}
  B.~Mele, S.~Petrarca and A.~Soddu,
  %``A new evaluation of the t --> c H decay width in the standard model,''
  Phys.\ Lett.\  B {\bf 435} (1998) 401
  [arXiv:hep-ph/9805498].
  %%CITATION = PHLTA,B435,401;%%

%%%%%%% exp. sensitivity at ATLAS, for CMS see their TDR %%

\bibitem{Carvalho:2007yi}
  J.~Carvalho {\it et al.}  [ATLAS Collaboration],
  %``Study of ATLAS sensitivity to FCNC top decays,''
  Eur.\ Phys.\ J.\  C {\bf 52} (2007) 999
  [arXiv:0712.1127 [hep-ex]].
  %%CITATION = EPHJA,C52,999;%%


%\cite{Hill:2002ap}
\bibitem{Hill:2002ap}
  C.~T.~Hill and E.~H.~Simmons,
  %``Strong dynamics and electroweak symmetry breaking,''
  Phys.\ Rept.\  {\bf 381} (2003) 235
  [Erratum-ibid.\  {\bf 390} (2004) 553]
  [arXiv:hep-ph/0203079].
  %%CITATION = PRPLC,381,235;%%

%\cite{Valenzuela:2005xr}
\bibitem{Valenzuela:2005xr}
  C.~Valenzuela,
  %``Spontaneous CP symmetry breaking at the electroweak scale,''
  Phys.\ Rev.\  D {\bf 71} (2005) 095014
  [arXiv:hep-ph/0503111].
  %%CITATION = PHRVA,D71,095014;%%

%%% the following ref.  is a review %%%%%%%%%
%\cite{AguilarSaavedra:2004wm}
\bibitem{AguilarSaavedra:2004wm}
  J.~A.~Aguilar-Saavedra,
  %``Top flavour-changing neutral interactions: Theoretical expectations and
  %experimental detection,''
  Acta Phys.\ Polon.\  B {\bf 35} (2004) 2695
  [arXiv:hep-ph/0409342].
  %%CITATION = APPOA,B35,2695;%%


%%%%%%%%%%%%%%%%%%% FCNC t decays in 2HDM and MSSM %%%%%%%%%%%%%%

%%%%%%% type 3 2HDM  %%%%%%%%%%%%
%\cite{Luke:1993cy}
\bibitem{Luke:1993cy}
  M.~E.~Luke and M.~J.~Savage,
  %``Flavor changing neutral currents in 
% the Higgs sector and rare top decays,''
  Phys.\ Lett.\  B {\bf 307} (1993) 387
  [arXiv:hep-ph/9303249].
  %%CITATION = PHLTA,B307,387;%%

%\cite{Atwood:1996vj}
\bibitem{Atwood:1996vj}
  D.~Atwood, L.~Reina and A.~Soni,
  %``Phenomenology of two Higgs doublet models with flavor changing neutral
  %currents,''
  Phys.\ Rev.\  D {\bf 55} (1997) 3156
  [arXiv:hep-ph/9609279].
  %%CITATION = PHRVA,D55,3156;%%


%%%%%%%%%%%%%%%%%% MSSM %%%%%%%%%%%%%
%%%%%%%%%%%%% t -> c h %%%%%%%%%%%%%%%%%%%

%\cite{Yang:1993rb}
\bibitem{Yang:1993rb}
  J.~M.~Yang and C.~S.~Li,
  %``Top quark rare decay t $\to$ c H(i) in the minimal supersymmetric model,''
  Phys.\ Rev.\  D {\bf 49} (1994) 3412
  [Erratum-ibid.\  D {\bf 51} (1995) 3974].
  %%CITATION = PHRVA,D49,3412;%%

%\cite{Guasch:1999jp}
\bibitem{Guasch:1999jp}
  J.~Guasch and J.~Sola,
  %``FCNC top quark decays: A door to SUSY physics in high luminosity
  %colliders?,''
  Nucl.\ Phys.\  B {\bf 562} (1999) 3
  [arXiv:hep-ph/9906268].
  %%CITATION = NUPHA,B562,3;%%

%\cite{Delepine:2004hr}
\bibitem{Delepine:2004hr}
  D.~Delepine and S.~Khalil,
  %``Top flavour violating decays in general supersymmetric models,''
  Phys.\ Lett.\  B {\bf 599} (2004) 62
  [arXiv:hep-ph/0406264].
  %%CITATION = PHLTA,B599,62;%%

%%%%%%%%%% MSSM: t -> c V %%%%%%%%%%%%%%%%%%
%\cite{Liu:2004qw}
\bibitem{Liu:2004qw}
  J.~J.~Liu, C.~S.~Li, L.~L.~Yang and L.~G.~Jin,
  %``t --> c V via SUSY FCNC couplings in the unconstrained MSSM,''
  Phys.\ Lett.\  B {\bf 599} (2004) 92
  [arXiv:hep-ph/0406155].
  %%CITATION = PHLTA,B599,92;%%

%%%%%%%%% this ref. contains also t production %%%
%\cite{Cao:2007dk}
\bibitem{Cao:2007dk}
  J.~J.~Cao, G.~Eilam, M.~Frank, K.~Hikasa, G.~L.~Liu, I.~Turan and J.~M.~Yang,
  %``SUSY-induced FCNC top-quark processes at the Large Hadron Collider,''
  Phys.\ Rev.\  D {\bf 75} (2007) 075021
  [arXiv:hep-ph/0702264].
  %%CITATION = PHRVA,D75,075021;%%



%%%%%%%%%%%%%%%%%%%%%%%%%% R-parity violating SUSY %%%%%%%%%%

%\cite{Eilam:2001dh}
\bibitem{Eilam:2001dh}
  G.~Eilam, A.~Gemintern, T.~Han, J.~M.~Yang and X.~Zhang,
  %``Top quark rare decay t --> c h in R-parity-violating SUSY,''
  Phys.\ Lett.\  B {\bf 510} (2001) 227
  [arXiv:hep-ph/0102037].
  %%CITATION = PHLTA,B510,227;%%

%\cite{Abraham:2000kx}
\bibitem{Abraham:2000kx}
  K.~J.~Abraham, K.~Whisnant, J.~M.~Yang and B.~L.~Young,
  %``Probing R-violating top quark decays at hadron colliders,''
  Phys.\ Rev.\  D {\bf 63} (2001) 034011
  [arXiv:hep-ph/0007280].
  %%CITATION = PHRVA,D63,034011;%%


%\cite{Lu:2003yr}
\bibitem{Lu:2003yr}
  G.~r.~Lu, F.~r.~Yin, X.~l.~Wang and L.~d.~Wan,
  %``The rare top quark decays t --> c V in the topcolor-assisted  technicolor
  %model,''
  Phys.\ Rev.\  D {\bf 68} (2003) 015002
  [arXiv:hep-ph/0303122].
  %%CITATION = PHRVA,D68,015002;%%




%%%  more exotic models, Q=2/3 singlets  %%%%%%%%%%%%

%
\bibitem{AguilarSaavedra:2002kr}
  J.~A.~Aguilar-Saavedra,
  %``Effects of mixing with quark singlets,''
  Phys.\ Rev.\  D {\bf 67} (2003) 035003
  [Erratum-ibid.\  D {\bf 69} (2004) 099901]
  [arXiv:hep-ph/0210112].
  %%CITATION = PHRVA,D67,035003;%%

%\cite{BarShalom:1998uq}
\bibitem{BarShalom:1998uq}
  S.~Bar-Shalom, G.~Eilam and A.~Soni,
  %``The flavor changing top decay t --> c sneutrino or  sneutrino --> t anti-c
  %in the MSSM without R-parity,''
  Phys.\ Rev.\  D {\bf 60} (1999) 035007
  [arXiv:hep-ph/9812518].
  %%CITATION = PHRVA,D60,035007;%%



%\cite{Abe:1997fz}
\bibitem{Abe:1997fz}
  F.~Abe {\it et al.}  [CDF Collaboration],
  %``Search for flavor-changing neutral current decays of the top quark in $p
  %\bar{p}$ collisions at $\sqrt{s} = 1.8$ TeV,''
  Phys.\ Rev.\ Lett.\  {\bf 80} (1998) 2525.
  %%CITATION = PRLTA,80,2525;%%

%\cite{Achard:2002vv}
\bibitem{Achard:2002vv}
  P.~Achard {\it et al.}  [L3 Collaboration],
  %``Search for single top production at LEP,''
  Phys.\ Lett.\  B {\bf 549} (2002) 290
  [arXiv:hep-ex/0210041].
  %%CITATION = PHLTA,B549,290;%%

%\cite{Chekanov:2003yt}
\bibitem{Chekanov:2003yt}
  S.~Chekanov {\it et al.}  [ZEUS Collaboration],
  %``Search for single-top production in e p collisions at HERA,''
  Phys.\ Lett.\  B {\bf 559} (2003) 153
  [arXiv:hep-ex/0302010].
  %%CITATION = PHLTA,B559,153;%%


%\cite{:2008be}
\bibitem{:2008be}
  V.~M.~Abazov {\it et al.}  [D0 Collaboration],
  %``Search for production of single top quarks via $tcg$ and $tug$
  %flavor-changing neutral current couplings,''
  Phys.\ Rev.\ Lett.\  {\bf 99} (2007) 191802
  [arXiv:0801.2556 [hep-ex]].
  %%CITATION = PRLTA,99,191802;%%

\bibitem{cdf-fcnc-note}
CDF Collaboration,
``Search for the Flavor Changing Neutral Current Decay $t\to Zq$
in $p {\bar p}$ collisions at $ \sqrt{s} = 1.96$ TeV
with 1.9 ${\rm fb}^{-1}$ of CDF-II data,'' CDF Note 9202 (2008).

%%%%%%%%%%%%%%%%%%%%% top mass %%%%%%%%%%%%%%%


%%%%% ATLAS ttbar corsssection %%%%%%%%%%%

%\cite{Borjanovic:2004ce}
\bibitem{Borjanovic:2004ce}
  I.~Borjanovic {\it et al.},
  %``Investigation of top mass measurements with the ATLAS detector at LHC,''
  Eur.\ Phys.\ J.\  C {\bf 39S2} (2005) 63
  [arXiv:hep-ex/0403021].
  %%CITATION = EPHJA,C39S2,63;%%

%\cite{D'hondt:2007aj}
\bibitem{D'hondt:2007aj}
  J.~D'hondt,
  ``Top Quark Physics at the LHC,''
  arXiv:0707.1247 [hep-ph].
  %%CITATION = ARXIV:0707.1247;%%



%%%%%%%%%%%%%%%%%%%%%%%%%%%%%%%%%%%%%%%%%%%%%%%%%%%%%%%%%%%%%%%%%
%%%%   ttbar cross section at NLO QCD + resummation %%%%%%%%%%%%%%%%%%%%
%%%%%%%%%%%%%%%%%%%%%%%%%%%%%%%%%%%%%%%%%%%%%%%%%%%%%%%%%%%%%%

%\cite{Nason:1987xz}
\bibitem{Nason:1987xz}
  P.~Nason, S.~Dawson and R.~K.~Ellis,
  %``The Total Cross-Section for the Production of Heavy Quarks in Hadronic
  %Collisions,''
  Nucl.\ Phys.\  B {\bf 303} (1988) 607.
  %%CITATION = NUPHA,B303,607;%%

%\cite{Beenakker:1988bq}
\bibitem{Beenakker:1988bq}
  W.~Beenakker, H.~Kuijf, W.~L.~van Neerven and J.~Smith,
  %``QCD Corrections to Heavy Quark Production in p anti-p Collisions,''
  Phys.\ Rev.\  D {\bf 40} (1989) 54.
  %%CITATION = PHRVA,D40,54;%%

%\cite{Nason:1989zy}
\bibitem{Nason:1989zy}
  P.~Nason, S.~Dawson and R.~K.~Ellis,
  %``The One Particle Inclusive Differential Cross-Section for Heavy Quark
  %Production in Hadronic Collisions,''
  Nucl.\ Phys.\  B {\bf 327} (1989) 49
  [Erratum-ibid.\  B {\bf 335} (1990) 260].
  %%CITATION = NUPHA,B327,49;%%


%\cite{Beenakker:1990maa}
\bibitem{Beenakker:1990maa}
  W.~Beenakker, W.~L.~van Neerven, R.~Meng, G.~A.~Schuler and J.~Smith,
  %``QCD corrections to heavy quark production in hadron hadron collisions,''
  Nucl.\ Phys.\  B {\bf 351} (1991) 507.
  %%CITATION = NUPHA,B351,507;%%

%\cite{Mangano:1991jk}
\bibitem{Mangano:1991jk}
  M.~L.~Mangano, P.~Nason and G.~Ridolfi,
  %``Heavy quark correlations in hadron collisions at next-to-leading order,''
  Nucl.\ Phys.\  B {\bf 373} (1992) 295.
  %%CITATION = NUPHA,B373,295;%%

%\cite{Frixione:1995fj}
\bibitem{Frixione:1995fj}
  S.~Frixione, M.~L.~Mangano, P.~Nason and G.~Ridolfi,
  %``Top quark distributions in hadronic collisions,''
  Phys.\ Lett.\  B {\bf 351} (1995) 555
  [arXiv:hep-ph/9503213].
  %%CITATION = PHLTA,B351,555;%%

%%%%%%%%%%% resumm. %%%%%%%%%%%

%\cite{Sterman:1986aj}
\bibitem{Sterman:1986aj}
  G.~Sterman,
  %``Summation of Large Corrections to Short 
 % Distance Hadronic Cross-Sections,''
  Nucl.\ Phys.\  B {\bf 281} (1987) 310.
  %%CITATION = NUPHA,B281,310;%%

%\cite{Catani:1989ne}
\bibitem{Catani:1989ne}
  S.~Catani and L.~Trentadue,
  %``Resummation Of The QCD Perturbative Series For Hard Processes,''
  Nucl.\ Phys.\  B {\bf 327} (1989) 323.
  %%CITATION = NUPHA,B327,323;%%

%\cite{Kidonakis:1997gm}
\bibitem{Kidonakis:1997gm}
  N.~Kidonakis and G.~Sterman,
  %``Resummation for QCD hard scattering,''
  Nucl.\ Phys.\  B {\bf 505} (1997) 321
  [arXiv:hep-ph/9705234].
  %%CITATION = NUPHA,B505,321;%%

%\cite{Kidonakis:1998nf}
%\bibitem{Kidonakis:1998nf}
%  N.~Kidonakis, G.~Oderda and G.~Sterman,
  %``Evolution of color exchange in {QCD} hard scattering,''
%  Nucl.\ Phys.\  B {\bf 531} (1998) 365
%  [arXiv:hep-ph/9803241].
  %%CITATION = NUPHA,B531,365;%%

%\cite{Bonciani:1998vc}
\bibitem{Bonciani:1998vc}
  R.~Bonciani, S.~Catani, M.~L.~Mangano and P.~Nason,
  %``NLL resummation of the heavy-quark hadroproduction cross-section,''
  Nucl.\ Phys.\  B {\bf 529} (1998) 424
  [arXiv:hep-ph/9801375].
  %%CITATION = NUPHA,B529,424;%%

%\cite{Kidonakis:2001nj}
\bibitem{Kidonakis:2001nj}
  N.~Kidonakis, E.~Laenen, S.~Moch and R.~Vogt,
  %``Sudakov resummation and finite order expansions of heavy quark
  %hadroproduction cross sections,''
  Phys.\ Rev.\  D {\bf 64} (2001) 114001
  [arXiv:hep-ph/0105041].
  %%CITATION = PHRVA,D64,114001;%%

%\cite{Kidonakis:2003qe}
\bibitem{Kidonakis:2003qe}
  N.~Kidonakis and R.~Vogt,
  %``Next-to-next-to-leading order soft-gluon corrections in top quark
  %hadroproduction,''
  Phys.\ Rev.\  D {\bf 68} (2003) 114014
  [arXiv:hep-ph/0308222].
  %%CITATION = PHRVA,D68,114014;%%

%\cite{Banfi:2004xa}
\bibitem{Banfi:2004xa}
  A.~Banfi and E.~Laenen,
  %``Joint resummation for heavy quark production,''
  Phys.\ Rev.\  D {\bf 71} (2005) 034003
  [arXiv:hep-ph/0411241].
  %%CITATION = PHRVA,D71,034003;%%

%%%%%%%%%%%%%%%%%%%%%%%%%%%%%% weak corrections to ttbar %%%%%%%%%%%%%%%%
%%%%%%%%%%%%%%%%%%%%%%%%%%%%%% %%%%%%%%%%%%%%%%%%%%%%%%%%%%%%%%%%%%%%%%%%%%%

%\cite{Beenakker:1993yr}
\bibitem{Beenakker:1993yr}
  W.~Beenakker, A.~Denner, W.~Hollik, R.~Mertig, T.~Sack and D.~Wackeroth,
  %``Electroweak one loop contributions to top pair production in hadron
  %colliders,''
  Nucl.\ Phys.\  B {\bf 411} (1994) 343.
  %%CITATION = NUPHA,B411,343;%%

%\cite{Bernreuther:2005is}
\bibitem{Bernreuther:2005is}
  W.~Bernreuther, M.~Fuecker and Z.~G.~Si,
  %``Mixed QCD and weak corrections to top quark pair production at hadron
  %colliders,''
  Phys.\ Lett.\  B {\bf 633} (2006) 54
  [arXiv:hep-ph/0508091].
  %%CITATION = PHLTA,B633,54;%%


%\cite{Kuhn:2005it}
\bibitem{Kuhn:2005it}
  J.~H.~K\"uhn, A.~Scharf and P.~Uwer,
  %``Electroweak corrections to top-quark pair production in quark-antiquark
  %annihilation,''
  Eur.\ Phys.\ J.\  C {\bf 45} (2006) 139
  [arXiv:hep-ph/0508092].
  %%CITATION = EPHJA,C45,139;%%



%\cite{Bernreuther:2006vg}
\bibitem{Bernreuther:2006vg}
  W.~Bernreuther, M.~Fuecker and Z.~G.~Si,
  %``Weak interaction corrections to hadronic top quark pair production,''
  Phys.\ Rev.\  D {\bf 74} (2006) 113005
  [arXiv:hep-ph/0610334].
 %%CITATION = PHRVA,D74,113005;%%

%\cite{Bernreuther:2008md}
\bibitem{Bernreuther:2008md}
  W.~Bernreuther, M.~Fuecker and Z.~G.~Si,
  ``Weak interaction corrections to hadronic top quark pair production:
  contributions from quark-gluon and $b \bar b$ induced reactions,''
  arXiv:0804.1237 [hep-ph].
  %%CITATION = ARXIV:0804.1237;%%


%\cite{Kuhn:2006vh}
\bibitem{Kuhn:2006vh}
  J.~H.~K\"uhn, A.~Scharf and P.~Uwer,
  %``Electroweak effects in top-quark pair production at hadron colliders,''
  Eur.\ Phys.\ J.\  C {\bf 51} (2007) 37
  [arXiv:hep-ph/0610335].
  %%CITATION = EPHJA,C51,37;%%


%\cite{Moretti:2006nf}
\bibitem{Moretti:2006nf}
  S.~Moretti, M.~R.~Nolten and D.~A.~Ross,
  %``Weak corrections to gluon-induced top-antitop hadro-production,''
  Phys.\ Lett.\  B {\bf 639} (2006) 513
  [Erratum-ibid.\  B {\bf 660} (2008) 607]
  [arXiv:hep-ph/0603083].
  %%CITATION = PHLTA,B639,513;%%


%\cite{Hollik:2007sw}
\bibitem{Hollik:2007sw}
  W.~Hollik and M.~Kollar,
  %``NLO QED contributions to top-pair production at hadron collider,''
  Phys.\ Rev.\  D {\bf 77} (2008) 014008
  [arXiv:0708.1697 [hep-ph]].
  %%CITATION = PHRVA,D77,014008;%%



%%%%%%%%%%%%%%%% ttbar jet%%%%%%%%%%%%%%%%%%%%%%%%%%%%%%%%%%%%
%%%%%%%%%%%%%%%% %%%%%%%%%%%%%%%%%%%%%%%%%%%%%%%%%%%%%%%%%%%%%
%%%%%%%%%%%%%%%%%%%%%%%%%%%%%%%%%%%%%%%%%%%%%%%%%%%%%%%%%%%%

 %%\cite{Dittmaier:2007wz}
\bibitem{Dittmaier:2007wz}
  S.~Dittmaier, P.~Uwer and S.~Weinzierl,
  %``NLO QCD corrections to t anti-t + jet production at hadron colliders,''
  Phys.\ Rev.\ Lett.\  {\bf 98} (2007) 262002
  [arXiv:hep-ph/0703120].
  %%CITATION = PRLTA,98,262002;%%

%\cite{Bernreuther:2001rq}
\bibitem{Bernreuther:2001rq}
  W.~Bernreuther, A.~Brandenburg, Z.~G.~Si and P.~Uwer,
  %``Top quark spin correlations at hadron colliders: Predictions at
  %next-to-leading order QCD,''
  Phys.\ Rev.\ Lett.\  {\bf 87} (2001) 242002
  [arXiv:hep-ph/0107086].
  %%CITATION = PRLTA,87,242002;%%

%\cite{Bernreuther:2004jv}
\bibitem{Bernreuther:2004jv}
  W.~Bernreuther, A.~Brandenburg, Z.~G.~Si and P.~Uwer,
  %``Top quark pair production and decay at hadron colliders,''
  Nucl.\ Phys.\  B {\bf 690} (2004) 81
  [arXiv:hep-ph/0403035].
  %%CITATION = NUPHA,B690,81;%%


%%%%%%%%%%%%%%%%%%%%%%%%%%%%%%%%%%%%%%%%%%%%%%%%%%%%%%%%%%
%%%%%%%%%%%%%%%%%%%%%%%%%%%% ttbar cross section at NNLO %%%%%%%%%%%
%%%%%%%%%%%%%%%%%%%%%%%%%%%%%%%%%%%%%%%%%%%%%%%%%%%%%%%%%%%%%%%%

%%%%%%%%% nonfactorizable QCD corrections %%%%%%%%%%%%%%%%

%\cite{Beenakker:1999ya}
\bibitem{Beenakker:1999ya}
  W.~Beenakker, F.~A.~Berends and A.~P.~Chapovsky,
  %``One-loop {QCD} interconnection effects in pair production of top quarks,''
  Phys.\ Lett.\  B {\bf 454} (1999) 129
  [arXiv:hep-ph/9902304].
  %%CITATION = PHLTA,B454,129;%%

\bibitem{Meyer}
L. Meyer, diploma thesis, RWTH Aachen (2005).

%\cite{Czakon:2007ej}
\bibitem{Czakon:2007ej}
  M.~Czakon, A.~Mitov and S.~Moch,
  %``Heavy-quark production in massless quark scattering at two loops in QCD,''
  Phys.\ Lett.\  B {\bf 651} (2007) 147
  [arXiv:0705.1975 [hep-ph]].
  %%CITATION = PHLTA,B651,147;%%

%\cite{Czakon:2007wk}
\bibitem{Czakon:2007wk}
  M.~Czakon, A.~Mitov and S.~Moch,
  ``Heavy-quark production in gluon fusion at two loops in QCD,''
  arXiv:0707.4139 [hep-ph].
  %%CITATION = ARXIV:0707.4139;%%

%\cite{Czakon:2008zu}
\bibitem{Czakon:2008zu}
  M.~Czakon,
   ``Mass effects and four-particle amplitudes at the two-loop level in QCD,''
  arXiv:0803.1414 [hep-ph].
  %%CITATION = ARXIV:0803.1414;%%

%\cite{Korner:2008bn}
\bibitem{Korner:2008bn}
  J.~G.~K\"orner, Z.~Merebashvili and M.~Rogal,
  ``NNLO ${\cal O}(\alpha_s^4)$ results for heavy quark pair production in
  quark--antiquark collisions: The one-loop squared contributions,''
  arXiv:0802.0106 [hep-ph].
  %%CITATION = ARXIV:0802.0106;%%


%\cite{Fadin:1993dz}
\bibitem{Fadin:1993dz}
  V.~S.~Fadin, V.~A.~Khoze and A.~D.~Martin,
  %``Interference radiative phenomena in the production of heavy unstable
  %particles,''
  Phys.\ Rev.\  D {\bf 49} (1994) 2247.
  %%CITATION = PHRVA,D49,2247;%%


%\cite{Melnikov:1993np}
\bibitem{Melnikov:1993np}
  K.~Melnikov and O.~I.~Yakovlev,
  %``Top near threshold: All alpha-S corrections are trivial,''
  Phys.\ Lett.\  B {\bf 324} (1994) 217
  [arXiv:hep-ph/9302311].
  %%CITATION = PHLTA,B324,217;%%

%\cite{Kauer:2001sp}
\bibitem{Kauer:2001sp}
  N.~Kauer and D.~Zeppenfeld,
  %``Finite-width effects in top quark production at hadron colliders,''
  Phys.\ Rev.\  D {\bf 65} (2002) 014021
  [arXiv:hep-ph/0107181].
  %%CITATION = PHRVA,D65,014021;%%



%\cite{Kauer:2002sn}
\bibitem{Kauer:2002sn}
  N.~Kauer,
  %``Top pair production beyond double-pole approximation: p p, p anti-p --> 6
  %fermions and 0, 1 or 2 additional partons,''
  Phys.\ Rev.\  D {\bf 67} (2003) 054013
  [arXiv:hep-ph/0212091].
  %%CITATION = PHRVA,D67,054013;%%


%%%%%%%%%%%%%%%%%%%%%%%%%%%%%%%%%%%%%%%%%%%%%%%%%%%%%%%%%%%%%%%%
%%%%%%%%%%%%%%%%%%%%%%%%%%%%%% Monte Carlo %%%%%%%%%%%%%%%%%%%%55
%%%%%%%%%%%%%%%%%%%%%%%%%%%%%%%%%%%%%%%%%%%%%%%%%%%%%%%%%%%%%5

%%%% review %%
%\cite{Mangano:2005dj}
\bibitem{Mangano:2005dj}
  M.~L.~Mangano and T.~J.~Stelzer,
  %``Tools For The Simulation Of Hard Hadronic Collisions,''
  Ann.\ Rev.\ Nucl.\ Part.\ Sci.\  {\bf 55} (2005) 555.
  %%CITATION = ARNUA,55,555;%%


%\cite{Sjostrand:2003wg}
\bibitem{Sjostrand:2003wg}
  T.~Sjostrand, L.~Lonnblad, S.~Mrenna and P.~Skands,
  ``PYTHIA 6.3: Physics and manual,''
  arXiv:hep-ph/0308153.
  %%CITATION = HEP-PH/0308153;%%

%\cite{Sjostrand:2007gs}
\bibitem{Sjostrand:2007gs}
  T.~Sjostrand, S.~Mrenna and P.~Skands,
  ``A Brief Introduction to PYTHIA 8.1,''
  arXiv:0710.3820 [hep-ph].
  %%CITATION = ARXIV:0710.3820;%%

%\cite{Corcella:2002jc}
\bibitem{Corcella:2002jc}
  G.~Corcella {\it et al.},
  ``HERWIG 6.5 release note,''
  arXiv:hep-ph/0210213.
  %%CITATION = HEP-PH/0210213;%%

%\cite{Bahr:2008pv}
\bibitem{Bahr:2008pv}
  M.~Bahr {\it et al.},
  ``Herwig++ Physics and Manual,''
  arXiv:0803.0883 [hep-ph].
  %%CITATION = ARXIV:0803.0883;%%

\bibitem{cedar}
CEDAR: {\tt http:/www.cedar.ac.uk/hepcode/}

%\cite{Slabospitsky:2002ag}
\bibitem{Slabospitsky:2002ag}
  S.~R.~Slabospitsky and L.~Sonnenschein,
  %``TopReX generator (version 3.25): Short manual,''
  Comput.\ Phys.\ Commun.\  {\bf 148} (2002) 87
  [arXiv:hep-ph/0201292].
  %%CITATION = CPHCB,148,87;%%

%\cite{Mangano:2002ea}
\bibitem{Mangano:2002ea}
  M.~L.~Mangano, M.~Moretti, F.~Piccinini, R.~Pittau and A.~D.~Polosa,
  %``ALPGEN, a generator for hard multiparton processes in hadronic
  %collisions,''
  JHEP {\bf 0307} (2003) 001
  [arXiv:hep-ph/0206293].
  %%CITATION = JHEPA,0307,001;%%

%\cite{Maltoni:2002qb}
\bibitem{Maltoni:2002qb}
  F.~Maltoni and T.~Stelzer,
  %``MadEvent: Automatic event generation with MadGraph,''
  JHEP {\bf 0302} (2003) 027
  [arXiv:hep-ph/0208156].
  %%CITATION = JHEPA,0302,027;%%

%\cite{Tsuno:2006cu}
\bibitem{Tsuno:2006cu}
  S.~Tsuno, T.~Kaneko, Y.~Kurihara, S.~Odaka and K.~Kato,
  %``GR@PPA 2.7 event generator for p p / p anti-p collisions,''
  Comput.\ Phys.\ Commun.\  {\bf 175} (2006) 665
  [arXiv:hep-ph/0602213].
  %%CITATION = CPHCB,175,665;%%

%\cite{Frixione:2006gn}
\bibitem{Frixione:2006gn}
  S.~Frixione and B.~R.~Webber,
  ``The MC@NLO 3.3 event generator,''
  arXiv:hep-ph/0612272.
  %%CITATION = HEP-PH/0612272;%%

\bibitem{CambEll}
J.~Campbell, R.~K.~Ellis,
``MCFM - Monte Carlo for FeMtobarn processes'', 
{\tt http://mcfm.fnal.gov/}

%\cite{Ellis:2006ar}
\bibitem{Ellis:2006ar}
  R.~K.~Ellis,
  %``An update on the next-to-leading order 
  % Monte Carlo MCFM,''
  Nucl.\ Phys.\ Proc.\ Suppl.\  {\bf 160} (2006) 170.
  %%CITATION = NUPHZ,160,170;%%

%\cite{Frixione:2007nw}
\bibitem{Frixione:2007nw}
  S.~Frixione, P.~Nason and G.~Ridolfi,
  %``A Positive-Weight Next-to-Leading-Order Monte Carlo 
  % for Heavy Flavour
  %Hadroproduction,''
  JHEP {\bf 0709} (2007) 126
  [arXiv:0707.3088 [hep-ph]].
  %%CITATION = JHEPA,0709,126;%%

%\cite{Frixione:2007nu}
\bibitem{Frixione:2007nu}
  S.~Frixione, P.~Nason and G.~Ridolfi,
  %``The POWHEG-hvq manual version 1.0,''
  arXiv:0707.3081 [hep-ph].
  %%CITATION = ARXIV:0707.3081;%%

%\cite{Boos:2006af}
\bibitem{Boos:2006af}
  E.~E.~Boos, V.~E.~Bunichev, L.~V.~Dudko, V.~I.~Savrin and A.~V.~Sherstnev,
  %``Method for simulating electroweak top-quark production events in the NLO
  %approximation: SingleTop event generator,''
  Phys.\ Atom.\ Nucl.\  {\bf 69} (2006) 1317
  [Yad.\ Fiz.\  {\bf 69} (2006) 1352].
  %%CITATION = YAFIA,69,1352;%%





%%%%%%%%%%%%%%%%%%%%%%%%%%%%%%%%%%%%%%%%%%%%
%%%%%%%%%%%%%%%%%%%%%%%% pdf %%%%%%%%%%%%%%%%%%%%

%\cite{Pumplin:2002vw}
\bibitem{Pumplin:2002vw}
  J.~Pumplin, D.~R.~Stump, J.~Huston, H.~L.~Lai, P.~Nadolsky and W.~K.~Tung,
  %``New generation of parton distributions with uncertainties from global  QCD
  %analysis,''
  JHEP {\bf 0207} (2002) 012
  [arXiv:hep-ph/0201195].
  %%CITATION = JHEPA,0207,012;%%


%\cite{Martin:2002aw}
\bibitem{Martin:2002aw}
  A.~D.~Martin, R.~G.~Roberts, W.~J.~Stirling and R.~S.~Thorne,
  %``Uncertainties of predictions from parton distributions. I: Experimental
  %errors. ((T)),''
  Eur.\ Phys.\ J.\  C {\bf 28} (2003) 455
  [arXiv:hep-ph/0211080].
  %%CITATION = EPHJA,C28,455;%%

%\cite{Martin:2003sk}
\bibitem{Martin:2003sk}
  A.~D.~Martin, R.~G.~Roberts, W.~J.~Stirling and R.~S.~Thorne,
  %``Uncertainties of predictions from parton distributions. II: Theoretical
  %errors,''
  Eur.\ Phys.\ J.\  C {\bf 35} (2004) 325
  [arXiv:hep-ph/0308087].
  %%CITATION = EPHJA,C35,325;%%


\bibitem{MochUw08}
S.~Moch and P.~Uwer,
  ``Theoretical status and prospects for top-quark pair production at hadron
  colliders,''
  arXiv:0804.1476 [hep-ph].
  %%CITATION = ARXIV:0804.1476;%%

%\cite{Cacciari:2008zb}
\bibitem{Cacciari:2008zb}
  M.~Cacciari, S.~Frixione, M.~M.~Mangano, P.~Nason and G.~Ridolfi,
  ``Updated predictions for the total production cross sections of top and of
   heavier quark pairs at the Tevatron and at the LHC,''
  arXiv:0804.2800 [hep-ph].
  %%CITATION = ARXIV:0804.2800;%%

%%%%%%%%%% tevatron cross section, resummed %%%%%%%%%
%\cite{Cacciari:2003fi}
\bibitem{Cacciari:2003fi}
  M.~Cacciari, S.~Frixione, M.~L.~Mangano, P.~Nason and G.~Ridolfi,
  %``The t anti-t cross-section at 1.8-TeV and 1.96-TeV: A study of the
  %systematics due to parton densities and scale dependence,''
  JHEP {\bf 0404} (2004) 068
  [arXiv:hep-ph/0303085].
  %%CITATION = JHEPA,0404,068;%%

%\cite{Catani:1996dj}
\bibitem{Catani:1996dj}
  S.~Catani, M.~L.~Mangano, P.~Nason and L.~Trentadue,
  %``The Top Cross Section in Hadronic Collisions,''
  Phys.\ Lett.\  B {\bf 378} (1996) 329
  [arXiv:hep-ph/9602208].
  %%CITATION = PHLTA,B378,329;%%

%\cite{Tung:2006tb}
\bibitem{Tung:2006tb}
  W.~K.~Tung, H.~L.~Lai, A.~Belyaev, J.~Pumplin, D.~Stump and C.~P.~Yuan,
  %``Heavy quark mass effects in deep inelastic scattering and global QCD
  %analysis,''
  JHEP {\bf 0702} (2007) 053
  [arXiv:hep-ph/0611254].
  %%CITATION = JHEPA,0702,053;%%

%\cite{Martin:2007bv}
\bibitem{Martin:2007bv}
  A.~D.~Martin, W.~J.~Stirling, R.~S.~Thorne and G.~Watt,
  %``Update of Parton Distributions at NNLO,''
  Phys.\ Lett.\  B {\bf 652} (2007) 292
  [arXiv:0706.0459 [hep-ph]].
  %%CITATION = PHLTA,B652,292;%%



%%%%%% exp. ttbar cross section at Tevatron %%%

%\cite{Abulencia:2006in}
\bibitem{Abulencia:2006in}
  A.~Abulencia {\it et al.}  [CDF Collaboration],
  %``Measurement of the tanti-t Production Cross Section in $p$ anti-ptnipbar
  %Collisions at $\sqrt{s}$ = 1.96-TeV,''
  Phys.\ Rev.\ Lett.\  {\bf 97} (2006) 082004
  [arXiv:hep-ex/0606017].
  %%CITATION = PRLTA,97,082004;%%

%\cite{:2007qf}
\bibitem{:2007qf}
  T.~Aaltonen {\it et al.}  [CDF Collaboration],
  %``Measurement of the $p \bar{p} \to t \bar{t}$ production cross- section and
  %the top quark mass at $\sqrt{s}$ = 1.96-TeV in 
% the all-hadronic decay mode,''
  Phys.\ Rev.\  D {\bf 76} (2007) 072009
  [arXiv:0706.3790 [hep-ex]].
  %%CITATION = PHRVA,D76,072009;%%




%\cite{Nadolsky:2008zw}
\bibitem{Nadolsky:2008zw}
  P.~M.~Nadolsky {\it et al.},
  ``Implications of CTEQ global analysis for collider observables,''
  arXiv:0802.0007 [hep-ph].
  %%CITATION = ARXIV:0802.0007;%%



%%%%%%%%%%%% top mass, exp. Tevatron %%%%%%%%%%

%\cite{:2007jw}
\bibitem{:2007jw}
  T.~Aaltonen {\it et al.}  [CDF Collaboration],
  %``Cross Section Constrained Top Quark Mass Measurement from Dilepton Events
  %at the Tevatron,''
  Phys.\ Rev.\ Lett.\  {\bf 100} (2008) 062005
  [arXiv:0710.4037 [hep-ex]].
  %%CITATION = PRLTA,100,062005;%%
%%%%%%%%%%%%%%%%%%% top in e+e- %%%%%%%%%%%%%%%%%%%%%%%%%%%%%%%%%%%



%\cite{AguilarSaavedra:2001rg}
\bibitem{AguilarSaavedra:2001rg}
  J.~A.~Aguilar-Saavedra {\it et al.}  [ECFA/DESY LC Physics Working Group],
  ``TESLA Technical Design Report Part III: Physics at an e+e- Linear
  Collider,''
  arXiv:hep-ph/0106315.
  %%CITATION = HEP-PH/0106315;%%

%\cite{Hoang:2000yr}
\bibitem{Hoang:2000yr}
  A.~H.~Hoang {\it et al.},
  %``Top-antitop pair production close to threshold: Synopsis of recent NNLO
  %results,''
  Eur.\ Phys.\ J.\ direct C {\bf 2} (2000) 1
  [arXiv:hep-ph/0001286].
  %%CITATION = EPHJD,C2,1;%%



%\cite{Fleming:2007qr}
\bibitem{Fleming:2007qr}
  S.~Fleming, A.~H.~Hoang, S.~Mantry and I.~W.~Stewart,
  ``Jets from Massive Unstable Particles: Top-Mass Determination,''
  arXiv:hep-ph/0703207.
  %%CITATION = HEP-PH/0703207;%%

%%%%%%%%%%%%%%%%%%%%%%%%%%%%%%%%%%%%%

%\cite{Aaltonen:2007xx}
\bibitem{Aaltonen:2007xx}
  T.~Aaltonen {\it et al.}  [CDF Collaboration],
  %``Measurement of the top-quark mass using missing $E_T$ + jets events with
  %secondary vertex $b-$tagging at CDF II,''
  Phys.\ Rev.\  D {\bf 75} (2007) 111103
  [arXiv:0705.1594 [hep-ex]].
  %%CITATION = PHRVA,D75,111103;%%



%\cite{Abazov:2007rk}
\bibitem{Abazov:2007rk}
  V.~M.~Abazov {\it et al.}  [D0 Collaboration],
  %``Measurement of the top quark mass in the lepton + jets channel using the
  %ideogram method,''
  Phys.\ Rev.\  D {\bf 75}, 092001 (2007)
  [arXiv:hep-ex/0702018].
  %%CITATION = PHRVA,D75,092001;%%


%\cite{Skands:2007zg}
\bibitem{Skands:2007zg}
  P.~Skands and D.~Wicke,
  %``Non-perturbative QCD effects and the top mass at the Tevatron,''
  Eur.\ Phys.\ J.\  C {\bf 52} (2007) 133
  [arXiv:hep-ph/0703081].
  %%CITATION = EPHJA,C52,133;%%

%%%%%%%%%%%%%%%%%%%% method for top mass %%%%%%%%%%%%

%\cite{Kharchilava:1999yj}
\bibitem{Kharchilava:1999yj}
  A.~Kharchilava,
  %``Top mass determination in leptonic final states with J/psi,''
  Phys.\ Lett.\  B {\bf 476} (2000) 73
  [arXiv:hep-ph/9912320].
  %%CITATION = PHLTA,B476,73;%%

%\cite{Hill:2005zy}
\bibitem{Hill:2005zy}
  C.~S.~Hill, J.~R.~Incandela and J.~M.~Lamb,
  %``A method for measurement of the top quark 
% mass using the mean decay  length
  %of b hadrons in t anti-t events,''
  Phys.\ Rev.\  D {\bf 71} (2005) 054029
  [arXiv:hep-ex/0501043].
  %%CITATION = PHRVA,D71,054029;%%

%\cite{Frederix:2007gi}
\bibitem{Frederix:2007gi}
  R.~Frederix and F.~Maltoni,
  ``Top pair invariant mass distribution: a window on new physics,''
  arXiv:0712.2355 [hep-ph].
  %%CITATION = ARXIV:0712.2355;%%


%\cite{Hagiwara:2008df}
\bibitem{Hagiwara:2008df}
  K.~Hagiwara, Y.~Sumino and H.~Yokoya,
  ``Bound-state Effects on Top Quark Production at Hadron Colliders,''
  arXiv:0804.1014 [hep-ph].
  %%CITATION = ARXIV:0804.1014;%%


%\cite{Baur:2007ck}
\bibitem{Baur:2007ck}
  U.~Baur and L.~H.~Orr,
  %``High p_T Top Quarks at the Large Hadron Collider,''
  Phys.\ Rev.\  D {\bf 76} (2007) 094012
  [arXiv:0707.2066 [hep-ph]].
  %%CITATION = PHRVA,D76,094012;%%



%%%%%%%%%%%%%%%%% charge asymmetry %%%%%%%%%%

%\cite{Halzen:1987xd}
\bibitem{Halzen:1987xd}
  F.~Halzen, P.~Hoyer and C.~S.~Kim,
  %``FORWARD - BACKWARD ASYMMETRY OF HADROPRODUCED HEAVY QUARKS IN QCD,''
  Phys.\ Lett.\  B {\bf 195} (1987) 74.
  %%CITATION = PHLTA,B195,74;%%

%\cite{Kuhn:1998kw}
\bibitem{Kuhn:1998kw}
  J.~H.~K\"uhn and G.~Rodrigo,
  %``Charge asymmetry of heavy quarks at hadron colliders,''
  Phys.\ Rev.\  D {\bf 59} (1999) 054017
  [arXiv:hep-ph/9807420].
  %%CITATION = PHRVA,D59,054017;%%

%\cite{Bowen:2005ap}
\bibitem{Bowen:2005ap}
  M.~T.~Bowen, S.~D.~Ellis and D.~Rainwater,
  %``Standard model top quark asymmetry at the Fermilab Tevatron,''
  Phys.\ Rev.\  D {\bf 73} (2006) 014008
  [arXiv:hep-ph/0509267].
  %%CITATION = PHRVA,D73,014008;%%

%\cite{Antunano:2007da}
\bibitem{Antunano:2007da}
  O.~Antunano, J.~H.~K\"uhn and G.~V.~Rodrigo,
  %``Top Quarks, Axigluons and Charge Asymmetries at Hadron Colliders,''
  Phys.\ Rev.\  D {\bf 77} (2008) 014003
  [arXiv:0709.1652 [hep-ph]].
  %%CITATION = PHRVA,D77,014003;%%

%\cite{Atwood:2000tu}
\bibitem{Atwood:2000tu}
  D.~Atwood, S.~Bar-Shalom, G.~Eilam and A.~Soni,
  %``CP violation in top physics,''
  Phys.\ Rept.\  {\bf 347} (2001) 1
  [arXiv:hep-ph/0006032].
  %%CITATION = PRPLC,347,1;%%

%\cite{:2007qb}
\bibitem{:2007qb}
   {\it et al.}  [D0 Collaboration],
  ``First measurement of the forward-backward charge asymmetry in top quark
   pair production,''
  arXiv:0712.0851 [hep-ex].
  %%CITATION = ARXIV:0712.0851;%%

\bibitem{cdf-afb-note}
CDF Collaboration,
``Measurement of the Charge Asymmetry in Top Pair Production
using 1.9 ${\rm fb}^{-1}$,'' CDF Note 9156 (2007);
``Measurement of the Forward Backward Asymmetry in Top Pair
Production,'' CDF Note 9169 (2007).


%\cite{Sehgal:1987wi}
\bibitem{Sehgal:1987wi}
  L.~M.~Sehgal and M.~Wanninger,
  %``FORWARD - BACKWARD ASYMMETRY IN TWO JET EVENTS: SIGNATURE OF AXIGLUONS IN
  %PROTON - ANTI-PROTON COLLISIONS,''
  Phys.\ Lett.\  B {\bf 200} (1988) 211.
  %%CITATION = PHLTA,B200,211;%%

%%%%%%%%%%% transverse pol. and spin correlations %%%%%%%%%%%%%%
%%%%%%%%%%%%%%%%%%%%%%%%%%%%%%%%%%%%%%%%%%%%%%%%%%%%%%%%%%%

%\cite{Bernreuther:1995cx}
\bibitem{Bernreuther:1995cx}
  W.~Bernreuther, A.~Brandenburg and P.~Uwer,
  %``Transverse Polarization of Top Quark Pairs at the Tevatron and the Large
  %Hadron Collider,''
  Phys.\ Lett.\  B {\bf 368} (1996) 153
  [arXiv:hep-ph/9510300].
  %%CITATION = PHLTA,B368,153;%%

\bibitem{Dharmaratna:1996xd}
  W.~G.~D.~Dharmaratna and G.~R.~Goldstein,
  %``Single quark polarization in quantum chromodynamics subprocesses,''
  Phys.\ Rev.\ D {\bf 53} (1996) 1073.
  %%CITATION = PHRVA,D53,1073;%%



%%%%%%%%%%%%%%%%  spin correl. Born %%%%%%%%%%%%%%%%

\bibitem{Barger:1988jj}
  V.~D.~Barger, J.~Ohnemus and R.~J.~N.~Phillips,
  %``SPIN CORRELATION EFFECTS IN THE HADROPRODUCTION AND 
  % DECAY OF VERY HEAVY TOP
  %QUARK PAIRS,''
  Int.\ J.\ Mod.\ Phys.\  A {\bf 4} (1989) 617.
  %%CITATION = IMPAE,A4,617;%%

%\cite{Arens:1992fg}
\bibitem{Arens:1992fg}
  T.~Arens and L.~M.~Sehgal,
  %``Azimuthal correlation of charged leptons produced in p anti-p $\to$ t
  %anti-t + ..,''
  Phys.\ Lett.\  B {\bf 302} (1993) 501.
  %%CITATION = PHLTA,B302,501;%%

%\cite{Stelzer:1995gc}
\bibitem{Stelzer:1995gc}
  T.~Stelzer and S.~Willenbrock,
  %``Spin Correlation in Top-Quark Production at Hadron Colliders,''
  Phys.\ Lett.\  B {\bf 374} (1996) 169
  [arXiv:hep-ph/9512292].
  %%CITATION = PHLTA,B374,169;%%

%\cite{Brandenburg:1996df}
\bibitem{Brandenburg:1996df}
  A.~Brandenburg,
  %``Spin-Spin Correlations of Top Quark Pairs at Hadron Colliders,''
  Phys.\ Lett.\  B {\bf 388} (1996) 626
  [arXiv:hep-ph/9603333].
  %%CITATION = PHLTA,B388,626;%%

%\cite{Chang:1995ay}
\bibitem{Chang:1995ay}
  D.~Chang, S.~C.~Lee and A.~Sumarokov,
  %``On the Observation of Top Spin Correlation Effect at Tevatron,''
  Phys.\ Rev.\ Lett.\  {\bf 77} (1996) 1218
  [arXiv:hep-ph/9512417].
  %%CITATION = PRLTA,77,1218;%%

%\cite{Mahlon:1995zn}
\bibitem{Mahlon:1995zn}
  G.~Mahlon and S.~J.~Parke,
  %``Angular Correlations in Top Quark Pair Production and Decay at Hadron
  %Colliders,''
  Phys.\ Rev.\  D {\bf 53} (1996) 4886
  [arXiv:hep-ph/9512264].
  %%CITATION = PHRVA,D53,4886;%%

%\cite{Mahlon:1997uc}
\bibitem{Mahlon:1997uc}
  G.~Mahlon and S.~J.~Parke,
  %``Maximizing spin correlations in top quark pair production at the
  %Tevatron,''
  Phys.\ Lett.\  B {\bf 411} (1997) 173
  [arXiv:hep-ph/9706304].
  %%CITATION = PHLTA,B411,173;%%

%\cite{Uwer:2004vp}
\bibitem{Uwer:2004vp}
  P.~Uwer,
  %``Maximizing the spin correlation of top quark pairs produced at the LHC,''
  Phys.\ Lett.\  B {\bf 609} (2005) 271
  [arXiv:hep-ph/0412097].
  %%CITATION = PHLTA,B609,271;%%

%%%% spin correl. NLO QCD %%%%%%%%%%%




%%%%%%%%%% more complicated spin correl, Born %%%%%%%%

%\cite{Nelson:2005jp}
\bibitem{Nelson:2005jp}
  C.~A.~Nelson, E.~G.~Barbagiovanni, J.~J.~Berger, 
  E.~K.~Pueschel and J.~R.~Wickman,
  %``Use of W-boson longitudinal-transverse interference in top quark
  %spin-correlation functions,''
  Eur.\ Phys.\ J.\  C {\bf 45} (2006) 121
  [arXiv:hep-ph/0506240].
  %%CITATION = EPHJA,C45,121;%%


%\cite{Nelson:2005wh}
\bibitem{Nelson:2005wh}
  C.~A.~Nelson, J.~J.~Berger and J.~R.~Wickman,
  %``Use of W-boson longitudinal-transverse interference in top quark
  %spin-correlation functions: II,''
  Eur.\ Phys.\ J.\  C {\bf 46} (2006) 385
  [arXiv:hep-ph/0510348].
  %%CITATION = EPHJA,C46,385;%%

%%%%%%% das folgende Papier wurde schon oben zitiert %%%%%
%\cite{Nelson:1997xd}

%%%%%%%%% exp. det. of helicity, anomalous couplings
%%%        and spin correl. at LHC %%%%%%%%%%



%%%%%%%%%%%%%%%%%%%%%%%% top charge, anomalous couplings %%%%%%%%%%%%%%%%%


%\cite{Baur:2004uw}
\bibitem{Baur:2004uw}
  U.~Baur, A.~Juste, L.~H.~Orr and D.~Rainwater,
  %``Probing electroweak top quark couplings at hadron colliders,''
  Phys.\ Rev.\  D {\bf 71} (2005) 054013
  [arXiv:hep-ph/0412021].
  %%CITATION = PHRVA,D71,054013;%%

%\cite{Bernreuther:2005gq}
\bibitem{Bernreuther:2005gq}
  W.~Bernreuther, R.~Bonciani, T.~Gehrmann, R.~Heinesch,
  T.~Leineweber, P.~Mastrolia and E.~Remiddi,
  %``QCD corrections to static heavy quark form factors,''
  Phys.\ Rev.\ Lett.\  {\bf 95} (2005) 261802
  [arXiv:hep-ph/0509341].
  %%CITATION = PRLTA,95,261802;%%

%\cite{Hollik:1988ii}
\bibitem{Hollik:1988ii}
  W.~F.~L.~Hollik,
  %``Radiative Corrections in the Standard Model and their Role for Precision
  %Tests of the Electroweak Theory,''
  Fortsch.\ Phys.\  {\bf 38} (1990) 165.
  %%CITATION = FPYKA,38,165;%%

%\cite{Berger:2005ht}
\bibitem{Berger:2005ht}
  C.~F.~Berger, M.~Perelstein and F.~Petriello,
  ``Top quark properties in little Higgs models,''
  arXiv:hep-ph/0512053.
  %%CITATION = ECONF,C0508141,ALCPG0428;%%

%\cite{Bernreuther:1992dz}
\bibitem{Bernreuther:1992dz}
  W.~Bernreuther, T.~Schr\"oder and T.~N.~Pham,
  %``CP Violating Dipole Form-Factors In E+ E- $\to$ Anti-T T,''
  Phys.\ Lett.\  B {\bf 279} (1992) 389.
  %%CITATION = PHLTA,B279,389;%%

%\cite{Baur:2001si}
\bibitem{Baur:2001si}
  U.~Baur, M.~Buice and L.~H.~Orr,
  %``Direct measurement of the top quark charge at hadron colliders,''
  Phys.\ Rev.\  D {\bf 64} (2001) 094019
  [arXiv:hep-ph/0106341].
  %%CITATION = PHRVA,D64,094019;%%



%\cite{Baur:2005wi}
\bibitem{Baur:2005wi}
  U.~Baur, A.~Juste, D.~Rainwater and L.~H.~Orr,
  %``Improved measurement of t t Z couplings at the LHC,''
  Phys.\ Rev.\  D {\bf 73} (2006) 034016
  [arXiv:hep-ph/0512262].
  %%CITATION = PHRVA,D73,034016;%%


%%%%%%%%%%%%%%%%%%%%%%%% ttbar Higgs %%%%%%%%%%%%%%%%%%%%%%%%%%%%
%%%%%%%%%%%%%%%%%%%%%%%%%%%%%%%%%%%%%%%%%%%%%%%%%%%%%%%%%%%%%%%%%

%\cite{Beenakker:2001rj}
\bibitem{Beenakker:2001rj}
  W.~Beenakker, S.~Dittmaier, M.~Kr\"amer, B.~Pl\"umper, 
  M.~Spira and P.~M.~Zerwas,
  %``Higgs radiation off top quarks at the Tevatron and the LHC,''
  Phys.\ Rev.\ Lett.\  {\bf 87} (2001) 201805
  [arXiv:hep-ph/0107081].
  %%CITATION = PRLTA,87,201805;%%

%\cite{Beenakker:2002nc}
\bibitem{Beenakker:2002nc}
  W.~Beenakker, S.~Dittmaier, M.~Kr\"amer, B.~Pl\"umper, 
  M.~Spira and P.~M.~Zerwas,
  %``NLO QCD corrections to t anti-t H production in hadron 
  % collisions. ((U)),''
  Nucl.\ Phys.\  B {\bf 653} (2003) 151
  [arXiv:hep-ph/0211352].
  %%CITATION = NUPHA,B653,151;%%

%\cite{Dawson:2002tg}
\bibitem{Dawson:2002tg}
  S.~Dawson, L.~H.~Orr, L.~Reina and D.~Wackeroth,
  %``Associated top quark Higgs boson production at the LHC,''
  Phys.\ Rev.\  D {\bf 67} (2003) 071503
  [arXiv:hep-ph/0211438].
  %%CITATION = PHRVA,D67,071503;%%

%\cite{Dawson:2003zu}
\bibitem{Dawson:2003zu}
  S.~Dawson, C.~Jackson, L.~H.~Orr, L.~Reina and D.~Wackeroth,
  %``Associated Higgs production with top quarks at the Large Hadron  Collider:
  %NLO QCD corrections,''
  Phys.\ Rev.\  D {\bf 68} (2003) 034022
  [arXiv:hep-ph/0305087].
  %%CITATION = PHRVA,D68,034022;%%



%%%%%%%%%%%%% ttbar in BSM %%%%%%%%%%%%%%%%%%

%\cite{Stange:1993td}
\bibitem{Stange:1993td}
  A.~Stange and S.~Willenbrock,
  %``Yukawa correction to top quark production at the Tevatron,''
  Phys.\ Rev.\  D {\bf 48} (1993) 2054
  [arXiv:hep-ph/9302291].
  %%CITATION = PHRVA,D48,2054;%%

%\cite{Zhou:1996dx}
\bibitem{Zhou:1996dx}
  H.~Y.~Zhou, C.~S.~Li and Y.~P.~Kuang,
  %``Yukawa corrections to top quark production at the 
  % LHC in  two-Higgs-doublet
  %models,''
  Phys.\ Rev.\  D {\bf 55} (1997) 4412
  [arXiv:hep-ph/9603435].
  %%CITATION = PHRVA,D55,4412;%%

%\cite{Hollik:1997hm}
\bibitem{Hollik:1997hm}
  W.~Hollik, W.~M.~M\"osle and D.~Wackeroth,
  %``Top pair production at hadron colliders in non-minimal standard models,''
  Nucl.\ Phys.\  B {\bf 516} (1998) 29
  [arXiv:hep-ph/9706218].
  %%CITATION = NUPHA,B516,29;%%


%\cite{Kao:1999kj}
\bibitem{Kao:1999kj}
  C.~Kao and D.~Wackeroth,
  %``Parity violating asymmetries in top pair 
  % production at hadron  colliders,''
  Phys.\ Rev.\  D {\bf 61} (2000) 055009
  [arXiv:hep-ph/9902202].
  %%CITATION = PHRVA,D61,055009;%%



%\cite{Berge:2007dz}
\bibitem{Berge:2007dz}
  S.~Berge, W.~Hollik, W.~M.~M\"osle and D.~Wackeroth,
  %``SUSY QCD one-loop effects in (un)polarized top-pair production at hadron
  %colliders,''
  Phys.\ Rev.\  D {\bf 76} (2007) 034016
  [arXiv:hep-ph/0703016].
  %%CITATION = PHRVA,D76,034016;%%


%\cite{Alam:1996mh}
\bibitem{Alam:1996mh}
  S.~Alam, K.~Hagiwara, S.~Matsumoto, K.~Hagiwara and S.~Matsumoto,
  %``One loop supersymmetric QCD radiative corrections to the top quark
  %production in p anti-p collisions. (Revised version),''
  Phys.\ Rev.\  D {\bf 55} (1997) 1307
  [arXiv:hep-ph/9607466].
  %%CITATION = PHRVA,D55,1307;%%

%\cite{Sullivan:1996ry}
\bibitem{Sullivan:1996ry}
  Z.~Sullivan,
  %``Supersymmetric QCD correction to top-quark production at the Tevatron,''
  Phys.\ Rev.\  D {\bf 56} (1997) 451
  [arXiv:hep-ph/9611302].
  %%CITATION = PHRVA,D56,451;%%

%\cite{Kim:1996nza}
\bibitem{Kim:1996nza}
  J.~Kim, J.~L.~Lopez, D.~V.~Nanopoulos and R.~Rangarajan,
  %``Enhanced supersymmetric corrections to top-quark production at the
  %Tevatron,''
  Phys.\ Rev.\  D {\bf 54} (1996) 4364
  [arXiv:hep-ph/9605419].
  %%CITATION = PHRVA,D54,4364;%%

%\cite{Yang:1996dma}
\bibitem{Yang:1996dma}
  J.~M.~Yang and C.~S.~Li,
  %``Top Squark Mixing Effects In The Supersymmetric Electroweak Corrections To
  %Top Quark Production At The Tevatron,''
  Phys.\ Rev.\  D {\bf 54} (1996) 4380
  [arXiv:hep-ph/9603442].
  %%CITATION = PHRVA,D54,4380;%%


%\cite{Ross:2007ez}
\bibitem{Ross:2007ez}
  D.~A.~Ross and M.~Wiebusch,
  %``MSSM Effects in Top-antitop Production at the LHC,''
  JHEP {\bf 0711} (2007) 041
  [arXiv:0707.4402 [hep-ph]].
  %%CITATION = JHEPA,0711,041;%%

%\cite{Li:1997gh}
%\bibitem{Li:1997gh}
%  C.~S.~Li, C.~P.~Yuan and H.~Y.~Zhou,
  %``Supersymmetric QCD parity nonconservation in top quark pairs at the
  %Tevatron,''
%  Phys.\ Lett.\ B {\bf 424} (1998) 76
%  [arXiv:hep-ph/9709275].
  %%CITATION = HEP-PH 9709275;%%


%\cite{Beccaria:2004sx}
%\bibitem{Beccaria:2004sx}
%  M.~Beccaria, S.~Bentvelsen, M.~Cobal, F.~M.~Renard and C.~Verzegnassi,
  %``Special supersymmetric features of large invariant mass unpolarized and
  %polarized top antitop production at LHC,''
%  Phys.\ Rev.\  D {\bf 71} (2005) 073003
%  [arXiv:hep-ph/0412249].
  %%CITATION = PHRVA,D71,073003;%%


%\cite{Pilaftsis:1998dd}
\bibitem{Pilaftsis:1998dd}
  A.~Pilaftsis,
  %``Higgs scalar-pseudoscalar mixing in the minimal supersymmetric standard
  %model,''
  Phys.\ Lett.\  B {\bf 435} (1998) 88
  [arXiv:hep-ph/9805373].
  %%CITATION = PHLTA,B435,88;%%

%\cite{Schmidt:1992et}
\bibitem{Schmidt:1992et}
  C.~R.~Schmidt and M.~E.~Peskin,
  %``A Probe of CP violation in top quark pair production at hadron
  %supercolliders,''
  Phys.\ Rev.\ Lett.\  {\bf 69} (1992) 410.
  %%CITATION = PRLTA,69,410;%%

%\cite{Bernreuther:1993df}
\bibitem{Bernreuther:1993df}
  W.~Bernreuther and A.~Brandenburg,
  %``Signatures of Higgs sector CP violation in top quark pair production at
  %proton proton supercolliders,''
  Phys.\ Lett.\  B {\bf 314} (1993) 104.
  %%CITATION = PHLTA,B314,104;%%

%\cite{Bernreuther:1993hq}
\bibitem{Bernreuther:1993hq}
  W.~Bernreuther and A.~Brandenburg,
  %``Tracing CP violation in the production of top quark pairs by multiple TeV
  %proton proton collisions,''
  Phys.\ Rev.\  D {\bf 49} (1994) 4481
  [arXiv:hep-ph/9312210].
  %%CITATION = PHRVA,D49,4481;%%


%\cite{Zhou:1998wz}
\bibitem{Zhou:1998wz}
  H.~Y.~Zhou,
  %``CP violation in top quark pair production at hadron colliders,''
  Phys.\ Rev.\  D {\bf 58} (1998) 114002
  [arXiv:hep-ph/9805358].
  %%CITATION = PHRVA,D58,114002;%%

%\cite{Khater:2003wq}
\bibitem{Khater:2003wq}
  W.~Khater and P.~Osland,
  %``CP violation in top quark production at the LHC and two-Higgs-doublet
  %models,''
  Nucl.\ Phys.\  B {\bf 661} (2003) 209
  [arXiv:hep-ph/0302004].
  %%CITATION = NUPHA,B661,209;%%



%\cite{Schmidt:1992kt}
\bibitem{Schmidt:1992kt}
  C.~R.~Schmidt,
  %``A Top quark CP violating asymmetry in supersymmetric models,''
  Phys.\ Lett.\  B {\bf 293} (1992) 111.
  %%CITATION = PHLTA,B293,111;%%
 


%\cite{Atwood:1994vm}
\bibitem{Atwood:1994vm}
  D.~Atwood, A.~Kagan and T.~G.~Rizzo,
  %``Constraining Anomalous Top Quark Couplings At The Tevatron,''
  Phys.\ Rev.\  D {\bf 52} (1995) 6264
  [arXiv:hep-ph/9407408].
  %%CITATION = PHRVA,D52,6264;%%

%\cite{Haberl:1995ek}
\bibitem{Haberl:1995ek}
  P.~Haberl, O.~Nachtmann and A.~Wilch,
  %``Top Production In Hadron Hadron Collisions And Anomalous Top - Gluon
  %Couplings,''
  Phys.\ Rev.\  D {\bf 53} (1996) 4875
  [arXiv:hep-ph/9505409].
  %%CITATION = PHRVA,D53,4875;%%

%\cite{Cheung:1996kc}
\bibitem{Cheung:1996kc}
  K.~m.~Cheung,
  %``Probing non-standard top couplings using spin-correlation,''
  Phys.\ Rev.\  D {\bf 55} (1997) 4430
  [arXiv:hep-ph/9610368].
  %%CITATION = PHRVA,D55,4430;%%




%\cite{Atwood:1992vj}
\bibitem{Atwood:1992vj}
  D.~Atwood, A.~Aeppli and A.~Soni,
  %``Extracting anomalous gluon - top effective couplings at the
  %supercolliders,''
  Phys.\ Rev.\ Lett.\  {\bf 69} (1992) 2754.
  %%CITATION = PRLTA,69,2754;%%


\bibitem{brama}
 A.~Brandenburg and J.~P.~Ma,
  %``CP odd observables for the top - anti-top system produced at proton -
  %anti-proton and proton proton colliders,''
  Phys.\ Lett.\ B {\bf 298} (1993) 211.
  %%CITATION = PHLTA,B298,211;%%


%\cite{Choi:1997ie}
\bibitem{Choi:1997ie}
  S.~Y.~Choi, C.~S.~Kim and J.~Lee,
  %``T-odd gluon top-quark effective couplings at the CERN Large Hadron
  %Collider,''
  Phys.\ Lett.\  B {\bf 415} (1997) 67
  [arXiv:hep-ph/9706379].
  %%CITATION = PHLTA,B415,67;%%

%\cite{Grzadkowski:1997yi}
\bibitem{Grzadkowski:1997yi}
  B.~Grzadkowski, B.~Lampe and K.~J.~Abraham,
  %``CP violation, top quarks and the Tevatron upgrade,''
  Phys.\ Lett.\ B {\bf 415} (1997) 193
  [arXiv:hep-ph/9706489].
  %%CITATION = HEP-PH 9706489;%% 




%\cite{Eriksson:2007fx}
\bibitem{Eriksson:2007fx}
  D.~Eriksson, G.~Ingelman, J.~Rathsman and O.~Stal,
  %``New angles on top quark decay to a charged Higgs,''
  JHEP {\bf 0801} (2008) 024
  [arXiv:0710.5906 [hep-ph]].
  %%CITATION = JHEPA,0801,024;%%




%%%%%%%%%%%%%%%%%%%%% heavy Higgs and CPV, heavy resonances %%%%%%%%%%%%%%%%%%%%55
%%%%%%%%%%%%%%%%%%%%%%%%%%%%%%%%%%%%%%%%%%%%%%%%%%%%%%%%%%%%%%%%%%%%%%%%%%

%\cite{Hill:1991at}
\bibitem{Hill:1991at}
  C.~T.~Hill,
  %``Topcolor: Top Quark Condensation In A Gauge Extension Of The Standard
  %Model,''
  Phys.\ Lett.\  B {\bf 266} (1991) 419.
  %%CITATION = PHLTA,B266,419;%%

%\cite{Hill:1993hs}
\bibitem{Hill:1993hs}
  C.~T.~Hill and S.~J.~Parke,
  %``Top production: Sensitivity to new physics,''
  Phys.\ Rev.\  D {\bf 49} (1994) 4454
  [arXiv:hep-ph/9312324].
  %%CITATION = PHRVA,D49,4454;%%

%\cite{ArkaniHamed:2001nc}
\bibitem{ArkaniHamed:2001nc}
  N.~Arkani-Hamed, A.~G.~Cohen and H.~Georgi,
  %``Electroweak symmetry breaking from dimensional deconstruction,''
  Phys.\ Lett.\  B {\bf 513} (2001) 232
  [arXiv:hep-ph/0105239].
  %%CITATION = PHLTA,B513,232;%%

%\cite{Schmaltz:2005ky}
\bibitem{Schmaltz:2005ky}
  M.~Schmaltz and D.~Tucker-Smith,
  %``Little Higgs review,''
  Ann.\ Rev.\ Nucl.\ Part.\ Sci.\  {\bf 55} (2005) 229
  [arXiv:hep-ph/0502182].
  %%CITATION = ARNUA,55,229;%%

%\cite{Antoniadis:1990ew}
\bibitem{Antoniadis:1990ew}
  I.~Antoniadis,
  %``A Possible new dimension at a few TeV,''
  Phys.\ Lett.\  B {\bf 246} (1990) 377.
  %%CITATION = PHLTA,B246,377;%%

%\cite{ArkaniHamed:1998rs}
\bibitem{ArkaniHamed:1998rs}
  N.~Arkani-Hamed, S.~Dimopoulos and G.~R.~Dvali,
  %``The hierarchy problem and new dimensions at a millimeter,''
  Phys.\ Lett.\  B {\bf 429} (1998) 263
  [arXiv:hep-ph/9803315].
  %%CITATION = PHLTA,B429,263;%%

%\cite{Randall:1999vf}
\bibitem{Randall:1999vf}
  L.~Randall and R.~Sundrum,
  %``An alternative to compactification,''
  Phys.\ Rev.\ Lett.\  {\bf 83} (1999) 4690
  [arXiv:hep-th/9906064].
  %%CITATION = PRLTA,83,4690;%%

%\cite{Agashe:2003zs}
\bibitem{Agashe:2003zs}
  K.~Agashe, A.~Delgado, M.~J.~May and R.~Sundrum,
  %``RS1, custodial isospin and precision tests,''
  JHEP {\bf 0308} (2003) 050
  [arXiv:hep-ph/0308036].
  %%CITATION = JHEPA,0308,050;%%

%\cite{Fitzpatrick:2007qr}
\bibitem{Fitzpatrick:2007qr}
  A.~L.~Fitzpatrick, J.~Kaplan, L.~Randall and L.~T.~Wang,
  %``Searching for the Kaluza-Klein graviton in bulk RS models,''
  JHEP {\bf 0709} (2007) 013
  [arXiv:hep-ph/0701150].
  %%CITATION = JHEPA,0709,013;%%

%\cite{Lillie:2007yh}
\bibitem{Lillie:2007yh}
  B.~Lillie, L.~Randall and L.~T.~Wang,
  %``The Bulk RS KK-gluon at the LHC,''
  JHEP {\bf 0709} (2007) 074
  [arXiv:hep-ph/0701166].
  %%CITATION = JHEPA,0709,074;%%



%\cite{Djouadi:2007eg}
\bibitem{Djouadi:2007eg}
  A.~Djouadi, G.~Moreau and R.~K.~Singh,
  ``Kaluza--Klein excitations of gauge bosons at the LHC,''
  arXiv:0706.4191 [hep-ph].
  %%CITATION = ARXIV:0706.4191;%%


%\cite{Lillie:2007ve}
\bibitem{Lillie:2007ve}
  B.~Lillie, J.~Shu and T.~M.~P.~Tait,
  %``Kaluza-Klein Gluons as a Diagnostic of Warped Models,''
  Phys.\ Rev.\  D {\bf 76} (2007) 115016
  [arXiv:0706.3960 [hep-ph]].
  %%CITATION = PHRVA,D76,115016;%%


%\cite{Gaemers:1984sj}
\bibitem{Gaemers:1984sj}
  K.~J.~F.~Gaemers and F.~Hoogeveen,
  %``Higgs Production And Decay Into Heavy Flavors With The Gluon Fusion
  %Mechanism,''
  Phys.\ Lett.\  B {\bf 146} (1984) 347.
  %%CITATION = PHLTA,B146,347;%%

%\cite{Dicus:1994bm}
\bibitem{Dicus:1994bm}
  D.~Dicus, A.~Stange and S.~Willenbrock,
  %``Higgs decay to top quarks at hadron colliders,''
  Phys.\ Lett.\ B {\bf 333} (1994) 126.
  %%CITATION = HEP-PH 9404359;%%

 
%\cite{Bernreuther:1997gs}
\bibitem{Bernreuther:1997gs}
  W.~Bernreuther, M.~Flesch and P.~Haberl,
  %``Signatures of Higgs bosons in the top quark decay channel at hadron
  %colliders,''
  Phys.\ Rev.\  D {\bf 58} (1998) 114031
  [arXiv:hep-ph/9709284].
  %%CITATION = PHRVA,D58,114031;%%

%\cite{Eichten:1994nc}
\bibitem{Eichten:1994nc}
  E.~Eichten and K.~D.~Lane,
  %``Multiscale technicolor and top production,''
  Phys.\ Lett.\  B {\bf 327} (1994) 129
  [arXiv:hep-ph/9401236].
  %%CITATION = PHLTA,B327,129;%%

%\cite{Choudhury:2007ux}
\bibitem{Choudhury:2007ux}
  D.~Choudhury, R.~M.~Godbole, R.~K.~Singh and K.~Wagh,
  %``Top production at the Tevatron / LHC and nonstandard, strongly interacting
  %spin one particles,''
  Phys.\ Lett.\  B {\bf 657} (2007) 69
  [arXiv:0705.1499 [hep-ph]].
  %%CITATION = PHLTA,B657,69;%%

%\cite{Arai:2004yd}
\bibitem{Arai:2004yd}
  M.~Arai, N.~Okada, K.~Smolek and V.~Simak,
  %``Top spin correlations in theories with large 
% extra-dimensions at the  Large
  %Hadron Collider,''
  Phys.\ Rev.\ D {\bf 70} (2004) 115015
  [{\tt hep-ph/0409273}].
  %%CITATION = HEP-PH 0409273;%%

%\cite{Arai:2007ts}
\bibitem{Arai:2007ts}
  M.~Arai, N.~Okada, K.~Smolek and V.~Simak,
  %``Top quark spin correlations in the Randall-Sundrum model 
  % at the CERN Large
  %Hadron Collider,''
  Phys.\ Rev.\  D {\bf 75} (2007) 095008
  [arXiv:hep-ph/0701155].
  %%CITATION = PHRVA,D75,095008;%%


%\cite{Antoniadis:1999bq}
\bibitem{Antoniadis:1999bq}
  I.~Antoniadis, K.~Benakli and M.~Quiros,
  %``Direct collider signatures of large extra dimensions,''
  Phys.\ Lett.\  B {\bf 460} (1999) 176
  [arXiv:hep-ph/9905311].
  %%CITATION = PHLTA,B460,176;%%

%
%\cite{Agashe:2007ki}
\bibitem{Agashe:2007ki}
  K.~Agashe {\it et al.},
  %``LHC Signals for Warped Electroweak Neutral Gauge Bosons,''
  Phys.\ Rev.\  D {\bf 76} (2007) 115015
  [arXiv:0709.0007 [hep-ph]].
  %%CITATION = PHRVA,D76,115015;%%


%\cite{Burdman:2006gy}
\bibitem{Burdman:2006gy}
  G.~Burdman, B.~A.~Dobrescu and E.~Ponton,
  %``Resonances from Two Universal Extra Dimensions,''
  Phys.\ Rev.\  D {\bf 74} (2006) 075008
  [arXiv:hep-ph/0601186].
  %%CITATION = PHRVA,D74,075008;%%

%\cite{Baur:2008uv}
\bibitem{Baur:2008uv}
  U.~Baur and L.~H.~Orr,
  ``Searching for $\ttbar$ Resonances at the Large Hadron Collider,''
  arXiv:0803.1160 [hep-ph].
  %%CITATION = ARXIV:0803.1160;%%

%\cite{Barger:2006hm}
\bibitem{Barger:2006hm}
  V.~Barger, T.~Han and D.~G.~E.~Walker,
  %``Top Quark Pairs at High Invariant Mass - A Model-Independent Discriminator
  %of New Physics at the LHC,''
  Phys.\ Rev.\ Lett.\  {\bf 100} (2008) 031801
  [arXiv:hep-ph/0612016].
  %%CITATION = PRLTA,100,031801;%%

%\cite{Han:2004zh}
\bibitem{Han:2004zh}
  T.~Han, G.~Valencia and Y.~Wang,
  %``Hadron collider signatures for new interactions of top and bottom
  %quarks,''
  Phys.\ Rev.\  D {\bf 70} (2004) 034002
  [arXiv:hep-ph/0405055].
  %%CITATION = PHRVA,D70,034002;%%

%\cite{Schwinn:2005qa}
\bibitem{Schwinn:2005qa}
  C.~Schwinn,
  %``Unitarity constraints on top quark signatures of Higgsless models,''
  Phys.\ Rev.\  D {\bf 71} (2005) 113005
  [arXiv:hep-ph/0504240].
  %%CITATION = PHRVA,D71,113005;%%


%\cite{Lillie:2007hd}
\bibitem{Lillie:2007hd}
  B.~Lillie, J.~Shu and T.~M.~P.~Tait,
  ``Top Compositeness at the Tevatron and LHC,''
  arXiv:0712.3057 [hep-ph].
  %%CITATION = ARXIV:0712.3057;%%

%\cite{Schwanenberger:2006tv}
\bibitem{Schwanenberger:2006tv}
  C.~Schwanenberger  [CDF and D0 Collaboration],
  %``Search for a new resonance decaying into top-antitop at Tevatron,''
  PoS {\bf HEP2005} (2006) 349
  [arXiv:hep-ex/0602048].
  %%CITATION = POSCI,HEP2005,349;%%

%\cite{Cabrera:2007ad}
\bibitem{Cabrera:2007ad}
  S.~Cabrera  [CDF collaboration],
  ``Looking for signals of New Physics in the Top quark samples with the CDF
   detector,''
  arXiv:0709.2264 [hep-ex].
  %%CITATION = ARXIV:0709.2264;%%


%\cite{Collins:1977iv}
\bibitem{Collins:1977iv}
  J.~C.~Collins and D.~E.~Soper,
  %``Angular Distribution Of Dileptons In High-Energy Hadron Collisions,''
  Phys.\ Rev.\  D {\bf 16} (1977) 2219.
  %%CITATION = PHRVA,D16,2219;%%

%\cite{Bernreuther:1998qv}
\bibitem{Bernreuther:1998qv}
  W.~Bernreuther, A.~Brandenburg and M.~Flesch,
  ``Effects of Higgs sector CP violation in top-quark pair production at  the
   LHC,''
  arXiv:hep-ph/9812387.
  %%CITATION = HEP-PH/9812387;%%


%\cite{Cohen:1993nk}
\bibitem{Cohen:1993nk}
  A.~G.~Cohen, D.~B.~Kaplan and A.~E.~Nelson,
  %``Progress in electroweak baryogenesis,''
  Ann.\ Rev.\ Nucl.\ Part.\ Sci.\  {\bf 43} (1993) 27
  [arXiv:hep-ph/9302210].
  %%CITATION = ARNUA,43,27;%%

%\cite{Accomando:2006ga}
\bibitem{Accomando:2006ga}
  E.~Accomando {\it et al.},
  ``Workshop on CP studies and non-standard Higgs physics,''
  arXiv:hep-ph/0608079.
  %%CITATION = HEP-PH/0608079;%%



%%%%%%%%%%%%%%%%%%%%%%%%%%%%%%%%%%%%%%%% single top %%%%%%%%%%%
%%%%%%%%%%%%%%%%
%%%%%%%%%%%%%%%%%%%%%%%%%%%%%%%%%%%%%%%%
%%%%%%%%%%%%%%%%%%%%%%%%%%%%%%%%%%%%%%%%
%%%%%%%%%%%%%%%%%%%%%%%%%%%%%%%%%%%%%%%%
%%%%%%%%%%%%%%%%%%%%%%%%%%%%%%%%%%%%%%%% %%%%%%%%%%%%%%%%%%%%%%%%%%%%%%%


%\cite{Aivazis:1993pi}
\bibitem{Aivazis:1993pi}
  M.~A.~G.~Aivazis, J.~C.~Collins, F.~I.~Olness and W.~K.~Tung,
  %``Leptoproduction of heavy quarks. 2. A Unified QCD formulation of charged
  %and neutral current processes from fixed target to collider energies,''
  Phys.\ Rev.\  D {\bf 50}, 3102 (1994)
  [arXiv:hep-ph/9312319].
  %%CITATION = PHRVA,D50,3102;%%


%%%%%%%%%%%%%%%%%%%%%%%%%%%%%%%%% single t production in SM %%%%%%%%%
%%%%%%%%%%%%%%%%%%%%%%%% 

%\cite{Stelzer:1995mi}
%\bibitem{Stelzer:1995mi}
%  T.~Stelzer and S.~Willenbrock,
  %``Single top quark production via q anti-q $\to$ t anti-b,''
%  Phys.\ Lett.\  B {\bf 357} (1995) 125
%  [arXiv:hep-ph/9505433].
  %%CITATION = PHLTA,B357,125;%%



%%%%%%%%%%%%%%%%%%%%% t channel %%%%%%%%%%%

%\cite{Bordes:1994ki}
\bibitem{Bordes:1994ki}
  G.~Bordes and B.~van Eijk,
  %``Calculating QCD corrections to single top production in hadronic
  %interactions,''
  Nucl.\ Phys.\  B {\bf 435} (1995) 23.
  %%CITATION = NUPHA,B435,23;%%

%\cite{Stelzer:1997ns}
\bibitem{Stelzer:1997ns}
  T.~Stelzer, Z.~Sullivan and S.~Willenbrock,
  %``Single-top-quark production via W-gluon fusion at next-to-leading
  % order,''
  Phys.\ Rev.\  D {\bf 56} (1997) 5919
  [arXiv:hep-ph/9705398].
  %%CITATION = PHRVA,D56,5919;%%

%\cite{Stelzer:1998ni}
\bibitem{Stelzer:1998ni}
  T.~Stelzer, Z.~Sullivan and S.~Willenbrock,
  %``Single top quark production at hadron colliders,''
  Phys.\ Rev.\  D {\bf 58} (1998) 094021
  [arXiv:hep-ph/9807340].
  %%CITATION = PHRVA,D58,094021;%%


%\cite{Harris:2002md}
\bibitem{Harris:2002md}
  B.~W.~Harris, E.~Laenen, L.~Phaf, Z.~Sullivan and S.~Weinzierl,
  %``The fully differential single top quark cross section in  next-to-leading
  %order QCD,''
  Phys.\ Rev.\  D {\bf 66} (2002) 054024
  [arXiv:hep-ph/0207055].
  %%CITATION = PHRVA,D66,054024;%%


%\cite{Sullivan:2004ie}
\bibitem{Sullivan:2004ie}
  Z.~Sullivan,
  %``Understanding single-top-quark production and jets at hadron
%colliders,''
Phys.\ Rev.\  D {\bf 70} (2004) 114012
  [arXiv:hep-ph/0408049].
  %%CITATION = PHRVA,D70,114012;%%


  %\cite{Sullivan:2005ar}
\bibitem{Sullivan:2005ar}
  Z.~Sullivan,
  %``Angular correlations in single-top-quark and W j j production at
  %next-to-leading order,''
  Phys.\ Rev.\  D {\bf 72} (2005) 094034
  [arXiv:hep-ph/0510224].
  %%CITATION = PHRVA,D72,094034;%%



%%%% s, t channel prod. and decay  %%%%%%%%

%\cite{Campbell:2004ch}
\bibitem{Campbell:2004ch}
  J.~Campbell, R.~K.~Ellis and F.~Tramontano,
  %``Single top production and decay at next-to-leading order,''
  Phys.\ Rev.\  D {\bf 70} (2004) 094012
  [arXiv:hep-ph/0408158].
  %%CITATION = PHRVA,D70,094012;%%


%\cite{Cao:2004ky}
\bibitem{Cao:2004ky}
  Q.~H.~Cao and C.~P.~Yuan,
  %``Single top quark production and decay at next-to-leading order in  hadron
  %collision,''
  Phys.\ Rev.\  D {\bf 71} (2005) 054022
  [arXiv:hep-ph/0408180].
  %%CITATION = PHRVA,D71,054022;%%


%\cite{Cao:2005pq}
\bibitem{Cao:2005pq}
  Q.~H.~Cao, R.~Schwienhorst, J.~A.~Benitez, R.~Brock and C.~P.~Yuan,
  %``Next-to-leading order corrections to single top 
  % quark production and  decay
  %at the Tevatron. II: t-channel process,''
  Phys.\ Rev.\  D {\bf 72} (2005) 094027
  [arXiv:hep-ph/0504230].
  %%CITATION = PHRVA,D72,094027;%%




 %\cite{Frixione:2005vw}
\bibitem{Frixione:2005vw}
  S.~Frixione, E.~Laenen, P.~Motylinski and B.~R.~Webber,
  %``Single-top production in MC@NLO,''
  JHEP {\bf 0603} (2006) 092
  [arXiv:hep-ph/0512250].
  %%CITATION = JHEPA,0603,092;%%

%\cite{Beccaria:2008av}
\bibitem{Beccaria:2008av}
  M.~Beccaria et al.,
  ``A complete one-loop calculation of electroweak supersymmetric effects in
   $t$-channel single top production at LHC,''
  arXiv:0802.1994 [hep-ph].
  %%CITATION = ARXIV:0802.1994;%%

%\cite{Beccaria:2006ir}
\bibitem{Beccaria:2006ir}
  M.~Beccaria, G.~Macorini, F.~M.~Renard and C.~Verzegnassi,
  %``Single top production in the t-channel at LHC: A realistic test of
  %electroweak models,''
  Phys.\ Rev.\  D {\bf 74} (2006) 013008
  [arXiv:hep-ph/0605108].
  %%CITATION = PHRVA,D74,013008;%%



%\cite{Kidonakis:2006bu}
\bibitem{Kidonakis:2006bu}
  N.~Kidonakis,
  %``Single top production at the Tevatron: Threshold resummation and
  %finite-order soft gluon corrections,''
  Phys.\ Rev.\  D {\bf 74} (2006) 114012
  [arXiv:hep-ph/0609287].
  %%CITATION = PHRVA,D74,114012;%%


%\cite{Kidonakis:2007ej}
\bibitem{Kidonakis:2007ej}
  N.~Kidonakis,
  %``Higher-order soft gluon corrections 
% in single top quark production at   the
  %LHC,''
  Phys.\ Rev.\  D {\bf 75} (2007) 071501
  [arXiv:hep-ph/0701080].
  %%CITATION = PHRVA,D75,071501;%%



%%%%%%%% s channel %%%%%%%%%%%%%

%\cite{Smith:1996ij}
\bibitem{Smith:1996ij}
  M.~C.~Smith and S.~Willenbrock,
  %``QCD and Yukawa Corrections to Single-Top-Quark Production via q qbar -> t
  %bbar,''
  Phys.\ Rev.\  D {\bf 54} (1996) 6696
  [arXiv:hep-ph/9604223].
  %%CITATION = PHRVA,D54,6696;%%


%\cite{Cao:2004ap}
\bibitem{Cao:2004ap}
  Q.~H.~Cao, R.~Schwienhorst and C.~P.~Yuan,
  %``Next-to-leading order corrections to single top quark 
% production and  decay
  %at Tevatron. I: s-channel process,''
  Phys.\ Rev.\  D {\bf 71} (2005) 054023
  [arXiv:hep-ph/0409040].
  %%CITATION = PHRVA,D71,054023;%%


%\cite{Mrenna:1997wp}
\bibitem{Mrenna:1997wp}
  S.~Mrenna and C.~P.~Yuan,
  %``Effects of QCD resummation on W+ h and t anti-b production at the
  %Tevatron,''
  Phys.\ Lett.\  B {\bf 416} (1998) 200
  [arXiv:hep-ph/9703224].
  %%CITATION = PHLTA,B416,200;%%


%%%%%%%%%% t W mode %%%%%%%%%%%%

%\cite{Belyaev:1998dn}
\bibitem{Belyaev:1998dn}
  A.~S.~Belyaev, E.~E.~Boos and L.~V.~Dudko,
  %``Single top quark at future hadron colliders: Complete signal and
  %background study,''
  Phys.\ Rev.\  D {\bf 59} (1999) 075001
  [arXiv:hep-ph/9806332].
  %%CITATION = PHRVA,D59,075001;%%

%\cite{Belyaev:2000me}
\bibitem{Belyaev:2000me}
  A.~Belyaev and E.~Boos,
  %``Single top quark t W + X production at the LHC: A closer look,''
  Phys.\ Rev.\  D {\bf 63} (2001) 034012
  [arXiv:hep-ph/0003260].
  %%CITATION = PHRVA,D63,034012;%%

%\cite{Tait:1999cf}
\bibitem{Tait:1999cf}
  T.~M.~P.~Tait,
  %``The t W- mode of single top production,''
  Phys.\ Rev.\  D {\bf 61} (2000) 034001
  [arXiv:hep-ph/9909352].
  %%CITATION = PHRVA,D61,034001;%%


%\cite{Campbell:2005bb}
\bibitem{Campbell:2005bb}
  J.~Campbell and F.~Tramontano,
  %``Next-to-leading order corrections to W t production and decay,''
  Nucl.\ Phys.\  B {\bf 726} (2005) 109
  [arXiv:hep-ph/0506289].
  %%CITATION = NUPHA,B726,109;%%


%\cite{Giele:1995kr}
\bibitem{Giele:1995kr}
  W.~T.~Giele, S.~Keller and E.~Laenen,
  %``QCD Corrections to W Boson plus Heavy Quark Production at the Tevatron,''
  Phys.\ Lett.\  B {\bf 372} (1996) 141
  [arXiv:hep-ph/9511449].
  %%CITATION = PHLTA,B372,141;%%

%\cite{Zhu:2002uj}
\bibitem{Zhu:2002uj}
  S.~Zhu,
  %``Next-to-leading order QCD corrections to b g $\to$ t W- at the CERN Large
  %Hadron Collider,''
  Phys.\ Lett.\  B {\bf 524} (2002) 283
  [Erratum-ibid.\  B {\bf 537} (2002) 351].
  %%CITATION = PHLTA,B524,283;%%



%%%%%%%%%%%%%%%%%%%%%%%%% single top, spin %%%%%%%%%%%%%%%%%

%\cite{Mahlon:1996pn}
\bibitem{Mahlon:1996pn}
  G.~Mahlon and S.~J.~Parke,
  %``Improved spin basis for angular correlation studies in single top quark
  %production at the Tevatron,''
  Phys.\ Rev.\  D {\bf 55} (1997) 7249
  [arXiv:hep-ph/9611367].
  %%CITATION = PHRVA,D55,7249;%%


%\cite{Mahlon:1999gz}
\bibitem{Mahlon:1999gz}
  G.~Mahlon and S.~J.~Parke,
  %``Single top quark production at the LHC: Understanding spin,''
  Phys.\ Lett.\  B {\bf 476} (2000) 323
  [arXiv:hep-ph/9912458].
  %%CITATION = PHLTA,B476,323;%%

%\cite{Boos:2002xw}
\bibitem{Boos:2002xw}
  E.~E.~Boos and A.~V.~Sherstnev,
  %``Spin effects in processes of single top quark production at hadron
  %colliders,''
  Phys.\ Lett.\  B {\bf 534} (2002) 97
  [arXiv:hep-ph/0201271].
  %%CITATION = PHLTA,B534,97;%%


%%%%%%%%%%%%%%%%%%%%% new physics in single top %%%%%%%%%%%%%%%%

%\cite{AguilarSaavedra:2008gt}
\bibitem{AguilarSaavedra:2008gt}
  J.~A.~Aguilar-Saavedra,
  ``Single top quark production at LHC with anomalous Wtb couplings,''
  arXiv:0803.3810 [hep-ph].
  %%CITATION = ARXIV:0803.3810;%%


%\cite{Yue:2006qx}
\bibitem{Yue:2006qx}
  C.~X.~Yue, L.~Zhou and S.~Yang,
  %``Little Higgs models and single top quark production at the LHC,''
  Eur.\ Phys.\ J.\  C {\bf 48} (2006) 243
  [arXiv:hep-ph/0604001].
  %%CITATION = EPHJA,C48,243;%%

%\cite{Cao:2007ea}
\bibitem{Cao:2007ea}
  Q.~H.~Cao, J.~Wudka and C.~P.~Yuan,
  %``Search for New Physics via Single Top Production at the LHC,''
  Phys.\ Lett.\  B {\bf 658}, 50 (2007)
  [arXiv:0704.2809 [hep-ph]].
  %%CITATION = PHLTA,B658,50;%%

%\cite{Zhang:2006cx}
\bibitem{Zhang:2006cx}
  J.~J.~Zhang, C.~S.~Li, Z.~Li and L.~L.~Yang,
  %``Supersymmetric QCD corrections to single top quark production at hadron
  %colliders,''
  Phys.\ Rev.\  D {\bf 75} (2007) 014020
  [arXiv:hep-ph/0610087].
  %%CITATION = PHRVA,D75,014020;%%

%\cite{Beccaria:2007tc}
\bibitem{Beccaria:2007tc}
  M.~Beccaria et al.,
  %``A complete one-loop description of associated tW production at LHC and a
  %search for possible genuine supersymmetric effects,''
  Eur.\ Phys.\ J.\  C {\bf 53} (2008) 257
  [arXiv:0705.3101 [hep-ph]].
  %%CITATION = EPHJA,C53,257;%%

%%%%%%%%%%%%%%% single top   %%%%%%%%%%%%%%%%%%%%%


%\cite{Pati:1974yy}
\bibitem{Pati:1974yy}
  J.~C.~Pati and A.~Salam,
  %``Lepton Number As The Fourth Color,''
  Phys.\ Rev.\  D {\bf 10} (1974) 275
  [Erratum-ibid.\  D {\bf 11} (1975) 703].
  %%CITATION = PHRVA,D10,275;%%

%\cite{Mohapatra:1974hk}
\bibitem{Mohapatra:1974hk}
  R.~N.~Mohapatra and J.~C.~Pati,
  %``Left-Right Gauge Symmetry And An Isoconjugate Model Of CP Violation,''
  Phys.\ Rev.\  D {\bf 11} (1975) 566.
  %%CITATION = PHRVA,D11,566;%%

%\cite{Mohapatra:1974gc}
\bibitem{Mohapatra:1974gc}
  R.~N.~Mohapatra and J.~C.~Pati,
  %``A Natural Left-Right Symmetry,''
  Phys.\ Rev.\  D {\bf 11} (1975) 2558.
  %%CITATION = PHRVA,D11,2558;%%


%\cite{Simmons:1996ws}
\bibitem{Simmons:1996ws}
  E.~H.~Simmons,
  %``New gauge interactions and single top quark production,''
  Phys.\ Rev.\  D {\bf 55} (1997) 5494
  [arXiv:hep-ph/9612402].
  %%CITATION = PHRVA,D55,5494;%%

%\cite{He:1998ie}
\bibitem{He:1998ie}
  H.~J.~He and C.~P.~Yuan,
  %``New method for detecting charged (pseudo-)scalars at colliders,''
  Phys.\ Rev.\ Lett.\  {\bf 83} (1999) 28
  [arXiv:hep-ph/9810367].
  %%CITATION = PRLTA,83,28;%%

%\cite{Tait:2000sh}
\bibitem{Tait:2000sh}
  T.~Tait and C.~P.~Yuan,
  %``Single top quark production as a window to physics beyond the Standard
  %Model,''
  Phys.\ Rev.\  D {\bf 63} (2001) 014018
  [arXiv:hep-ph/0007298].
  %%CITATION = PHRVA,D63,014018;%%


%\cite{Boos:2006xe}
\bibitem{Boos:2006xe}
  E.~Boos, V.~Bunichev, L.~Dudko and M.~Perfilov,
  %``Interference between W' and W in single-top quark production processes,''
  Phys.\ Lett.\  B {\bf 655} (2007) 245
  [arXiv:hep-ph/0610080].
  %%CITATION = PHLTA,B655,245;%%

%\cite{Magass:2007zz}
\bibitem{Magass:2007zz}
  C.~M.~Magass,
  ``Search for new heavy charged gauge bosons,''
   doctoral thesis, RWTH Aachen (2007).
  %%CITATION = FERMILAB-THESIS-2007-33;%%


%%%%%%%%%%%% t + charged Higgs %%%%%%%%%%

%\cite{Maltoni:2001hu}
\bibitem{Maltoni:2001hu}
  F.~Maltoni, K.~Paul, T.~Stelzer and S.~Willenbrock,
  %``Associated production of Higgs and single top at hadron colliders,''
  Phys.\ Rev.\  D {\bf 64} (2001) 094023
  [arXiv:hep-ph/0106293].
  %%CITATION = PHRVA,D64,094023;%%




%%%%%%%%%%%%%%%%%%%%%%%%%%%%%%%% FCNC in (single) top %%%%%%%%%%%%%%
%%%                       and in tt production  %%%%%%%%%%%%%%%%%%%%

%\cite{Han:1995pk}
\bibitem{Han:1995pk}
  T.~Han, R.~D.~Peccei and X.~Zhang,
  %``Top Quark Decay Via Flavor Changing Neutral 
  % Currents At Hadron Colliders,''
  Nucl.\ Phys.\  B {\bf 454} (1995) 527
  [arXiv:hep-ph/9506461].
  %%CITATION = NUPHA,B454,527;%%

%% chromomagnetic anomlous FCNC coulings%



%\cite{Han:1998tp}
\bibitem{Han:1998tp}
  T.~Han, M.~Hosch, K.~Whisnant, B.~L.~Young and X.~Zhang,
  %``Single top quark production via FCNC couplings at hadron colliders,''
  Phys.\ Rev.\  D {\bf 58} (1998) 073008
  [arXiv:hep-ph/9806486].
  %%CITATION = PHRVA,D58,073008;%%




%%% this is model-idendendent incl. QCD coorections 


%%%%%%%%%%%%%%%%%%%%%%%%%%%%%%%%%%%%


%\cite{Larios:2003jq}
\bibitem{Larios:2003jq}
  F.~Larios and F.~Penunuri,
  %``FCNC production of same sign top quark pairs at the LHC,''
  J.\ Phys.\ G {\bf 30} (2004) 895
  [arXiv:hep-ph/0311056].
  %%CITATION = JPHGB,G30,895;%%

%\cite{Liu:2005dp}
\bibitem{Liu:2005dp}
  J.~J.~Liu, C.~S.~Li, L.~L.~Yang and L.~G.~Jin,
  %``Next-to-leading order QCD corrections to the direct top quark  production
  %via model-independent FCNC couplings at hadron colliders,''
  Phys.\ Rev.\  D {\bf 72} (2005) 074018
  [arXiv:hep-ph/0508016].
  %%CITATION = PHRVA,D72,074018;%%


%\cite{Liu:2004bb}
\bibitem{Liu:2004bb}
  J.~J.~Liu, C.~S.~Li, L.~L.~Yang and L.~G.~Jin,
  %``Single top quark production via SUSY-QCD FCNC couplings at 
% the CERN LHC  in
  %the unconstrained MSSM,''
  Nucl.\ Phys.\  B {\bf 705} (2005) 3
  [arXiv:hep-ph/0404099].
  %%CITATION = NUPHA,B705,3;%%


%\cite{LopezVal:2007rc}
\bibitem{LopezVal:2007rc}
  D.~Lopez-Val, J.~Guasch and J.~Sola,
  %``Single top-quark production by strong and electroweak supersymmetric
  %flavor-changing interactions at the LHC,''
  JHEP {\bf 0712} (2007) 054
  [arXiv:0710.0587 [hep-ph]].
  %%CITATION = JHEPA,0712,054;%%


%\cite{Cao:2007bx}
\bibitem{Cao:2007bx}
  J.~j.~Cao, G.~l.~Liu, J.~M.~Yang and H.~j.~Zhang,
  %``Top-quark FCNC productions at LHC in topcolour-assisted 
% technicolor  model,''
  Phys.\ Rev.\  D {\bf 76} (2007) 014004
  [arXiv:hep-ph/0703308].
  %%CITATION = PHRVA,D76,014004;%%

%\cite{Gouz:1998rk}
\bibitem{Gouz:1998rk}
  Yu.~P.~Gouz and S.~R.~Slabospitsky,
  %``Double top production at hadronic colliders,''
  Phys.\ Lett.\  B {\bf 457} (1999) 177
  [arXiv:hep-ph/9811330].
  %%CITATION = PHLTA,B457,177;%%



\end{thebibliography}
\end{document}